\documentclass{LMCS}

\def\dOi{10(1:16)2014}
\lmcsheading%
{\dOi}
{1--42}
{}
{}
{Mar.~\phantom04, 2013}
{Mar.~\phantom03, 2014}
{}

\subjclass{Models of Computation, Probabilistic Computation,
  Concurrency, Process Calculi}

\ACMCCS{[{\bf Theory of computation}]: Models of
  Computation---Probabilistic Computation;  Models of
  computation---Concurrency---Process calculi} 

\usepackage{amsmath,amsfonts,amstext,amssymb,mathrsfs,stmaryrd}
\usepackage{latexsym}
\usepackage{graphicx}
\usepackage{color}
\usepackage{proof}
\usepackage{hyperref}

\def\ms#1{\null\ifmmode\mathord{\mathcode`-="702D\it #1\mathcode`\-="2200}%
	\else$\mathord{\mathcode`-="702D\it #1\mathcode`\-="2200}$\fi}

\newcommand{\cws}[2]
	{\\ \centerline{$#2$} \\[-#1pt]}

\newlength{\spacelen}

\newcommand{\bibtrick}[1]
	{}

\newcommand{\cala}
        {\mathcal{A}}

\newcommand{\calc}
        {\mathcal{C}}

\newcommand{\calcc}
        {\mathcal{CC}}

\newcommand{\cald}
        {\mathcal{D}}

\newcommand{\calfcc}
        {\mathcal{FCC}}

\newcommand{\calftcc}
        {\mathcal{FTCC}}

\newcommand{\cali}
        {\mathcal{I}}

\newcommand{\call}
        {\mathcal{L}}

\newcommand{\calrtcc}
        {\mathcal{RTCC}}

\newcommand{\calsc}
        {\mathcal{SC}}

\newcommand{\calscc}
        {\mathcal{SCC}}

\newcommand{\calt}
        {\mathcal{T}}

\newcommand{\calz}
        {\mathcal{Z}}

\newcommand{\natns}
	{\mathbb{N}}

\newcommand{\realns}
	{\mathbb{R}}

\newcommand{\step}[2]
        {\, {\auxstep\limits^{#1}}_{#2} \,}
\newcommand{\auxstep}
	{\mathop{-\hspace{-0.15cm}\mapsto}}

\newcommand{\arrow}[2]
        {\, {\auxarrow\limits^{#1}}_{#2} \,}
\newcommand{\auxarrow}
	{\mathop{\longrightarrow}}

\newcommand{\sbis}[1]
	{\sim_{#1}}

\newcommand{\pco}[1]
	{\mathop{\Vert_{#1}}}

\newcommand{\fullbox}
	{{\mbox{}\nolinebreak\hfill{$\rule{2mm}{2mm}$}}}

\newcommand{\ultras}
	{{\sc ULTraS}}

\newif\ifwithtikz
\withtikzfalse
\ifwithtikz
\usepackage{tikz}
\usetikzlibrary{calc,arrows,shapes,chains,matrix,positioning,scopes,decorations.pathmorphing,shadows,fit,
		mindmap,backgrounds}
\usetikzlibrary{external}
\fi

\begin{document}

\title[Revisiting Trace and Testing Equivalences for Nondet.\ and Prob.\ Processes]
      {Revisiting Trace and Testing Equivalences \\
       for Nondeterministic and Probabilistic Processes}

\author[M.~Bernardo]{Marco Bernardo\rsuper a}
\address{{\lsuper a}Dipartimento di Scienze di Base e Fondamenti -- Universit\`a di Urbino -- Italy}
\email{marco.bernardo@uniurb.it}

\author[R.~De~Nicola]{Rocco De Nicola\rsuper b}
\address{{\lsuper b}IMT -- Institute for Advanced Studies Lucca -- Italy}
\email{rocco.denicola@imtlucca.it}

\author[M.~Loreti]{Michele Loreti\rsuper c}
\address{{\lsuper c}Dipartimento di Statistica, Informatica, Applicazioni -- Universit\`a di Firenze -- Italy}
\email{michele.loreti@unifi.it}


\begin{abstract}
  Two of the most studied extensions of trace and testing equivalences
  to nondeterministic and probabilistic processes induce distinctions
  that have been questioned and lack properties that are desirable.
  Probabilistic trace-distribution equivalence differentiates systems
  that can perform the same set of traces with the same probabilities,
  and is not a congruence for parallel composition. Probabilistic
  testing equivalence, which relies only on extremal success
  probabilities, is backward compatible with testing equivalences for
  restricted classes of processes, such as fully nondeterministic
  processes or generative/reactive probabilistic processes, only if
  specific sets of tests are admitted. In this paper, new versions of
  probabilistic trace and testing equivalences are presented for the
  general class of nondeterministic and probabilistic processes. The
  new trace equivalence is coarser because it compares execution
  probabilities of single traces instead of entire trace
  distributions, and turns out to be compositional. The new testing
  equivalence requires matching all resolutions of nondeterminism on
  the basis of their success probabilities, rather than comparing only
  extremal success probabilities, and considers success probabilities
  in a trace-by-trace fashion, rather than cumulatively on entire
  resolutions. It is fully backward compatible with testing
  equivalences for restricted classes of processes; as a consequence,
  the trace-by-trace approach uniformly captures the standard
  probabilistic testing equivalences for generative and reactive
  probabilistic processes. The paper discusses in full details the new
  equivalences and provides a simple spectrum that relates them with
  existing ones in the setting of nondeterministic and probabilistic
  processes.
\end{abstract}


\maketitle

%
%
\section{Introduction}
\label{sec:intro}
%
%

Modeling and abstraction are two key concepts of computer science that go hand in hand. If we wish to model
a computer system for the purpose of (computer-aided) analysis, it is essential that the right level of
abstraction is chosen when describing system behaviors. Operational models based on variants of automata or
labeled transition systems very often provide descriptions that are too detailed; it is then necessary to
resort to additional machineries to abstract from unwanted details. Behavioral equivalences are one of such
machineries and indeed many equivalences have been proposed depending on the specific aspects of systems
descriptions to ignore or the specific properties to capture. Equivalences are used to assess the
relationships between different views of the same system. If both the specification and the implementation
of a concurrent system are described via the same formalism, then the correctness of the latter with respect
to the former can be established by studying their behavioral relationships.

Behavioral equivalences were first of all defined for labeled transition systems (LTS -- set of states
related via transitions each labeled with the action that gives rise to the state change~\cite{Kel76}) that
were used as models of nonderministic processes. Then, they have been extended/adapted to generalizations of
such models to take into account probabilistic, stochastic, or timed behaviors.

Among the most important equivalences defined for abstracting unnecessary details of nondeterministic
processes modeled as LTS, we would like to mention the following three:

	\begin{itemize}

\item \emph{trace equivalence}, equating systems performing the same sequences of actions, 

\item \emph{testing equivalence}, equating systems reacting similarly to external experiments by peer
systems, and

\item \emph{bisimulation equivalence}, equating systems performing the same sequences of actions and
recursively exhibiting the same behavior after them.

	\end{itemize}

\noindent
Studies about their relationships have shown that the first equivalence is coarser than the second one,
which in turn is coarser than the third one. A coarser equivalence provides a more abstract view of a system
and produces more identifications.

When probabilities enter the game and probabilistic extensions of LTS are considered, the possible
alternatives in choosing what to observe and compare, in deciding how to resolve nondeterminism, or in
assembling the results of the observations are very many and the different choices can give rise to
significantly different behavioral relations. Indeed, many proposals have been put forward and discussion is
still going on about whether the identifications that these relations induce do capture the intuition one
has in mind about the wanted behavior of probabilistic descriptions.

In this paper, we would like to concentrate on probabilistic trace and testing equivalences for processes
described by means of an extension of the LTS model that combines nondeterminism and probabilities. The
extended model, which we have thus called NPLTS, is such that every action-labeled transition goes from a
source state to a probability distribution over target states -- in the style of~\cite{LS91,Seg95a} --
rather than to a single target state.

The most used definition of probabilistic trace equivalence for nondeterministic and probabilistic processes
is the one provided in~\cite{Seg95b}. To resolve nondeterminism, it resorts to the notion of
\emph{scheduler} (or adversary), which can be viewed as an external entity that selects the next action to
perform according to the current state and the past history. When a scheduler is applied to a process, a
fully probabilistic model called a \emph{resolution} is obtained. Two processes are considered trace
equivalent if, for each resolution of any of the two processes, there exists a resolution of the other
process such that the probability of \emph{each trace} is the same in the two resolutions. In other words,
the two resolutions must exhibit the \emph{same trace distribution}. We shall denote this equivalence by
$\sbis{\rm PTr,dis}$.

Testing equivalence for the same class of processes has been studied in~\cite{YL92,JY95,Seg96,DGHM08}. It
considers the probability of performing computations along which the same tests are passed, called
successful computations. Due to the possible presence of equally labeled transitions departing from the same
state, there is not necessarily a single probability value with which a nondeterministic and probabilistic
process passes a test. Given two states $s_{1}$ and $s_{2}$ and the initial state $o$ of an observer, this
testing equivalence computes the probability of performing a successful computation from $(s_{1}, o)$ and
$(s_{2}, o)$ in every maximal resolution of the interaction system resulting from the parallel composition
of each process with the observer. Then, it compares \emph{only extremal success probabilities}, i.e., the
suprema ($\sqcup$) and the infima ($\sqcap$) of the success probabilities over all maximal resolutions of
the two interaction systems. We shall denote this equivalence by $\sbis{\textrm{PTe-}\sqcup\sqcap}$.

After examining the above mentioned trace and testing equivalences for nondeterministic and probabilistic
processes, we noticed that both equivalences induce differentiations that might be questionable and lack
properties that are in general desirable.

For the equivalence~$\sbis{\rm PTr,dis}$, we have that it considers as inequivalent the two processes in
Fig.~\ref{fig:counterex_ptrdis_ptr} (p.~\pageref{fig:counterex_ptrdis_ptr}), in spite of the fact that they
can undoubtedly exhibit the same set of traces with the same probabilities. Moreover, $\sbis{\rm PTr,dis}$
is not preserved by parallel composition. As shown in~\cite{Seg95b}, given two $\sbis{\rm
PTr,dis}$-equivalent processes and given a third process, it is not necessarily the case that the parallel
composition of the first process with the third one is $\sbis{\rm PTr,dis}$-equivalent to the parallel
composition of the second process with the third one.

The equivalence $\sbis{\textrm{PTe-}\sqcup\sqcap}$, instead, identifies the two processes in
Fig.~\ref{fig:counterex_ptesupinf_trace} (p.~\pageref{fig:counterex_ptesupinf_trace}) mainly because its
definition only considers extremal success probabilities. A consequence of such a choice is that this
testing equivalence, contrary to what happens for the purely nondeterministic case, does not imply the trace
equivalence $\sbis{\rm PTr,dis}$. Indeed, the two processes in Fig.~\ref{fig:counterex_ptesupinf_trace},
which are identified by $\sbis{\textrm{PTe-}\sqcup\sqcap}$, are distinguished by $\sbis{\rm PTr,dis}$.
Actually, the inclusion depends on the type of schedulers used for deriving resolutions of nondeterminism;
it holds if randomized schedulers are admitted for $\sbis{\rm PTr,dis}$ as in~\cite{Seg95b}, while it does
not hold if only deterministic schedulers are considered.

Another characteristic of $\sbis{\textrm{PTe-}\sqcup\sqcap}$ is that of being only \emph{partially} backward
compatible with existing testing equivalences for restricted classes of processes. Compatibility depends on
the set of admitted tests. For example, the two fully nondeterministic processes in
Fig.~\ref{fig:counterex_tefnd_testing} (p.~\pageref{fig:counterex_tefnd_testing}) are identified by the
original testing equivalence of~\cite{DH84}. The relation $\sbis{\textrm{PTe-}\sqcup\sqcap}$ equates them if
only fully nondeterministic tests are employed, but distinguishes them as soon as probabilities are admitted
within tests. Dually, following the terminology of~\cite{GSS95}, the two generative/reactive probabilistic
processes in Fig.~\ref{fig:counterex_tepr_testing} (p.~\pageref{fig:counterex_tepr_testing}), which are
identified by the generative probabilistic testing equivalence of~\cite{CDSY99} and the reactive
probabilistic testing equivalence of~\cite{KN98}, are equated by $\sbis{\textrm{PTe-}\sqcup\sqcap}$ if only
generative/reactive probabilistic tests are employed, but are told apart by the same relation as soon as
internal nondeterminism is admitted within tests.

Indeed, these two examples show that $\sbis{\textrm{PTe-}\sqcup\sqcap}$ is sensitive to the moment of
occurrence of internal choices when testing fully nondeterministic processes (resp.\ generative/reactive
probabilistic processes) with tests admitting probabilities (resp.\ internal nondeterminism), because it
becomes possible to make copies of intermediate states of the processes under test. As pointed out
in~\cite{Abr87}, this capability increases the distinguishing power of testing equivalence. In a
probabilistic setting, this may lead to questionable estimations of success probabilities (see~\cite{GA12}
and the references therein).

In this paper, we study new trace and testing equivalences (for nondeterministic and probabilistic
processes) that, different from the old ones, do possess the above-mentioned properties. We shall start by
defining a coarser probabilistic trace equivalence $\sbis{\rm PTr}$ that, instead of considering entire
trace distributions as in $\sbis{\rm PTr,dis}$, compares the execution probabilities of single traces.
Moreover, we shall define a finer probabilistic testing equivalence $\sbis{\textrm{PTe-}\forall\exists}$
that, instead of focussing only on the highest and the lowest probability of passing a test as in
$\sbis{\textrm{PTe-}\sqcup\sqcap}$, requires for each maximal resolution of the interaction system on one
side the existence of a maximal resolution of the interaction system on the other side that has the same
success probability.

While the new trace equivalence $\sbis{\rm PTr}$ reaches the goal of being compositional, the new testing
equivalence $\sbis{\textrm{PTe-}\forall\exists}$ is still not fully backward compatible with the testing
equivalences for restricted classes of processes. We shall however use $\sbis{\textrm{PTe-}\forall\exists}$
as a stepping stone to define another probabilistic testing equivalence, $\sbis{\textrm{PTe-tbt}}$, that
requires matching success probabilities of maximal resolutions in a \emph{trace-by-trace} fashion rather
than cumulatively over all successful computations of the maximal resolutions. This further testing
equivalence is a fully conservative extension of the ones in~\cite{DH84,CDSY99,KN98} and avoids questionable
estimations of success probabilities without resorting to model transformations as in~\cite{GA12}. Thus, the
trace-by-trace approach provides a uniform way of defining testing equivalence over different probabilistic
models. This means that the standard notions of testing equivalence for generative/reactive probabilistic
processes could be redefined by following the same trace-by-trace approach taken for the general model,
without altering their discriminating power. Interestingly, we shall see that $\sbis{\textrm{PTe-tbt}}$ is
comprised between $\sbis{\rm PTr}$ and a novel probabilistic failure equivalence $\sbis{\rm PF}$, which in
turn is comprised between $\sbis{\textrm{PTe-tbt}}$ and $\sbis{\textrm{PTe-}\forall\exists}$.

For each of the equivalences considered in the paper, we shall introduce the two variants determined by the
assumed nature of the schedulers used to resolve nondeterminism, namely \emph{deterministic schedulers} or
\emph{randomized schedulers}.

The rest of the paper, which is a revised and extended version of~\cite{BDL12}, is organized as follows.
Section~\ref{sec:nplts} presents the necessary definitions for the NPLTS model.
Section~\ref{sec:trace_equiv} introduces $\sbis{\rm PTr}$ and shows that it is a congruence with respect to
parallel composition. Sections~\ref{sec:testing_equiv} and~\ref{sec:tbt_testing_equiv} respectively deal
with $\sbis{\textrm{PTe-}\forall\exists}$ and $\sbis{\textrm{PTe-tbt}}$ by providing the necessary results
to relate them to the new trace equivalence (inclusion) and to testing equivalences for restricted classes
of processes (backward compatibility), emphasizing that the trace-by-trace approach unifies the testing
equivalences defined for subclasses of NPLTS models without internal nondeterminism.
Section~\ref{sec:spectrum} places in a spectrum old and new trace ad testing equivalences.
Section~\ref{sec:concl} draws some conclusions and suggests future works.

%
%
\section{Nondeterministic and Probabilistic Processes}
\label{sec:nplts}
%
%

Processes combining nondeterminism and probability are typically described by means of extensions of the LTS
model, in which every action-labeled transition goes from a source state to a \emph{probability distribution
over target states} rather than to a single target state. They are essentially Markov decision
processes~\cite{Der70} and are representative of a number of slightly different probabilistic computational
models including internal nondeterminism that have appeared in the literature with names such as, e.g.,
concurrent Markov chains~\cite{Var85}, alternating probabilistic models~\cite{HJ90,YL92,PLS00},
NP-systems~\cite{JHY94}, probabilistic automata in the sense of~\cite{Seg95a}, probabilistic processes in
the sense of~\cite{JY95}, denotational probabilistic models in the sense of~\cite{JSM97}, probabilistic
transition systems in the sense of~\cite{JY02}, and pLTS~\cite{DGHM08} (see~\cite{SD04} for an overview). We
formalize them as a variant of simple probabilistic automata~\cite{Seg95a} and give them the acronym NPLTS to
stress the possible simultaneous presence of nondeterminism (N) and probability (P) in the LTS-like model.

	\begin{figure}[tp]

\input{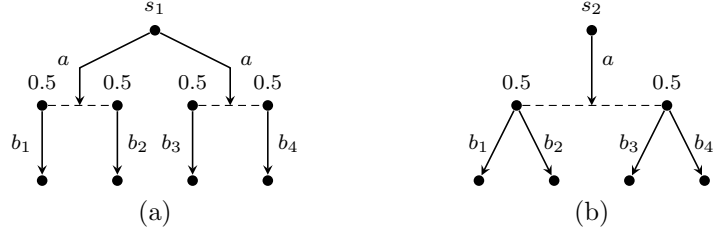}
\caption{Graphical representation of two example NPLTS models}
\label{fig:nplts_example}

	\end{figure}

  	\begin{defi}

A \emph{nondeterministic and probabilistic labeled transition system}, NPLTS for short, is a triple $(S, A,
\! \arrow{}{} \!)$ where:

		\begin{itemize}

\item $S$ is an at most countable set of states.

\item $A$ is a countable set of transition-labeling actions.

\item $\! \arrow{}{} \! \subseteq S \times A \times \ms{Distr}(S)$ is a transition relation, where
$\ms{Distr}(S)$ is the set of discrete probability distributions over $S$.
\fullbox

		\end{itemize}

	\end{defi}

A transition $(s, a, \cald)$ is written $s \arrow{a}{} \cald$. We say that $s' \in S$ is not reachable
from~$s$ via that $a$-transition if $\cald(s') = 0$, otherwise we say that it is reachable with probability
$p = \cald(s')$. The reachable states form the support of $\cald$, i.e., $\ms{supp}(\cald) = \{ s' \in S
\mid \cald(s') > 0 \}$. \linebreak The choice among all the transitions departing from $s$ is
nondeterministic and can be influenced by the external environment, while the choice of the target state for
a specific transition is probabilistic and takes place internally.

An NPLTS can be depicted as a directed graph-like structure in which vertices represent states and
action-labeled edges represent action-labeled transitions. Given a transition $s \arrow{a}{} \cald$, the
corresponding $a$-labeled edge goes from the vertex for state $s$ to a set of vertices linked by a dashed
line, each of which represents a state $s' \in \ms{supp}(\cald)$ and is labeled with $\cald(s')$ -- label
omitted if $\cald(s') = 1$. The graphical representation is exemplified in Fig.~\ref{fig:nplts_example}.

The NPLTS model embeds various less expressive models. In particular, it represents:

	\begin{enumerate}

\item A \emph{fully nondeterministic process} when every transition leads to a distribution that
concentrates all the probability mass into a single target state.

\item A \emph{fully probabilistic process} when every state has at most one outgoing transition.

\item A \emph{reactive probabilistic process}~\cite{GSS95} -- or probabilistic automaton in the sense
of~\cite{Rab63} -- when no state has two or more outgoing transitions labeled with the same action.

	\end{enumerate}

\noindent
The NPLTS in Fig.~\ref{fig:nplts_example}(a) mixes probability and internal nondeterminism, while the one in
Fig.~\ref{fig:nplts_example}(b) describes a reactive probabilistic process. An example of fully
probabilistic process can be obtained from the NPLTS in Fig.~\ref{fig:nplts_example}(a) by removing one of
its two $a$-transitions.

In this setting, a computation is a sequence of state-to-state steps, each denoted by $s \step{a}{} s'$ and
derived from a state-to-distribution transition $s \arrow{a}{} \cald$.

	\begin{defi}\label{def:computation}

Let $\call = (S, A, \! \arrow{}{} \!)$ be an NPLTS and $s, s' \in S$. We say that: \cws{0}{c \: \equiv \:
s_{0} \step{a_{1}}{} s_{1} \step{a_{2}}{} s_{2} \dots s_{n - 1} \step{a_{n}}{} s_{n}} is a
\emph{computation} of $\call$ of length $n$ from $s = s_{0}$ to $s' = s_{n}$ iff for all $i = 1, \dots, n$
there exists a transition $s_{i - 1} \arrow{a_{i}}{} \cald_{i}$ such that $s_{i} \in \ms{supp}(\cald_{i})$,
with $\cald_{i}(s_{i})$ being the execution probability of step $s_{i - 1} \step{a_{i}}{} s_{i}$ conditioned
on the selection of transition $s_{i - 1} \arrow{a_{i}}{} \cald_{i}$ of $\call$ at state~$s_{i - 1}$. We say
that $c$ is \emph{maximal} iff it is not a proper prefix of any other computation from $s$. \linebreak We
denote by $\calc_{\rm fin}(s)$ the set of finite-length computations from $s$.
\fullbox

	\end{defi}

A resolution of a state $s$ of an NPLTS $\call$ is the result of any possible way of resolving
nondeterminism starting from $s$. A resolution is a tree-like structure whose branching points represent
probabilistic choices. This is obtained by unfolding from $s$ the graph structure underlying~$\call$ and by
selecting at each state a single transition of~$\call$ (\emph{deterministic scheduler}) or a combined
transition of $\call$ (\emph{randomized scheduler}) among all the transitions that are possible from the
reached state. We shall consider only history-independent schedulers.

	\begin{figure}[tp]

\input{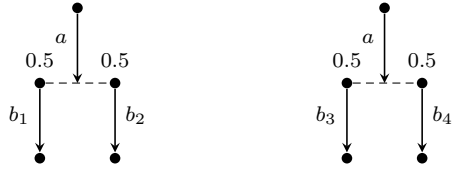}
\caption{The two maximal resolutions of the NPLTS in Fig.~\ref{fig:nplts_example}(a)}
\label{fig:nplts_resolutions}

	\end{figure}

Below, we formalize the notion of resolution arising from a deterministic scheduler as a fully probabilistic
NPLTS. Notice that, when $\call$ is fully nondeterministic, these resolutions coincide with the computations
of $\call$.

	\begin{defi}\label{def:resolution}

Let $\call = (S, A, \! \arrow{}{} \!)$ be an NPLTS and $s \in S$. We say that an NPLTS $\calz = (Z, A, \!
\arrow{}{\calz} \!)$ is a \emph{resolution} of~$s$ obtained via a \emph{deterministic scheduler} iff there
exists a state correspondence function $\ms{corr}_{\calz} : Z \rightarrow S$ such that $s =
\ms{corr}_{\calz}(z_{s})$, for some $z_{s} \in Z$, and for all $z \in Z$ it holds that:

		\begin{itemize}

\item If $z \arrow{a}{\calz} \cald$, then $\ms{corr}_{\calz}(z) \arrow{a}{} \cald'$ with $\cald(z') =
\cald'(\ms{corr}_{\calz}(z'))$ for all $z' \in Z$.

\item If $z \arrow{a_{1}}{\calz} \cald_{1}$ and $z \arrow{a_{2}}{\calz} \cald_{2}$, then $a_{1} = a_{2}$ and
$\cald_{1} = \cald_{2}$.

		\end{itemize}

\noindent
We say that $\calz$ is \emph{maximal} iff it cannot be further extended in accordance with the graph
structure of $\call$ and the constraints above. We denote by $\ms{Res}(s)$ and $\ms{Res}_{\rm max}(s)$ the
sets of resolutions and maximal resolutions of~$s$ obtained via deterministic schedulers.
\fullbox

	\end{defi}

Since $\calz \in \ms{Res}(s)$ is fully probabilistic, the probability $\ms{prob}(c)$ of executing $c \in
\calc_{\rm fin}(z_{s})$ can be defined as the product of the (no longer conditional) execution probabilities
of the individual steps of $c$, with $\ms{prob}(c)$ being always equal to $1$ if $\call$ is fully
nondeterministic. This notion is lifted to $\calc \subseteq \calc_{\rm fin}(z_{s})$ by letting
$\ms{prob}(\calc) = \sum_{c \in \calc} \ms{prob}(c)$ whenever none of the computations in $\calc$ is a
proper prefix of one of the others. The two maximal resolutions of the NPLTS in
Fig.~\ref{fig:nplts_example}(a) are shown in Fig.~\ref{fig:nplts_resolutions}; both of them possess two
maximal computations, each having probability $0.5$.

The transitions of a resolution obtained via a randomized scheduler are not necessarily ordinary transitions
of $\call$, but combined transitions derived as convex combinations of equally labeled transitions of the
original model. Formally, the first clause of Def.~\ref{def:resolution} changes as follows:

	\begin{itemize}

\item If $z \arrow{a}{\calz} \cald$, then there are $n \in \natns_{> 0}$, $(p_{i} \in \realns_{]0, 1]} \mid
1 \le i \le n)$, and $(\ms{corr}_{\calz}(z) \arrow{a}{} \cald_{i} \mid 1 \le i \le n)$ such that $\sum_{i =
1}^{n} p_{i} = 1$ and $\cald(z') = \sum_{i = 1}^{n} p_{i} \cdot \cald_{i}(\ms{corr}_{\calz}(z'))$ for all
$z' \in Z$.

	\end{itemize}

\noindent
It is worth noting that an ordinary transition is a combined transition in which $n = 1$ and $p_{1} = 1$ and
that, when $\call$ has no internal nondeterminism (like in the fully/reactive probabilistic case), a
resolution arising from randomized schedulers can only be originated by a convex combination of a transition
with itself. In the following, we use the shorthand ct for ``based on combined transitions''. We thus denote
by $\ms{Res}^{\rm ct}(s)$ and $\ms{Res}^{\rm ct}_{\rm max}(s)$ the sets of resolutions and maximal
resolutions of~$s$ obtained via randomized schedulers.

We finally introduce a parallel operator $\_ \pco{\cala} \_$ for NPLTS models that synchronize on a set of
actions $\cala$ and proceed independently of each other on any other action. The adoption of a CSP-like
parallel operator is by now standard in the definition of testing equivalences for probabilistic processes
(see, e.g., \cite{JY95,Seg96,CDSY99,DGHM08}). We have preferred using this operator rather than a CCS-like
parallel operator because the former embodies a mechanism for enforcing synchronizations, while the latter
does not and hence, when defining testing equivalences, requires either resorting to an additional operator
(e.g., restriction in a CCS setting as in~\cite{YL92}) or considering only computations whose steps are all
labeled with invisible $\tau$-actions stemming from the synchronization of an action with the corresponding
coaction (like in traditional testing theory~\cite{DH84}). We would, however, like to stress that, if we had
used a CCS-like parallel composition supporting $\tau$-labeled two-way synchronizations, the resulting
testing equivalences and the compositionality results would have been much the same.

	\begin{defi}\label{def:par_comp}

Let $\call_{i} = (S_{i}, A, \! \arrow{}{i} \!)$ be an NPLTS for $i = 1, 2$ and $\cala \subseteq A$. The
\emph{parallel composition} of $\call_{1}$ and $\call_{2}$ with synchronization on $\cala$ is the NPLTS
$\call_{1} \pco{\cala} \call_{2} = (S_{1} \times S_{2}, A, \! \arrow{}{} \!)$ where $\! \arrow{}{} \!
\subseteq (S_{1} \times S_{2}) \times A \times \ms{Distr}(S_{1} \times S_{2})$ is such that $(s_{1}, s_{2})
\arrow{a}{} \cald$ iff one of the following holds:

		\begin{itemize}

\item $a \! \in \! \cala$, $s_{1} \arrow{a}{1} \cald_{1}$, $s_{2} \arrow{a}{2} \cald_{2}$, and
$\cald(s'_{1}, s'_{2}) \! = \! \cald_{1}(s'_{1}) \cdot \cald_{2}(s'_{2})$ for all $(s'_{1}, s'_{2}) \! \in
\! S_{1} \times S_{2}$.

\item $a \! \notin \! \cala$, $s_{1} \arrow{a}{1} \cald_{1}$, $\cald(s'_{1}, s'_{2}) = \cald_{1}(s'_{1})$ if
$s'_{2} \! = \! s_{2}$, and $\cald(s'_{1}, s'_{2}) = 0$ if $s'_{2} \in S_{2} \! \setminus \! \{ s_{2} \}$.

\item $a \! \notin \! \cala$, $s_{2} \arrow{a}{2} \cald_{2}$, $\cald(s'_{1}, s'_{2}) = \cald_{2}(s'_{2})$ if
$s'_{1} \! = \! s_{1}$, and $\cald(s'_{1}, s'_{2}) = 0$ if $s'_{1} \in S_{1} \! \setminus \! \{ s_{1} \}$.
\fullbox

		\end{itemize}

	\end{defi}

	\begin{figure}[tp]

\input{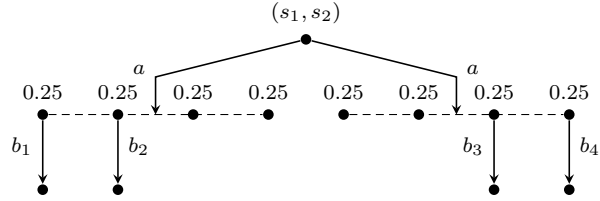}
\caption{Fully synchronous parallel composition of the two NPLTS models in Fig.~\ref{fig:nplts_example}}
\label{fig:nplts_parallel}

	\end{figure}

Throughout the paper, we shall use $\call_{1} \pco{} \call_{2}$ to denote the fully synchronous parallel
composition $\call_{1} \pco{A} \call_{2}$. Figure~\ref{fig:nplts_parallel} shows the NPLTS resulting from
the fully synchronous parallel composition of the two NPLTS models in Fig.~\ref{fig:nplts_example}. Note
that the two nondeterministic choices after the $a$-transition of the NPLTS in
Fig.~\ref{fig:nplts_example}(b) have disappeared in Fig.~\ref{fig:nplts_parallel}, because the
synchronization between a state with a single transition and a state with several differently labeled
transitions always results in a state with at most a single transition.

%
%
\section{Trace Equivalences for NPLTS Models}
\label{sec:trace_equiv}
%
%

Trace equivalences for NPLTS models examine the probability with which two states perform computations
labeled with the same action sequences, called traces, for each possible way of resolving nondeterminism.
We say that a finite-length computation is \emph{compatible} with a trace $\alpha \in A^{*}$ iff the
sequence of actions labeling the computation steps is equal to~$\alpha$. \linebreak Given an NPLTS $\call =
(S, A, \! \arrow{}{} \!)$ and a resolution $\calz$ of a state $s$, we denote by $\calcc(z_{s}, \alpha)$ the
set of $\alpha$-compatible computations in $\calc_{\rm fin}(z_{s})$. We now recall two variants of the
probabilistic trace-distribution equivalence introduced in~\cite{Seg95b} and further studied
in~\cite{CSV07,LSV03,PS04,CLSV06}.

	\begin{defi}\label{def:ptrdis} 

Let $(S, A, \! \arrow{}{} \!)$ be an NPLTS. We say that $s_{1}, s_{2} \in S$ are \emph{probabilistic
trace-distribution equivalent}, written $s_{1} \sbis{\rm PTr,dis} s_{2}$, iff:

		\begin{itemize}

\item For each $\calz_{1} \in \ms{Res}(s_{1})$ there exists $\calz_{2} \in \ms{Res}(s_{2})$ such that
\underline{for all $\alpha \in A^{*}$}:
\cws{10}{\hspace*{-1.2cm} \ms{prob}(\calcc(z_{s_{1}}, \alpha)) \: = \: \ms{prob}(\calcc(z_{s_{2}}, \alpha))}

\item For each $\calz_{2} \in \ms{Res}(s_{2})$ there exists $\calz_{1} \in \ms{Res}(s_{1})$ such that
\underline{for all $\alpha \in A^{*}$}:
\cws{12}{\hspace*{-1.2cm} \ms{prob}(\calcc(z_{s_{2}}, \alpha)) \: = \: \ms{prob}(\calcc(z_{s_{1}}, \alpha))}

		\end{itemize}

\noindent
We denote by $\sbis{\rm PTr,dis}^{\rm ct}$ the coarser variant based on randomized schedulers.
\fullbox

	\end{defi}

The relations $\sbis{\rm PTr,dis}$ and $\sbis{\rm PTr,dis}^{\rm ct}$ are quite discriminating because they
compare entire trace distributions and hence impose a constraint on the execution probability of \emph{all
the traces} of any pair of corresponding resolutions (\emph{fully matching resolutions}). For instance,
states $s_{1}$ and~$s_{2}$ in Fig.~\ref{fig:counterex_ptrdis_ptr} are distinguished by $\sbis{\rm PTr,dis}$
because neither of the two maximal resolutions of~$s_{1}$, which are depicted in
Fig.~\ref{fig:nplts_resolutions}, is matched according to Def.~\ref{def:ptrdis} by (i.e., has the same trace
distribution as) one of the two maximal resolutions of $s_{2}$.

However, $s_{1}$ and $s_{2}$ have exactly the same set of traces, which is $\{ \varepsilon, a, a \, b_{1}, a
\, b_{2}, a \, b_{3}, a \, b_{4} \}$, and each of these traces has the same probability of being performed
in both processes once nondeterminism has been resolved, hence it might seem reasonable to identify $s_{1}$
and $s_{2}$. \linebreak The constraint on trace distributions can indeed be relaxed by considering \emph{a
single trace at a time}, i.e., by anticipating the quantification over traces with respect to the
quantification over resolutions in Def.~\ref{def:ptrdis}. In this way, differently labeled computations of a
resolution are allowed to be matched by computations of different resolutions (\emph{partially matching
resolutions}), which leads to the following new probabilistic trace equivalences.

	\begin{defi}\label{def:ptr}

Let $(S, A, \! \arrow{}{} \!)$ be an NPLTS. We say that $s_{1}, s_{2} \in S$ are \emph{probabilistic trace
equivalent}, written $s_{1} \sbis{\rm PTr} s_{2}$, iff \underline{for all $\alpha \in A^{*}$} it holds that:

		\begin{itemize}

\item For each $\calz_{1} \in \ms{Res}(s_{1})$ there exists $\calz_{2} \in \ms{Res}(s_{2})$ such that:
\cws{10}{\hspace*{-1.2cm} \ms{prob}(\calcc(z_{s_{1}}, \alpha)) \: = \: \ms{prob}(\calcc(z_{s_{2}}, \alpha))}

\item For each $\calz_{2} \in \ms{Res}(s_{2})$ there exists $\calz_{1} \in \ms{Res}(s_{1})$ such that:
\cws{12}{\hspace*{-1.2cm} \ms{prob}(\calcc(z_{s_{2}}, \alpha)) \: = \: \ms{prob}(\calcc(z_{s_{1}}, \alpha))}

		\end{itemize}

\noindent
We denote by $\sbis{\rm PTr}^{\rm ct}$ the coarser variant based on randomized schedulers.
\fullbox

	\end{defi}

	\begin{thm}\label{thm:ptrdis_incl_ptr}

Let $(S, A, \! \arrow{}{} \!)$ be an NPLTS and $s_{1}, s_{2} \in S$. Then:
\cws{11}{\begin{array}{rcl}
s_{1} \sbis{\rm PTr,dis} s_{2} & \!\!\! \Longrightarrow \!\!\! & s_{1} \sbis{\rm PTr} s_{2} \\
s_{1} \sbis{\rm PTr,dis}^{\rm ct} s_{2} & \!\!\! \Longrightarrow \!\!\! & s_{1} \sbis{\rm PTr}^{\rm ct}
s_{2} \\
\end{array}}

\proof
If $s_{1} \sbis{\rm PTr,dis} s_{2}$ (resp.\ $s_{1} \sbis{\rm PTr,dis}^{\rm ct} s_{2}$), then $s_{1}
\sbis{\rm PTr} s_{2}$ (resp. $s_{1} \sbis{\rm PTr}^{\rm ct} s_{2}$) follows by taking the same fully
matching resolutions considered for $\sbis{\rm PTr,dis}$ (resp.\ $\sbis{\rm PTr,dis}^{\rm ct}$).
\qed

	\end{thm}

	\begin{figure}[tp]

\input{Pictures/counterex_ptrdis_ptr}
\caption{NPLTS models distinguished by $\sbis{\rm PTr,dis}$/$\sbis{\rm PTr,dis}^{\rm ct}$ and identified by
$\sbis{\rm PTr}$/$\sbis{\rm PTr}^{\rm ct}$}
\label{fig:counterex_ptrdis_ptr}

	\end{figure}

The implications in Thm.~\ref{thm:ptrdis_incl_ptr} cannot be reversed. For example, in
Fig.~\ref{fig:counterex_ptrdis_ptr} it holds that $s_{1} \sbis{\rm PTr} s_{2}$ because the leftmost maximal
resolution of $s_{1}$ is matched by the leftmost maximal resolution of $s_{2}$ with respect to trace $a \,
b_{1}$, and by the rightmost maximal resolution of $s_{2}$ with respect to trace $a \, b_{2}$.
Figures~\ref{fig:counterex_ptrdis_ptr} and~\ref{fig:counterex_ptesupinf_trace}
(p.~\pageref{fig:counterex_ptesupinf_trace}) together show that $\sbis{\rm PTr}$ and $\sbis{\rm
PTr,dis}^{\rm ct}$ are incomparable with each other.

All the four trace equivalences above are fully backward compatible with the two trace equivalences
respectively defined in~\cite{BHR84} for fully nondeterministic processes -- denoted by $\sbis{\rm Tr,fnd}$
-- and in~\cite{JS90} for fully probabilistic processes -- denoted by $\sbis{\rm Tr,fpr}$. Moreover, they
are partially backward compatible with the trace equivalence -- denoted by $\sbis{\rm Tr,rpr}$ -- that can
be defined for reactive probabilistic processes by following one of the approaches in~\cite{Sei95}.

	\begin{thm}\label{thm:trace_compat}

Let $\call = (S, A, \! \arrow{}{} \!)$ be an NPLTS and $s_{1}, s_{2} \in S$.

		\begin{enumerate}

\item If $\call$ is fully nondeterministic, then:
\cws{10}{\hspace*{-1.2cm} s_{1} \sbis{\rm PTr,dis} s_{2} \: \Leftrightarrow \: s_{1} \sbis{\rm PTr,dis}^{\rm
ct} s_{2} \: \Leftrightarrow \: s_{1} \sbis{\rm PTr} s_{2} \: \Leftrightarrow \: s_{1} \sbis{\rm PTr}^{\rm
ct} s_{2} \: \Leftrightarrow \: s_{1} \sbis{\rm Tr,fnd} s_{2}}

\item If $\call$ is fully probabilistic, then:
\cws{10}{\hspace*{-1.2cm} s_{1} \sbis{\rm PTr,dis} s_{2} \: \Leftrightarrow \: s_{1} \sbis{\rm PTr,dis}^{\rm
ct} s_{2} \: \Leftrightarrow \: s_{1} \sbis{\rm PTr} s_{2} \: \Leftrightarrow \: s_{1} \sbis{\rm PTr}^{\rm
ct} s_{2} \: \Leftrightarrow \: s_{1} \sbis{\rm Tr,fpr} s_{2}}

\item If $\call$ is reactive probabilistic, then:
\cws{10}{\hspace*{-1.2cm}\begin{array}{rcl}
s_{1} \sbis{\rm PTr,dis} s_{2} & \!\!\! \Rightarrow \!\!\! & s_{1} \sbis{\rm Tr,rpr} s_{2} \\
s_{1} \sbis{\rm PTr,dis}^{\rm ct} s_{2} & \!\!\! \Rightarrow \!\!\! & s_{1} \sbis{\rm Tr,rpr} s_{2} \\
s_{1} \sbis{\rm PTr} s_{2} & \!\!\! \Rightarrow \!\!\! & s_{1} \sbis{\rm Tr,rpr} s_{2} \\
s_{1} \sbis{\rm PTr}^{\rm ct} s_{2} & \!\!\! \Rightarrow \!\!\! & s_{1} \sbis{\rm Tr,rpr} s_{2} \\
\end{array}}

		\end{enumerate}

\proof
We proceed as follows:

		\begin{enumerate}

\item Suppose that $\call$ is fully nondeterministic. We recall from~\cite{BHR84} that $s_{1} \sbis{\rm
Tr,fnd} s_{2}$ means that, for all $\alpha \in A^{*}$, there is an $\alpha$-compatible computation
from~$s_{1}$ iff there is an \linebreak $\alpha$-compatible computation from $s_{2}$. The result is a
straightforward consequence of the fact that the resolutions of $\call$ coincide with the computations of
$\call$, hence the probability of performing within a resolution of $\call$ a computation compatible with a
given trace can only be $1$ or $0$. Note that randomized schedulers are not important in this setting
because, due to the absence of probabilistic choices, the model cannot contain submodels that arise from
convex combinations of other submodels.

\item Suppose that $\call$ is fully probabilistic. We recall from~\cite{JS90} that $s_{1} \sbis{\rm Tr,fpr}
s_{2}$ means that, for all $\alpha \in A^{*}$, $\ms{prob}(\calcc(s_{1}, \alpha)) = \ms{prob}(\calcc(s_{2},
\alpha))$. The result is a straightforward consequence of the fact that $\call$ has a single maximal
resolution, which coincides with $\call$ itself. Note that schedulers are not important in this setting
because there is no nondeterminism.

\item Suppose that $\call$ is reactive probabilistic. Due to the absence of internal nondeterminism,
$\sbis{\rm Tr,rpr}$ can be defined in the same way as $\sbis{\rm Tr,fpr}$ provided that, given $\alpha \in
A^{*}$, probabilities of the form $\ms{prob}(\calcc(s, \alpha))$ are viewed as being
conditional~\cite{Sei95} on selecting the maximal resolution of $s \in S$ that contains all the
$\alpha$-compatible computations from $s$ (this resolution is unique because $\call$ is reactive
probabilistic). The result immediately follows.
\qed

		\end{enumerate}

	\end{thm}

\noindent In the reactive probabilistic case, the first two implications cannot be reversed. If we consider a variant
of $s_{1}$ (resp.\ $s_{2}$) in Fig.~\ref{fig:counterex_ptrdis_ptr} having a single outgoing $a$-transition
reaching with probability $0.5$ a state with a $b_{1}$-transition and a $b_{2}$-transition (resp.\
$b_{3}$-transition) and with probability $0.5$ a state with a $b_{3}$-transition (resp.\ $b_{2}$-transition)
and a $b_{4}$-transition, then the two resulting states are related by $\sbis{\rm Tr,rpr}$ but distinguished
by $\sbis{\rm PTr,dis}$ and $\sbis{\rm PTr,dis}^{\rm ct}$.


Interestingly, $\sbis{\rm PTr}$ and $\sbis{\rm PTr}^{\rm ct}$ are congruences with respect to parallel
composition. This is quite surprising because, while $\sbis{\rm Tr,fnd}$ is compositional~\cite{BHR84}, all
probabilistic trace semantics proposed so far in the literature, i.e., $\sbis{\rm Tr,fpr}$, $\sbis{\rm
PTr,dis}$, and $\sbis{\rm PTr,dis}^{\rm ct}$, are \emph{not} compositional~\cite{JS90,Seg95b}. In
particular, in~\cite{LSV03} it was shown that the coarsest congruence contained in $\sbis{\rm PTr,dis}^{\rm
ct}$ is a variant of the simulation equivalence of~\cite{SL94}, while in~\cite{CLSV06} distributed
schedulers (as opposed to centralized ones) were introduced to achieve compositionality.

To prove preservation of $\sbis{\rm PTr}$ under parallel composition, we make use of an alternative
characterization of $\sbis{\rm PTr}$ itself based on \emph{weighted traces}, each of which is an element of
$A^{*} \times \realns_{]0, 1]}$. Before defining the function that associates the set of its weighted traces
with each state, we introduce the following auxiliary notation where $X, Y \subseteq A^{*} \times
\realns_{]0, 1]}$, $a \in A$, $\alpha \in A^{*}$, $p \in \realns_{]0, 1]}$, and $q \in \realns_{[0, 1]}$:

	\begin{itemize}

\item $X \: \vdash \: (\alpha, q)$ iff either $(\alpha, q) \in X$, or $q = 0$ and $(\alpha, p') \notin X$
for all $p' \in \realns_{]0, 1]}$.

\item $X + Y \: = \: \{ (\alpha, q_1 + q_2) \mid X \vdash (\alpha, q_1) \land Y \vdash (\alpha, q_2) \land
q_1 + q_2 > 0 \}$.

\item $a.X \: = \: \{ (a \, \alpha, p') \mid (\alpha, p') \in X \}$.

\item $p \cdot X \: = \: \{ (\alpha, p \cdot p') \mid (\alpha, p') \in X \}$.

	\end{itemize}

	\begin{defi}\label{def:wtset}

Let $(S, A, \! \arrow{}{} \!)$ be an NPLTS. The set of functions $\ms{traces}_{i} : S \rightarrow 2^{A^{*}
\times \realns_{]0, 1]}}$, $i \in \natns$, is inductively defined as follows:

		\begin{itemize}

\item $\ms{traces}_{0}(s) \: = \: \{ (\varepsilon, 1) \}$.

\item $\ms{traces}_{i + 1}(s) \: = \: \{ (\varepsilon, 1) \} \cup \bigcup\limits_{s \arrow{a}{} \cald} a.
(\sum\limits_{s' \in \ms{supp}(\cald)} \cald(s') \cdot \ms{traces}_{i}(s'))$.

		\end{itemize}

\noindent
We let $\ms{traces}(s) \: = \: \bigcup\limits_{i \in \natns} \ms{traces}_{i}(s)$.
\fullbox

	\end{defi}

For every $i \in \natns$, function $\ms{traces}_{i}$ maps each state $s$ to the set of weighted traces built
by considering only the computations from $s$ of length at most $i$. The set $\ms{traces}(s)$ is then
obtained by considering all finite-length computations from $s$. The following lemma guarantees that the
construction is monotonic.

	\begin{lem}\label{lemma:wtset_monotonicity}

Let $(S, A, \! \arrow{}{} \!)$ be an NPLTS. For all $s \in S$ and $i \in \natns$ it holds that:
\cws{12}{\ms{traces}_{i}(s) \: \subseteq \: \ms{traces}_{i + 1}(s)}

\proof
We prove that for all $s \in S$, $i \in \natns$, $\alpha \in A^{*}$, and $p \in \realns_{]0, 1]}$ it holds
that $(\alpha, p) \in \ms{traces}_{i}(s)$ implies $(\alpha, p) \in \ms{traces}_{i + 1}(s)$ by proceeding by
induction on the length of $\alpha$. \\
\emph{Base of induction:} Let $|\alpha| = 0$, i.e., $\alpha = \varepsilon$. Directly from
Def.~\ref{def:wtset}, for all $j \in \natns$ we have that $(\varepsilon, p) \in \ms {traces}_{j}(s)$ iff $p
= 1$. Hence, the result holds when $\alpha = \varepsilon$. \\
\emph{Induction hypothesis:} We assume that for all $s' \in S$, $j \in \natns$, $\alpha' \in A^{*}$, and $p'
\in \realns_{]0, 1]}$ it holds that $(\alpha', p') \in \ms{traces}_{j}(s')$ implies $(\alpha', p') \in
\ms{traces}_{j + 1}(s')$ when $|\alpha'| \leq n$ for some $n \in \natns$. \\
\emph{Induction step:} Let $\alpha = a \, \alpha'$ with $|\alpha'| = n$ and suppose that $(\alpha, p) \in
\ms{traces}_{i}(s)$. Then there exists a transition $s \arrow{a}{} \cald$ such that:
\cws{0}{(\alpha', p) \: \in \: \sum\limits_{s' \in \ms{supp}(\cald)} \cald(s') \cdot \ms{traces}_{i -
1}(s')}
Hence, for each $s' \in \ms{supp}(\cald)$ there exists $p_{s'} \in \realns_{[0, 1]}$ such that
$\ms{traces}_{i - 1}(s') \vdash (\alpha', p_{s'})$, and $p = \sum_{s' \in \ms{supp}(\cald)} \cald(s') \cdot
p_{s'}$. By the induction hypothesis, we have that $(\alpha', p_{s'}) \in \ms{traces}_{i}(s')$ for each $s'
\in \ms{supp}(\cald)$ such that $(\alpha', p_{s'}) \in \ms{traces}_{i - 1}(s')$. Therefore:
\cws{0}{(\alpha', p) \: \in \: \sum\limits_{s' \in \ms{supp}(\cald)} \cald(s') \cdot \ms{traces}_{i}(s')}
which implies $(\alpha, p) \in \ms{traces}_{i + 1}(s)$.
\qed

	\end{lem}

We now show that function $\ms{traces}$ can be used to provide an alternative definition of $\sbis{\rm
PTr}$, which will be exploited at the end of this section to prove that $\sbis{\rm PTr}$ is preserved under
parallel composition. The key property is that $(\alpha, p)$ is a weighted trace associated with a state $s$
iff there exists a resolution of $s$ where trace $\alpha$ can occur with probability $p$.

	\begin{lem}\label{lemma:wtset_vs_res_prob}

Let $(S, A, \! \arrow{}{} \!)$ be an NPLTS. For all $s \in S$, $\alpha \in A^{*}$, and $p \in \realns_{]0,
1]}$ it holds that:
\cws{12}{(\alpha, p) \in \ms{traces}(s) \: \Longleftrightarrow \: \exists \calz \in \ms{Res}(s) \ldotp
\ms{prob}(\calcc(z_{s}, \alpha)) = p}

\proof
We prove the result by proceeding by induction on the length of $\alpha$. \\
\emph{Base of induction:} Let $|\alpha| = 0$, i.e., $\alpha = \varepsilon$. Directly from
Def.~\ref{def:wtset}, for all $j \in \natns$ we have that $(\varepsilon, p) \in \ms{traces}_{j}(s)$ iff $p =
1$. Moreover, for each $\calz \in \ms{Res}(s)$ it holds that $\ms{prob}(\calcc(z_{s}, \varepsilon)) = 1$.
Hence, the result holds when $\alpha = \varepsilon$. \\
\emph{Induction hypothesis:} We assume that for all $s' \in S$, $\alpha' \in A^{*}$, and $p' \in
\realns_{]0, 1]}$ it holds that $(\alpha', p') \in \ms{traces}(s')$ iff there exists $\calz \in
\ms{Res}(s')$ such that $\ms{prob}(\calcc(z_{s'}, \alpha')) = p'$ when $|\alpha'| \leq n$ for some $n \in
\natns$. \\
\emph{Induction step:} Let $\alpha = a \, \alpha'$ with $|\alpha'| = n$. Suppose that $(\alpha, p) \in
\ms{traces}(s)$. This means that $(\alpha, p) \in \ms{traces}_{i}(s)$ for some $i \in \natns$. Then there
exists a transition $s \arrow{a}{} \cald$ such that:
\cws{0}{(\alpha', p) \: \in \: \sum\limits_{s' \in \ms{supp}(\cald)} \cald(s') \cdot \ms{traces}_{i -
1}(s')}
Hence, for each $s' \in \ms{supp}(\cald)$ there exists $p_{s'} \in \realns_{[0, 1]}$ such that
$\ms{traces}_{i - 1}(s') \vdash (\alpha', p_{s'})$, and $p = \sum_{s' \in \ms{supp}(\cald)} \cald(s') \cdot
p_{s'}$. Since $\ms{traces}_{i - 1}(s') \subseteq \ms{traces}(s')$, by the induction hypothesis we have that
there exists $\calz_{s'} \in \ms{Res}(s')$ such that $\ms{prob}(\calcc(z_{s'}, \alpha')) = p_{s'}$ for each
$s' \in \ms{supp}(\cald)$ such that $(\alpha', p_{s'}) \in \ms{traces}_{i - 1}(s')$. Therefore, if we
consider the resolution $\calz \in \ms{Res}(s)$ that first selects transition $s \arrow{a}{} \cald$ and then
behaves as $\calz_{s'}$ for each $s' \in \ms{supp}(\cald)$ such that $(\alpha', p_{s'}) \in \ms{traces}_{i -
1}(s')$ whereas it halts in each $s' \in \ms{supp}(\cald)$ such that $(\alpha', p_{s'}) \notin
\ms{traces}_{i - 1}(s')$, it is easy to see that $\ms{prob}(\calcc(z_{s}, \alpha)) = p$. \\
Suppose now that there exists $\calz = (Z, A, \! \arrow{}{\calz} \!) \in \ms{Res}(s)$ such that
$\ms{prob}(\calcc(z_{s}, \alpha)) = p$. Then there exists a transition $z_{s} \arrow{a}{\calz} \cald$ such
that:
\cws{0}{p \: = \: \sum\limits_{z' \in \ms{supp}(\cald)} \cald(z') \cdot \ms{prob}(\calcc(z', \alpha'))}
Hence, for each $z' \in \ms{supp}(\cald)$ there exists $p_{z'} \in \realns_{[0, 1]}$ such that $p_{z'} =
\ms{prob}(\calcc(z', \alpha'))$, and $p = \sum_{z' \in \ms{supp}(\cald)} \cald(z') \cdot p_{z'}$. Denoting
by $\ms{corr}_{\calz}$ the correspondence function for $\calz$, by the induction hypothesis we have that
$(\alpha', p_{z'}) \in \ms{traces}(\ms{corr}_{\calz}(z'))$ for each $z' \in \ms{supp}(\cald)$ such that
$p_{z'} > 0$. Due to Lemma~\ref{lemma:wtset_monotonicity}, for all $i \in \natns_{\ge |\alpha'|}$ it holds
that $(\alpha', p_{z'}) \in \ms{traces}_{i}(\ms{corr}_{\calz}(z'))$ for each $z' \in \ms{supp}(\cald)$ such
that $p_{z'} > 0$. Since there must exist a transition $s \arrow{a}{} \cald'$ such that $\cald(z') =
\cald'(\ms{corr}_{\calz}(z'))$ for all $z' \in Z$, it holds that:
\cws{0}{(\alpha', p) \: \in \: \sum\limits_{\ms{corr}_{\calz}(z') \in \ms{supp}(\cald')}
\cald'(\ms{corr}_{\calz}(z')) \cdot \ms{traces}_{|\alpha'|}(\ms{corr}_{\calz}(z'))}
and hence:
\cws{0}{(\alpha, p) \: \in \: a.(\sum\limits_{\ms{corr}_{\calz}(z') \in \ms{supp}(\cald')}
\cald'(\ms{corr}_{\calz}(z')) \cdot \ms{traces}_{|\alpha'|}(\ms{corr}_{\calz}(z'))) \: \subseteq \:
\ms{traces}_{|\alpha|}(s)}
which implies $(\alpha, p) \in \ms{traces}(s)$.
\qed

	\end{lem}

	\begin{thm}\label{thm:ptr_char_wtset}

Let $(S, A, \! \arrow{}{} \!)$ be an NPLTS and $s_{1}, s_{2} \in S$. Then:
\cws{12}{s_{1} \sbis{\rm PTr} s_{2} \: \Longleftrightarrow \: \ms{traces}(s_{1}) = \ms{traces}(s_{2})}

\proof
Directly from Def.~\ref{def:ptr} and Lemma~\ref{lemma:wtset_vs_res_prob}. Notice that, given $\alpha \in
A^{*}$, from the point of view of $\sbis{\rm PTr}$ a resolution $\calz \in \ms{Res}(s_{k})$, $k = 1, 2$,
such that $\ms{prob}(\calcc(z_{s_{k}}, \alpha)) = 0$ is always matched by the resolution of $s_{3 - k}$
having only the initial state. Therefore, the exclusion of weighted traces with weight $0$ from the set
resulting from the application of function $\ms{traces}$ does not violate the present characterization of
$\sbis{\rm PTr}$.
\qed

	\end{thm}

We finally exploit the result in Thm.~\ref{thm:ptr_char_wtset} to show that $\sbis{\rm PTr}$ is preserved
under parallel composition. This is an important and much wanted property that is essential for behavioral
equivalences to support the compositional analysis of system descriptions.

	\begin{thm}\label{thm:ptr_compos}

Let $\call_{k} = (S_{k}, A, \! \arrow{}{k} \!)$ be an NPLTS for $k = 0, 1, 2$ and consider $\call_{1}
\pco{\cala} \call_{0}$ and $\call_{2} \pco{\cala} \call_{0}$ for $\cala \subseteq A$. Let $s_{k} \in S_{k}$
for $k = 0, 1, 2$. Then:
\cws{12}{s_{1} \sbis{\rm PTr} s_{2} \: \Longrightarrow \: (s_{1}, s_{0}) \sbis{\rm PTr} (s_{2}, s_{0})}

\proof
For $\alpha_1, \alpha_2, \alpha \in A^{*}$, we let $\alpha_1 \otimes_{\cala} \alpha_2 \vdash \alpha$ denote
the smallest relation induced by the following inference rules:
\cws{0}{\begin{array}{c}
\infer{
\varepsilon \otimes_{\cala} \varepsilon \vdash \varepsilon}{} \quad  
\infer[a \in \cala]{
a \, \alpha_1 \otimes_{\cala} a \, \alpha_2 \vdash a \, \alpha
}{
\alpha_1 \otimes_{\cala} \alpha_2 \vdash \alpha
}\quad
\infer[a \notin \cala]{
a \, \alpha_1 \otimes_{\cala} \alpha_2 \vdash a \, \alpha
}{
\alpha_1 \otimes_{\cala} \alpha_2 \vdash \alpha
}\quad
\infer[a \notin \cala]{
\alpha_1 \otimes_{\cala} a \, \alpha_2 \vdash a \, \alpha
}{
\alpha_1 \otimes_{\cala} \alpha_2 \vdash \alpha
}
\end{array}}
Moreover, for $X, Y \subseteq A^{*} \times \realns_{]0, 1]}$ we let:
\cws{0}{X \otimes_{\cala} Y \: = \: \{ (\alpha, p_1 \cdot p_2) \mid (\alpha_1, p_1) \in X \land (\alpha_2,
p_2) \in Y \land \alpha_1 \otimes_{\cala} \alpha_2 \vdash \alpha \}}
In the rest of this proof, we show that $\ms{traces}(s_k, s_0) = \ms{traces}(s_k) \otimes_{\cala}
\ms{traces}(s_0)$ for $k = 1, 2$. This, together with Thm.~\ref{thm:ptr_char_wtset}, guarantees that if $s_1
\sbis{\rm PTr} s_2$ then $(s_1, s_0) \sbis{\rm PTr} (s_2, s_0)$. Indeed, if $s_1 \sbis{\rm PTr} s_2$, then
$\ms{traces}(s_1) = \ms{traces}(s_2)$ by Thm.~\ref{thm:ptr_char_wtset}. Thus:
\cws{0}{\ms{traces}(s_1, s_0) \: = \: \ms{traces}(s_1) \otimes_{\cala} \ms{traces}(s_0) \: = \:
\ms{traces}(s_2) \otimes_{\cala} \ms{traces}(s_0) \: = \: \ms{traces}(s_2, s_0)}
and hence $(s_1, s_0) \sbis{\rm PTr} (s_2, s_0)$ by Thm.~\ref{thm:ptr_char_wtset}. \\
To be precise, we show that for all $s \in S_{1} \cup S_{2}$, $s_{0} \in S_{0}$, $\alpha \in A^{*}$, and $p
\in \realns_{]0, 1]}$ there exists $i \in \natns$ such that $(\alpha, p) \in \ms{traces}_{i}(s, s_0)$ iff
there exist $j, h \leq i$ such that $(\alpha, p) \in \ms{traces}_{j}(s) \otimes_{\cala}
\ms{traces}_{h}(s_0)$ by proceeding by induction on the length of $\alpha$. \\
\emph{Base of induction:} Let $|\alpha| = 0$, i.e., $\alpha = \varepsilon$. In this case, the result follows
directly from the fact that: 
\cws{0}{\ms{traces}_{0}(s, s_0) \: = \: \{ (\varepsilon, 1) \} \: = \: \{ (\varepsilon, 1) \}
\otimes_{\cala} \{ (\varepsilon, 1) \} \: = \: \ms{traces}_{0}(s) \otimes_{\cala} \ms{traces}_{0}(s_0)}
\emph{Induction hypothesis:} We assume that for all $s' \in S_{1} \cup S_{2}$, $s'_{0} \in S_{0}$, $\alpha'
\in A^{*}$, and $p' \in \realns_{]0, 1]}$ there exists $i' \in \natns$ such that $(\alpha', p') \in
\ms{traces}_{i'}(s', s'_0)$ iff there exist $j', h' \leq i'$ such that $(\alpha', p') \in
\ms{traces}_{j'}(s') \otimes_{\cala} \ms{traces}_{h'}(s'_0)$ when $|\alpha| \le n$ for some $n \in \natns$.
\\
\emph{Induction step:} Let $\alpha = a \, \alpha'$ with $|\alpha'| = n$. The fact that $(\alpha, p) \in
\ms{traces}_{n + 1}(s, s_0)$ means that there exists a transition $(s, s_0) \arrow{a}{} \cald$ such that:
\cws{0}{(\alpha', p) \: \in \: \sum\limits_{(s', s'_0) \in \ms{supp}(\cald)} \cald(s', s'_0) \cdot
\ms{traces}_{n}(s', s'_0)}
where $\! \arrow{}{} \!$ is the transition relation of $\call_{1} \pco{\cala} \call_{0}$ or $\call_{2}
\pco{\cala} \call_{0}$ depending on whether $s$ belongs to $S_{1}$ or $S_{2}$. Similarly, we denote by $\!
\arrow{}{1, 2} \!$ the transition relation of $\call_{1}$ or $\call_{2}$ depending on whether $s$ belongs to
$S_{1}$ or $S_{2}$. \\
We distinguish two cases: $a \in \cala$ and $a \notin \cala$. If $a \in \cala$, then $(s, s_0) \arrow{a}{}
\cald$ means that $s \arrow{a}{1, 2} \cald'$, $s_0 \arrow{a}{0} \cald''$, and $\cald(s', s'_0) = \cald'(s')
\cdot \cald''(s'_0)$ for all $(s', s'_0) \in (S_{1} \cup S_{2}) \times S_{0}$, hence:
\cws{0}{(\alpha', p) \: \in \: \sum\limits_{s' \in \ms{supp}(\cald')} \, \sum\limits_{s'_0 \in
\ms{supp}(\cald'')} \cald'(s') \cdot \cald''(s'_0) \cdot \ms{traces}_{n}(s', s'_0)}
This means that for each $s' \in \ms{supp}(\cald')$ and $s'_0 \in \ms{supp}(\cald'')$ there exists $p_{(s',
s'_0)} \in \realns_{[0, 1]}$ such that $\ms{traces}_{n}(s', s'_0) \vdash (\alpha', p_{(s', s'_0)})$, and $p
= \sum_{s' \in \ms{supp}(\cald')} \, \sum_{s'_0 \in \ms{supp}(\cald'')} \cald'(s') \cdot \cald''(s'_0) \cdot
p_{(s', s'_0)}$. By applying the induction hypothesis to all $s' \in \ms{supp}(\cald')$ and $s'_0 \in
\ms{supp}(\cald'')$ such that $(\alpha', p_{(s', s'_0)}) \in \ms{traces}_{n}(s', s'_0)$ and exploiting
Lemma~\ref{lemma:wtset_monotonicity} so as to obtain a single pair from the various pairs $j_{(s', s'_0)},
h_{(s', s'_0)} \leq n$, it follows that the fact that $(\alpha, p) \in \ms{traces}_{n + 1}(s, s_0)$ means
that there exist $j, h \leq n$ such that:
\cws{0}{\begin{array}{rcl}
(\alpha, p) & \!\!\! \in \!\!\! & a. (\sum\limits_{s' \in \ms{supp}(\cald')} \, \sum\limits_{s'_0 \in
\ms{supp}(\cald'')} \cald'(s') \cdot \cald''(s'_0) \cdot (\ms{traces}_{j}(s') \otimes_{\cala}
\ms{traces}_{h}(s'_0))) \\[0.5cm]
& \!\!\! = \!\!\! & a. (\sum\limits_{s' \in \ms{supp}(\cald')} \, \sum\limits_{s'_0 \in \ms{supp}(\cald'')}
(\cald'(s') \cdot \ms{traces}_{j}(s')) \otimes_{\cala} (\cald''(s'_0) \cdot \ms{traces}_{h}(s'_0)))
\\[0.5cm]
& \!\!\! = \!\!\! & a. ((\sum\limits_{s' \in \ms{supp}(\cald')} \cald'(s') \cdot \ms{traces}_{j}(s'))
\otimes_{\cala} (\sum\limits_{s'_0 \in \ms{supp}(\cald'')} \cald''(s'_0) \cdot \ms{traces}_{h}(s'_0)))
\\[0.5cm]
& \!\!\! = \!\!\! & a. (\sum\limits_{s' \in \ms{supp}(\cald')} \cald'(s') \cdot \ms{traces}_{j}(s'))
\otimes_{\cala} a. (\sum\limits_{s'_0 \in \ms{supp}(\cald'')} \cald''(s'_0) \cdot \ms{traces}_{h}(s'_0))
\\[0.5cm]
& \!\!\! \subseteq \!\!\! & \ms{traces}_{j + 1}(s) \otimes_{\cala} \ms{traces}_{h + 1}(s_0) \\
\end{array}}
Similarly, if $a \notin \cala$, then $(s, s_0) \arrow{a}{} \cald$ means that either $s \arrow{a}{1, 2}
\cald'$ with $\cald(s', s'_0) = \cald'(s')$ if $s'_0 = s_{0}$ and $\cald(s', s'_0) = 0$ if $s'_0 \in S_{0}
\setminus \{ s_{0} \}$, or $s_0 \arrow{a}{0} \cald''$ with $\cald(s', s'_0) = \cald''(s'_{0})$ if $s' = s$
and $\cald(s', s'_0) = 0$ if $s' \in (S_{1} \cup S_{2}) \setminus \{ s \}$, hence:
\cws{0}{(\alpha', p) \: \in \: \sum\limits_{s' \in \ms{supp}(\cald')} \cald'(s') \cdot \ms{traces}_{n}(s',
s_0) \cup \sum\limits_{s'_0 \in \ms{supp}(\cald'')} \cald''(s'_0) \cdot \ms{traces}_{n}(s, s'_0)}
This means that (i) for each $s' \in \ms{supp}(\cald')$ there exists $p_{s'} \in \realns_{[0, 1]}$ such that
$\ms{traces}_{n}(s', s_0) \vdash (\alpha', p_{s'})$, (ii) for each $s'_0 \in \ms{supp}(\cald'')$ there
exists $p_{s'_0} \in \realns_{[0, 1]}$ such that $\ms{traces}_{n}(s, s'_0) \vdash (\alpha', p_{s'_0})$, and
(iii) either $p = \sum_{s' \in \ms{supp}(\cald')} \cald'(s') \cdot p_{s'}$ or $p = \sum_{s'_0 \in
\ms{supp}(\cald'')} \cald''(s'_0) \cdot p_{s'_0}$.
By applying the induction hypothesis to all $s' \in \ms{supp}(\cald')$ such that $(\alpha', p_{s'}) \in
\ms{traces}_{n}(s', s_0)$ and to all $s'_0 \in \ms{supp}(\cald'')$ such that $(\alpha', p_{s'_0}) \in
\ms{traces}_{n}(s, s'_0)$, and exploiting Lemma~\ref{lemma:wtset_monotonicity} so as to obtain a single pair
from the various pairs $j_{s'}, h_{s'} \leq n$ and $j_{s'_0}, h_{s'_0} \leq n$, it follows that the fact
that $(\alpha, p) \in \ms{traces}_{n + 1}(s, s_0)$ means that there exist $j, h \leq n$ such that:
\cws{0}{\begin{array}{rcl}
(\alpha, p) & \!\!\! \in \!\!\! & a. (\sum\limits_{s' \in \ms{supp}(\cald')} \cald'(s') \cdot
(\ms{traces}_{j}(s') \otimes_{\cala} \ms{traces}_{h}(s_0))) \; \cup \\[0.5cm]
& & a. (\sum\limits_{s'_0 \in \ms{supp}(\cald'')} \cald''(s'_0) \cdot (\ms{traces}_{j}(s) \otimes_{\cala}
\ms{traces}_{h}(s'_0))) \\[0.5cm]
& \!\!\! = \!\!\! & a. ((\sum\limits_{s' \in \ms{supp}(\cald')} \cald'(s') \cdot \ms{traces}_{j}(s'))
\otimes_{\cala} \ms{traces}_{h}(s_0)) \; \cup \\[0.5cm]
& & a. (\ms{traces}_{j}(s) \otimes_{\cala} (\sum\limits_{s'_0 \in \ms{supp}(\cald'')} \cald''(s'_0) \cdot
\ms{traces}_{h}(s'_0))) \\[0.5cm]
& \!\!\! = \!\!\! & (a. (\sum\limits_{s' \in \ms{supp}(\cald')} \cald'(s') \cdot \ms{traces}_{j}(s')))
\otimes_{\cala} \ms{traces}_{h}(s_0) \; \cup \\[0.5cm]
& & \ms{traces}_{j}(s) \otimes_{\cala} (a. (\sum\limits_{s'_0 \in \ms{supp}(\cald'')} \cald''(s'_0) \cdot
\ms{traces}_{h}(s'_0))) \\[0.5cm]
& \!\!\! \subseteq \!\!\! & \ms{traces}_{j + 1}(s) \otimes_{\cala} \ms{traces}_{h}(s_0) \; \cup \\
& & \ms{traces}_{j}(s) \otimes_{\cala} \ms{traces}_{h + 1}(s_0) \\
& \!\!\! \subseteq \!\!\! & \ms{traces}_{j + 1}(s) \otimes_{\cala} \ms{traces}_{h + 1}(s_0) \; \cup \\
& & \ms{traces}_{j + 1}(s) \otimes_{\cala} \ms{traces}_{h + 1}(s_0) \\
& \!\!\! = \!\!\! & \ms{traces}_{j + 1}(s) \otimes_{\cala} \ms{traces}_{h + 1}(s_0) \\
\end{array}}
where we have exploited again Lemma~\ref{lemma:wtset_monotonicity}.
\qed

	\end{thm}

It can be similarly proved that also $\sbis{\rm PTr}^{\rm ct}$ is a congruence with respect to parallel
composition if combined transitions are considered instead of ordinary ones in Def.~\ref{def:wtset}.

	\begin{thm}\label{thm:ptr_ct_compos}

Let $\call_{k} = (S_{k}, A, \! \arrow{}{k} \!)$ be an NPLTS for $k = 0, 1, 2$ and consider $\call_{1}
\pco{\cala} \call_{0}$ and $\call_{2} \pco{\cala} \call_{0}$ for $\cala \subseteq A$. Let $s_{k} \in S_{k}$
for $k = 0, 1, 2$. Then:
\cws{12}{s_{1} \sbis{\rm PTr}^{\rm ct} s_{2} \: \Longrightarrow \: (s_{1}, s_{0}) \sbis{\rm PTr}^{\rm ct}
(s_{2}, s_{0})}
\qed

	\end{thm}

%
%
\section{Testing Equivalences for NPLTS Models}
\label{sec:testing_equiv}
%
%

Testing equivalences for NPLTS models consider the probability of performing computations along which the
same tests are passed. Tests specify the actions a process can perform; in this setting, tests are
formalized as NPLTS models equipped with a success state. For the sake of simplicity, we restrict ourselves
to \emph{finite} tests, each of which has finitely many states, finitely many outgoing transitions from each
state, an acyclic graph structure, and hence finitely many computations leading to the success state.


	\begin{defi}

A \emph{nondeterministic and probabilistic test}, NPT for short, is a finite NPLTS $\calt = (O, A, \!
\arrow{}{} \!)$ where $O$~contains a distinguished success state denoted by $\omega$ with no outgoing
transitions. We say that a computation of~$\calt$ is \emph{successful} iff its last state is $\omega$.
\fullbox

	\end{defi}

	\begin{defi}

Let $\call = (S, A, \! \arrow{}{} \!)$ be an NPLTS and $\calt = (O, A, \! \arrow{}{\calt} \!)$ be an NPT.
The \emph{interaction system} of $\call$ and $\calt$ is the NPLTS $\cali(\call, \calt) = \call \pco{} \calt$
where:

		\begin{itemize}

\item Every element $(s, o) \in S \times O$ is called a \emph{configuration} and is said to be
\emph{successful} iff $o = \omega$.

\item A computation of $\cali(\call, \calt)$ is said to be \emph{successful} iff its last configuration is
successful. Given a resolution $\calz$ of $(s, o) \in S \times O$, we denote by $\calsc(z_{s, o})$ the set
of successful computations from the state $z_{s, o}$ of $\calz$ corresponding to the configuration $(s, o)$
of $\cali(\call, \calt)$.
\fullbox

		\end{itemize}

	\end{defi}

In the following, we shall consider only maximal resolutions of interaction systems because the non-maximal
ones do not expose all successful computations.

Due to the possible presence of equally labeled transitions departing from the same state, there is not
necessarily a single probability value with which an NPLTS passes a test. Thus, to compare two states
$s_{1}$ and $s_{2}$ of an NPLTS via an NPT with initial state $o$, we need to compute the probability of
performing a successful computation from the two configurations $(s_{1}, o)$ and $(s_{2}, o)$ in every
maximal resolution of the interaction system. As done in~\cite{YL92,JY95,Seg96,DGHM08}, one option is
comparing only the suprema ($\sqcup$) and the infima ($\sqcap$) of these success probabilities over all
maximal resolutions of the interaction systems. To avoid infima to be trivially zero, it is strictly
necessary to consider only maximal resolutions.

	\begin{defi}\label{def:ptesupinf}

Let $(S, A, \! \arrow{}{} \!)$ be an NPLTS. We say that $s_{1}, s_{2} \in S$ are \emph{probabilistic
\linebreak $\sqcup\sqcap$-testing equivalent}, written $s_{1} \sbis{\textrm{PTe-}\sqcup\sqcap} s_{2}$, iff
for every NPT $\calt = (O, A, \! \arrow{}{\calt} \!)$ with initial state $o \in O$ it holds that:
\cws{0}{\begin{array}{rcl}
\bigsqcup\limits_{\calz_{1} \in \ms{Res}_{\rm max}(s_{1}, o)} \ms{prob}(\calsc(z_{s_{1}, o})) & \!\!\! = 
\!\!\! & \bigsqcup\limits_{\calz_{2} \in \ms{Res}_{\rm max}(s_{2}, o)} \ms{prob}(\calsc(z_{s_{2}, o}))
\\[0.4cm]
\bigsqcap\limits_{\calz_{1} \in \ms{Res}_{\rm max}(s_{1}, o)} \ms{prob}(\calsc(z_{s_{1}, o})) & \!\!\! =
\!\!\! & \bigsqcap\limits_{\calz_{2} \in \ms{Res}_{\rm max}(s_{2}, o)} \ms{prob}(\calsc(z_{s_{2}, o})) \\
\end{array}}
We denote by $\sbis{\textrm{PTe-}\sqcup\sqcap}^{\rm ct}$ the variant based on randomized schedulers.
\fullbox

	\end{defi}

Following the structure of classical testing equivalence $\sbis{\rm Te,fnd}$ for fully nondeterministic
processes~\cite{DH84}, the constraint on suprema represents the may-part of
$\sbis{\textrm{PTe-}\sqcup\sqcap}$ while the constraint on infima represents the must-part of
$\sbis{\textrm{PTe-}\sqcup\sqcap}$. The probabilistic testing equivalences in~\cite{YL92,JY95,DGHM08} are
essentially defined as $\sbis{\textrm{PTe-}\sqcup\sqcap}$, while the one in~\cite{Seg96} resolves
nondeterminism through randomized schedulers instead of deterministic ones and makes use of countably many
success actions in place of a single one. Notably, a single success action suffices when testing finitary
processes, as proved in~\cite{DGMZ07}, and the use of different classes of schedulers does not change the
discriminating power, as we now show.

	\begin{thm}\label{thm:ptesupinf_sched}

Let $(S, A, \! \arrow{}{} \!)$ be an NPLTS and $s_{1}, s_{2} \in S$. Then:
\cws{12}{s_{1} \sbis{\textrm{{\rm PTe}-}\sqcup\sqcap}^{\rm ct} s_{2} \: \Longleftrightarrow \: s_{1}
\sbis{\textrm{{\rm PTe}-}\sqcup\sqcap} s_{2}}

\proof
The result follows from the fact that, given an arbitrary state $s \in S$ and an arbitrary NPT $\calt = (O,
A, \! \arrow{}{\calt} \!)$ with initial state $o \in O$, it holds that:
\cws{0}{\begin{array}{rcl}
\bigsqcup\limits_{\calz \in \ms{Res}^{\rm ct}_{\rm max}(s, o)} \ms{prob}(\calsc(z_{s, o})) & \!\!\! = \!\!\!
& \bigsqcup\limits_{\calz \in \ms{Res}_{\rm max}(s, o)} \ms{prob}(\calsc(z_{s, o})) \\[0.4cm]
\end{array}}
and an analogous equality holds for infima. In fact, first of all we note that:
\cws{0}{\bigsqcup\limits_{\calz \in \ms{Res}^{\rm ct}_{\rm max}(s, o)} \ms{prob}(\calsc(z_{s, o})) \: \ge \:
\bigsqcup\limits_{\calz \in \ms{Res}_{\rm max}(s, o)} \ms{prob}(\calsc(z_{s, o}))}
because a deterministic scheduler is a special case of randomized scheduler and hence the set of
probabilities on the left contains the set of probabilities on the right (a dual property based on $\le$
holds for infima). Therefore, it suffices to show that:
\cws{0}{\bigsqcup\limits_{\calz \in \ms{Res}^{\rm ct}_{\rm max}(s, o)} \ms{prob}(\calsc(z_{s, o})) \: \le \:
\bigsqcup\limits_{\calz \in \ms{Res}_{\rm max}(s, o)} \ms{prob}(\calsc(z_{s, o}))}
which we prove below by proceeding by induction on the length $n$ of the longest successful computation from
$(s, o)$, which is finite because $\calt$ is finite (a dual property based on $\ge$ can be established for
infima):

		\begin{itemize}

\item If $n = 0$, i.e., $o = \omega$, then:
\cws{4}{\hspace*{-1.2cm} \bigsqcup\limits_{\calz \in \ms{Res}^{\rm ct}_{\rm max}(s, o)}
\ms{prob}(\calsc(z_{s, o})) \: = \: 1 \: = \: \bigsqcup\limits_{\calz \in \ms{Res}_{\rm max}(s, o)}
\ms{prob}(\calsc(z_{s, o}))}

\item Let $n \in \natns_{> 0}$ and suppose that the property holds for all configurations from which the
longest successful computation has length $m = 0, \dots, n - 1$. Indicating with $(s, o) \arrow{a}{\rm c}
\cald_{\rm c}$ a combined transition from $(s, o)$ with $\cald_{\rm c} = \sum_{i = 1}^{m} p_{i} \cdot
\cald_{i}$, we have that:
\cws{0}{\hspace*{-1.2cm}\begin{array}{l}
\hspace*{-0.2cm} \bigsqcup\limits_{\calz \in \ms{Res}^{\rm ct}_{\rm max}(s, o)} \hspace{-0.7cm}
\ms{prob}(\calsc(z_{s, o})) \: = \\
\hspace*{0.4cm} = \: \bigsqcup\limits_{(s, o) \arrow{a}{\rm c} \cald_{\rm c}} \, \sum\limits_{(s', o') \in S
\times O} \left( \cald_{\rm c}(s', o') \cdot \bigsqcup\limits_{\calz' \in \ms{Res}^{\rm ct}_{\rm max}(s',
o')} \hspace{-0.7cm} \ms{prob}(\calsc(z'_{s', o'})) \right) \\
\hspace*{0.4cm} \le \: \bigsqcup\limits_{(s, o) \arrow{a}{\rm c} \cald_{\rm c}} \, \sum\limits_{(s', o') \in
S \times O} \left( \cald_{\rm c}(s', o') \cdot \bigsqcup\limits_{\calz' \in \ms{Res}_{\rm max}(s', o')}
\hspace{-0.7cm} \ms{prob}(\calsc(z'_{s', o'})) \right) \\
\hspace*{0.4cm} = \: \bigsqcup\limits_{(s, o) \arrow{a}{\rm c} \cald_{\rm c}} \, \sum\limits_{(s', o') \in S
\times O} \left( \sum\limits_{i = 1}^{m} (p_{i} \cdot \cald_{i}(s', o')) \cdot \bigsqcup\limits_{\calz' \in
\ms{Res}_{\rm max}(s', o')} \hspace{-0.7cm} \ms{prob}(\calsc(z'_{s', o'})) \right) \\
\hspace*{0.4cm} = \: \bigsqcup\limits_{(s, o) \arrow{a}{\rm c} \cald_{\rm c}} \, \sum\limits_{i = 1}^{m}
p_{i} \cdot \left( \sum\limits_{(s', o') \in S \times O} \left( \cald_{i}(s', o') \cdot
\bigsqcup\limits_{\calz' \in \ms{Res}_{\rm max}(s', o')} \hspace{-0.7cm} \ms{prob}(\calsc(z'_{s', o'}))
\right) \right) \\
\hspace*{0.4cm} \le \: \bigsqcup\limits_{(s, o) \arrow{a}{\rm c} \cald_{\rm c}} \, \sum\limits_{i = 1}^{m}
p_{i} \cdot \bigsqcup\limits_{i = 1}^{m} \left( \sum\limits_{(s', o') \in S \times O} \left( \cald_{i}(s',
o') \cdot \bigsqcup\limits_{\calz' \in \ms{Res}_{\rm max}(s', o')} \hspace{-0.7cm} \ms{prob}(\calsc(z'_{s',
o'})) \right) \right) \\
\hspace*{0.4cm} = \: \bigsqcup\limits_{(s, o) \arrow{a}{\rm c} \cald_{\rm c}} \, \bigsqcup\limits_{i =
1}^{m} \left( \sum\limits_{(s', o') \in S \times O} \left( \cald_{i}(s', o') \cdot \bigsqcup\limits_{\calz'
\in \ms{Res}_{\rm max}(s', o')} \hspace{-0.7cm} \ms{prob}(\calsc(z'_{s', o'})) \right) \right) \\
\hspace*{0.4cm} = \: \bigsqcup\limits_{(s, o) \arrow{a}{} \cald} \, \sum\limits_{(s', o') \in S \times O}
\left( \cald(s', o') \cdot \bigsqcup\limits_{\calz' \in \ms{Res}_{\rm max}(s', o')} \hspace{-0.7cm}
\ms{prob}(\calsc(z'_{s', o'})) \right) \\[0.6cm]
\hspace*{0.4cm} = \: \bigsqcup\limits_{\calz \in \ms{Res}_{\rm max}(s, o)} \hspace{-0.7cm}
\ms{prob}(\calsc(z_{s, o})) \\
\end{array}}
where in the third line we have exploited the induction hypothesis and in the seventh line the fact that
$\sum_{i = 1}^{m} p_{i} = 1$.
\qed

		\end{itemize}

	\end{thm}

The relation $\sbis{\textrm{PTe-}\sqcup\sqcap}$ does not enjoy the desirable property -- possessed by
$\sbis{\rm Te,fnd}$ -- of resulting in a testing semantics finer than the trace semantics for the same class
of processes. Whether $\sbis{\textrm{PTe-}\sqcup\sqcap}$ is included in the trace equivalences of
Sect.~\ref{sec:trace_equiv} depends on the type of schedulers that are considered on the trace semantics
side. In the case of randomized schedulers, as shown in~\cite{Seg96} it holds that
$\sbis{\textrm{PTe-}\sqcup\sqcap} \, \subseteq \, \sbis{\rm PTr,dis}^{\rm ct}$, and hence
$\sbis{\textrm{PTe-}\sqcup\sqcap} \, \subseteq \, \sbis{\rm PTr}^{\rm ct}$ by virtue of
Thm.~\ref{thm:ptrdis_incl_ptr}. However, inclusion no longer holds when only deterministic schedulers are
admitted. Let us consider the two NPLTS models in Fig.~\ref{fig:counterex_ptesupinf_trace}. We have that
$s_{1} \sbis{\textrm{PTe-}\sqcup\sqcap} s_{2}$ while $s_{1} \not\sbis{\rm PTr,dis} s_{2}$ and $s_{1}
\not\sbis{\rm PTr} s_{2}$. It holds that $s_{1} \sbis{\textrm{PTe-}\sqcup\sqcap} s_{2}$ because, for any
test, the central maximal resolution of $s_{1}$ always gives rise to a success probability comprised between
the success probabilities of the other two maximal resolutions of $s_{1}$, which correspond to the two
maximal resolutions of $s_{2}$. In contrast, $s_{1}$ and $s_{2}$ are not related by the two probabilistic
trace equivalences because the maximal resolution of $s_{1}$ starting with the central $a$-transition is not
matched by any of the two maximal resolutions of $s_{2}$.

	\begin{figure}[tp]

\input{Pictures/counterex_ptesupinf_trace}
\caption{NPLTS models identified by $\sbis{\rm PTr,dis}^{\rm ct}$/$\sbis{\textrm{PTe-}\sqcup\sqcap}$ and
told apart by $\sbis{\rm PTr,dis}$/$\sbis{\rm PTr}$}
\label{fig:counterex_ptesupinf_trace}

	\end{figure}

Under deterministic schedulers, inclusion can be achieved by considering $\sbis{\rm PTr}$ in lieu of the
finer $\sbis{\rm PTr,dis}$ and the new testing equivalence $\sbis{\textrm{PTe-}\forall\exists}$ introduced
by the next definition in lieu of the coarser~$\sbis{\textrm{PTe-}\sqcup\sqcap}$. Instead of focussing only
on extremal success probabilities, $\sbis{\textrm{PTe-}\forall\exists}$ requires matching the success
probabilities of \emph{all} maximal resolutions of the interaction systems. Interestingly, the variant of
$\sbis{\textrm{PTe-}\forall\exists}$ based on randomized schedulers coincides with
$\sbis{\textrm{PTe-}\sqcup\sqcap}$.

	\begin{defi}\label{def:pteallexists}

Let $(S, A, \! \arrow{}{} \!)$ be an NPLTS. We say that $s_{1}, s_{2} \in S$ are \emph{probabilistic
$\forall\exists$-testing equivalent}, written $s_{1} \sbis{\textrm{PTe-}\forall\exists} s_{2}$, iff for
every NPT $\calt = (O, A, \! \arrow{}{\calt} \!)$ with initial state $o \in O$ it holds that:

		\begin{itemize}

\item For each $\calz_{1} \in \ms{Res}_{\rm max}(s_{1}, o)$ there exists $\calz_{2} \in \ms{Res}_{\rm
max}(s_{2}, o)$ such that:
\cws{10}{\hspace*{-1.2cm} \ms{prob}(\calsc(z_{s_{1}, o})) \: = \: \ms{prob}(\calsc(z_{s_{2}, o}))}

\item For each $\calz_{2} \in \ms{Res}_{\rm max}(s_{2}, o)$ there exists $\calz_{1} \in \ms{Res}_{\rm
max}(s_{1}, o)$ such that:
\cws{10}{\hspace*{-1.2cm} \ms{prob}(\calsc(z_{s_{2}, o})) \: = \: \ms{prob}(\calsc(z_{s_{1}, o}))}

		\end{itemize}

\noindent
We denote by $\sbis{\textrm{PTe-}\forall\exists}^{\rm ct}$ the coarser variant based on randomized
schedulers.
\fullbox

	\end{defi}

	\begin{figure}[tp]

\input{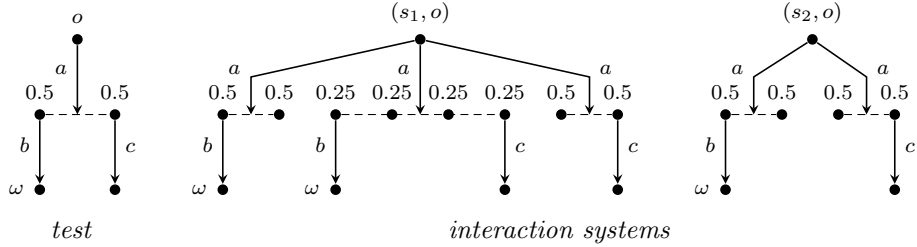}
\caption{A test showing that the two NPLTS models in Fig.~\ref{fig:counterex_ptesupinf_trace} are
distinguished by $\sbis{\textrm{PTe-}\forall\exists}$}
\label{fig:test_pteallexists}

	\end{figure}

	\begin{thm}\label{thm:pteallexists_incl_ptesupinf}

Let $(S, A, \! \arrow{}{} \!)$ be an image-finite NPLTS and $s_{1}, s_{2} \in S$. Then:
\cws{12}{\begin{array}{rcl}
s_{1} \sbis{\textrm{{\rm PTe}-}\forall\exists} s_{2} & \!\!\! \Longrightarrow \!\!\! & s_{1}
\sbis{\textrm{{\rm PTe}-}\sqcup\sqcap} s_{2} \\
s_{1} \sbis{\textrm{{\rm PTe}-}\forall\exists}^{\rm ct} s_{2} & \!\!\! \Longleftrightarrow \!\!\! & s_{1}
\sbis{\textrm{{\rm PTe}-}\sqcup\sqcap} s_{2} \\
\end{array}}

\proof
If $s_{1} \sbis{\textrm{PTe-}\forall\exists} s_{2}$, then we immediately derive that for every NPT $\calt =
(O, A, \! \arrow{}{\calt} \!)$ with initial state $o \in O$ it holds that:
\cws{0}{\begin{array}{rcl}
\{ \ms{prob}(\calsc(z_{s_{1}, o})) \mid \calz_{1} \in \ms{Res}_{\rm max}(s_{1}, o) \} & \!\!\! \subseteq
\!\!\! & \{ \ms{prob}(\calsc(z_{s_{2}, o})) \mid \calz_{2} \in \ms{Res}_{\rm max}(s_{2}, o) \} \\
\{ \ms{prob}(\calsc(z_{s_{2}, o})) \mid \calz_{2} \in \ms{Res}_{\rm max}(s_{2}, o) \} & \!\!\! \subseteq
\!\!\! & \{ \ms{prob}(\calsc(z_{s_{1}, o})) \mid \calz_{1} \in \ms{Res}_{\rm max}(s_{1}, o) \} \\
\end{array}}
As a consequence:
\cws{0}{\{ \ms{prob}(\calsc(z_{s_{1}, o})) \mid \calz_{1} \in \ms{Res}_{\rm max}(s_{1}, o) \} \: = \: \{
\ms{prob}(\calsc(z_{s_{2}, o})) \mid \calz_{2} \in \ms{Res}_{\rm max}(s_{2}, o) \}}
and hence:
\cws{0}{\begin{array}{rcl}
\bigsqcup\limits_{\calz_{1} \in \ms{Res}_{\rm max}(s_{1}, o)} \ms{prob}(\calsc(z_{s_{1}, o})) & \!\!\! =
\!\!\! & \bigsqcup\limits_{\calz_{2} \in \ms{Res}_{\rm max}(s_{2}, o)} \ms{prob}(\calsc(z_{s_{2}, o}))
\\[0.4cm]
\bigsqcap\limits_{\calz_{1} \in \ms{Res}_{\rm max}(s_{1}, o)} \ms{prob}(\calsc(z_{s_{1}, o})) & \!\!\! =
\!\!\! & \bigsqcap\limits_{\calz_{2} \in \ms{Res}_{\rm max}(s_{2}, o)} \ms{prob}(\calsc(z_{s_{2}, o})) \\
\end{array}}
which means that $s_{1} \sbis{\textrm{PTe-}\sqcup\sqcap} s_{2}$. \\
The fact that $s_{1} \sbis{\textrm{{\rm PTe}-}\forall\exists}^{\rm ct} s_{2} \: \Longrightarrow \: s_{1}
\sbis{\textrm{{\rm PTe}-}\sqcup\sqcap} s_{2}$ stems from $s_{1} \sbis{\textrm{{\rm
PTe}-}\forall\exists}^{\rm ct} s_{2} \: \Longrightarrow \: s_{1} \sbis{\textrm{{\rm PTe}-}\sqcup\sqcap}^{\rm
ct} s_{2}$ (as a consequence of the previous result) and $\sbis{\textrm{PTe-}\sqcup\sqcap}^{\rm ct} \: = \:
\sbis{\textrm{PTe-}\sqcup\sqcap}$ (by virtue of Thm.~\ref{thm:ptesupinf_sched}). \\
Suppose now that $s_{1} \sbis{\textrm{PTe-}\sqcup\sqcap} s_{2}$ and consider an arbitrary NPT $\calt = (O,
A, \! \arrow{}{\calt} \!)$ with initial state $o \in O$, so that:
\cws{0}{\begin{array}{rcccl}
\bigsqcup\limits_{\calz_{1} \in \ms{Res}_{\rm max}(s_{1}, o)} \ms{prob}(\calsc(z_{s_{1}, o})) & \!\!\! =
\!\!\! & p_{\sqcup} & \!\!\! = \!\!\! & \bigsqcup\limits_{\calz_{2} \in \ms{Res}_{\rm max}(s_{2}, o)}
\ms{prob}(\calsc(z_{s_{2}, o})) \\[0.4cm]
\bigsqcap\limits_{\calz_{1} \in \ms{Res}_{\rm max}(s_{1}, o)} \ms{prob}(\calsc(z_{s_{1}, o})) & \!\!\! =
\!\!\! & p_{\sqcap} & \!\!\! = \!\!\! & \bigsqcap\limits_{\calz_{2} \in \ms{Res}_{\rm max}(s_{2}, o)}
\ms{prob}(\calsc(z_{s_{2}, o})) \\
\end{array}}
If $p_{\sqcup} = p_{\sqcap}$, then all the maximal resolutions of $(s_{1}, o)$ and $(s_{2}, o)$ have the
same success probability, from which it trivially follows that $s_{1} \sbis{\textrm{PTe-}\forall\exists}
s_{2}$ and hence $s_{1} \sbis{\textrm{PTe-}\forall\exists}^{\rm ct} s_{2}$. \\
Recalling that the NPLTS is image finite and the test is finite so that $\ms{Res}_{\rm max}(s_{1}, o)$ and
$\ms{Res}_{\rm max}(s_{2}, o)$ are both finite, if $p_{\sqcup} > p_{\sqcap}$, then $p_{\sqcup}$ must be
achieved on $\calz_{1, \sqcup} \in \ms{Res}_{\rm max}(s_{1}, o)$ and $\calz_{2, \sqcup} \in \ms{Res}_{\rm
max}(s_{2}, o)$ exhibiting the same successful traces, otherwise -- observing that both resolutions must
have at least one successful trace, otherwise it would be $p_{\sqcup} = 0$ thus violating $p_{\sqcup} >
p_{\sqcap}$ -- states $s_{1}$ and $s_{2}$ would be distinguished with respect to
$\sbis{\textrm{PTe-}\sqcup\sqcap}$ by a test obtained from $\calt$ by making success reachable only along
the successful traces of the one of $\calz_{1, \sqcup}$ and $\calz_{2, \sqcup}$ having a successful trace
not possessed by the other, unless that resolution also contains all the successful traces of the other
resolution, in which case success must be made reachable only along the successful traces of the other
resolution in order to contradict $s_{1} \sbis{\textrm{PTe-}\sqcup\sqcap} s_{2}$. \\
Likewise, $p_{\sqcap}$ must be achieved on $\calz_{1, \sqcap} \in \ms{Res}_{\rm max}(s_{1}, o)$ and
$\calz_{2, \sqcap} \in \ms{Res}_{\rm max}(s_{2}, o)$ exhibiting the same unsuccessful maximal traces,
otherwise -- observing that both resolutions must have at least one unsuccessful maximal trace, otherwise it
would be $p_{\sqcap} = 1$ thus violating $p_{\sqcup} > p_{\sqcap}$ -- states $s_{1}$ and $s_{2}$ would be
distinguished with respect to $\sbis{\textrm{PTe-}\sqcup\sqcap}$ by a test obtained from $\calt$ by making
success reachable also along an unsuccessful maximal trace occurring only in either $\calz_{1, \sqcap}$ or
$\calz_{2, \sqcap}$. \\
By reasoning on the dual test $\calt'$ in which the final states of $\calt$ that are successful (resp.\
unsuccessful) are made unsuccessful (resp.\ successful), it turns out that $\calz_{1, \sqcup}$ and
$\calz_{2, \sqcup}$ must also exhibit the same unsuccessful maximal traces and that $\calz_{1, \sqcap}$ and
$\calz_{2, \sqcap}$ must also exhibit the same successful traces. \\
If $\calz_{1, \sqcup}$ and $\calz_{2, \sqcup}$ do not have sequences of initial transitions in common with
$\calz_{1, \sqcap}$ and $\calz_{2, \sqcap}$, \linebreak then $\calz_{1, \sqcup}$ and $\calz_{1, \sqcap}$ on
one side and $\calz_{2, \sqcup}$ and $\calz_{2, \sqcap}$ on the other side cannot generate via convex
combinations any new resolution that would arise from a randomized scheduler, otherwise they can generate
all such resolutions having a certain sequence of initial transitions, thus covering all the intermediate
success probabilities between $p_{\sqcup}$ and $p_{\sqcap}$ for that sequence of initial transitions. This
shows that for each $\calz_{1} \in \ms{Res}^{\rm ct}_{\rm max}(s_{1}, o)$ with that sequence of initial
transitions there exists $\calz_{2} \in \ms{Res}^{\rm ct}_{\rm max}(s_{2}, o)$ with that sequence of initial
transitions such that $\ms{prob}(\calsc(z_{s_{1}, o})) = \ms{prob}(\calsc(z_{s_{2}, o}))$, and vice versa.
\\
The same procedure can now be applied to the remaining resolutions in $\ms{Res}_{\rm max}(s_{1}, o)$ and
$\ms{Res}_{\rm max}(s_{2}, o)$ that are not convex combinations of previously considered resolutions,
starting from those among the remaining resolutions on which the maximal and minimal success probabilities
are achieved. We can thus conclude that $s_{1} \sbis{\textrm{PTe-}\forall\exists}^{\rm ct} s_{2}$.
\qed

	\end{thm}

	\begin{figure}[tp]

\input{Pictures/counterex_pteallexists_trace}
\caption{NPLTS models identified by $\sbis{\rm PTr}$ and told apart by $\sbis{\textrm{PTe-}\forall\exists}$}
\label{fig:counterex_pteallexists_trace}

	\end{figure}

The inclusion of $\sbis{\textrm{PTe-}\forall\exists}$ in $\sbis{\textrm{PTe-}\sqcup\sqcap}$ is strict.
Indeed, if we consider again the two $\sbis{\textrm{PTe-}\sqcup\sqcap}$-equivalent NPLTS models in
Fig.~\ref{fig:counterex_ptesupinf_trace} and we apply the test in Fig.~\ref{fig:test_pteallexists}, it turns
out that $s_{1} \not\sbis{\textrm{PTe-}\forall\exists} s_{2}$. For the two interaction systems in
Fig.~\ref{fig:test_pteallexists}, we have that the maximal resolution of $(s_{1}, o)$ starting with the
central $a$-transition gives rise to a success probability equal to~0.25 that is not matched by any of the
two maximal resolutions of $(s_{2}, o)$. These resolutions, which correspond to the maximal resolutions of
$(s_{1}, o)$ starting with the two outermost $a$-transitions, have success probability 0.5 and 0,
respectively.

	\begin{thm}\label{thm:pteallexists_incl_ptr}

Let $(S, A, \! \arrow{}{} \!)$ be an NPLTS and $s_{1}, s_{2} \in S$. Then:
\cws{12}{s_{1} \sbis{\textrm{{\rm PTe}-}\forall\exists} s_{2} \: \Longrightarrow \: s_{1} \sbis{\rm PTr}
s_{2}}

\proof
If $s_{1} \sbis{\textrm{PTe-}\forall\exists} s_{2}$, then in particular for every NPT $\calt_{\alpha} = (O,
A, \! \arrow{}{\calt_{\alpha}} \!)$ with initial state $o \in O$ having a single maximal computation that is
labeled with $\alpha \in A^{*}$ and reaches success, it holds that:

		\begin{itemize}

\item For each $\calz_{1} \in \ms{Res}_{\rm max}(s_{1}, o)$ there exists $\calz_{2} \in \ms{Res}_{\rm
max}(s_{2}, o)$ such that:
\cws{10}{\hspace*{-1.2cm} \ms{prob}(\calsc(z_{s_{1}, o})) \: = \: \ms{prob}(\calsc(z_{s_{2}, o}))}

\item For each $\calz_{2} \in \ms{Res}_{\rm max}(s_{2}, o)$ there exists $\calz_{1} \in \ms{Res}_{\rm
max}(s_{1}, o)$ such that:
\cws{10}{\hspace*{-1.2cm} \ms{prob}(\calsc(z_{s_{2}, o})) \: = \: \ms{prob}(\calsc(z_{s_{1}, o}))}

		\end{itemize}

\noindent
Since $\ms{prob}(\calsc^{\calz}(z_{s, o})) = \ms{prob}(\calcc^{\calz'}(z_{s}, \alpha))$ for all $s \in S$
due to the structure of~$\calt_{\alpha}$ -- where $\calz \in \ms{Res}_{\rm max}(s, o)$ and $\calz' \in
\ms{Res}(s)$ originates $\calz$ in the interaction with $\calt_{\alpha}$ -- we immediately derive that for
all $\alpha \in A^{*}$ it holds that:

		\begin{itemize}

\item For each $\calz_{1} \in \ms{Res}(s_{1})$ there exists $\calz_{2} \in \ms{Res}(s_{2})$ such that:
\cws{10}{\hspace*{-1.2cm} \ms{prob}(\calcc(z_{s_{1}}, \alpha)) \: = \: \ms{prob}(\calcc(z_{s_{2}}, \alpha))}

\item For each $\calz_{2} \in \ms{Res}(s_{2})$ there exists $\calz_{1} \in \ms{Res}(s_{1})$ such that:
\cws{10}{\hspace*{-1.2cm} \ms{prob}(\calcc(z_{s_{2}}, \alpha)) \: = \: \ms{prob}(\calcc(z_{s_{1}}, \alpha))}

		\end{itemize}

\noindent
This means that $s_{1} \sbis{\rm PTr} s_{2}$.	
\qed

	\end{thm}

The inclusion of $\sbis{\textrm{PTe-}\forall\exists}$ in $\sbis{\rm PTr}$ is strict. For instance, if we
consider the two NPLTS models in Fig.~\ref{fig:counterex_pteallexists_trace}, it turns out that $s_{1}
\sbis{\rm PTr} s_{2}$ while $s_{1} \not\sbis{\textrm{PTe-}\forall\exists} s_{2}$. In fact, the test in
Fig.~\ref{fig:counterex_pteallexists_trace} distinguishes~$s_{1}$ from~$s_{2}$ with respect to
$\sbis{\textrm{PTe-}\forall\exists}$ because -- looking at the two interaction systems also reported in the
figure -- the only maximal resolution of $(s_{1}, o)$ has a success probability equal to 1 that is not
matched by any of the two maximal resolutions of $(s_{2}, o)$, whose success probabilities are $p_{1}$ and
$p_{2}$, respectively.

Another desirable property of relations like $\sbis{\textrm{PTe-}\sqcup\sqcap}$ and
$\sbis{\textrm{PTe-}\forall\exists}$ that are defined over a general class of processes is that of being
backward compatible with analogous relations for restricted classes of processes. Specifically, we refer to
testing equivalences $\sbis{\rm Te,fnd}$ for fully nondeterministic processes~\cite{DH84}, $\sbis{\rm
Te,fpr}$ for fully probabilistic processes~\cite{CDSY99}, and $\sbis{\rm Te,rpr}$ for reactive probabilistic
processes inspired by~\cite{KN98}.

	\begin{figure}[tp]

\input{Pictures/counterex_tefnd_testing}
\caption{NPLTS models equated by $\sbis{\rm Te,fnd}$ and distinguished by
$\sbis{\textrm{PTe-}\sqcup\sqcap}$/$\sbis{\textrm{PTe-}\forall\exists}$}
\label{fig:counterex_tefnd_testing}

	\end{figure}

	\begin{figure}

\input{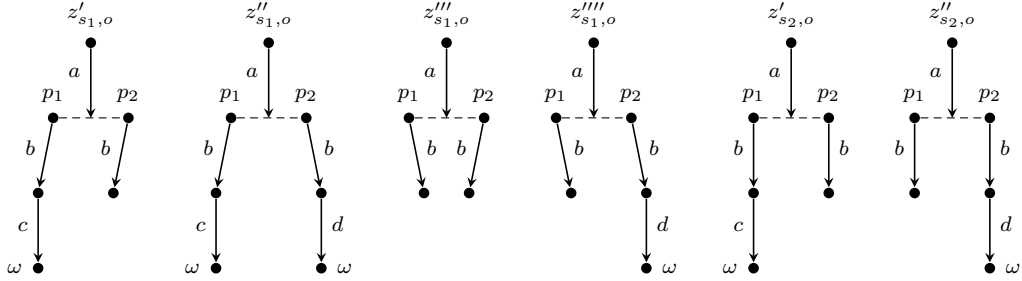}
\caption{Maximal resolutions of the two interaction systems in Fig.~\ref{fig:counterex_tefnd_testing}.}
\label{fig:counterex_tefnd_testing_max_res}

	\end{figure}

As we shall see by means of two counterexamples, backward compatibility is only \emph{partial} as it depends
on the set of tests that are used. Intuitively, $\sbis{\textrm{PTe-}\sqcup\sqcap}$ and
$\sbis{\textrm{PTe-}\forall\exists}$ become sensitive to the moment of occurrence of internal choices when
comparing fully nondeterministic processes (resp.\ fully/reactive probabilistic processes) on the basis of
tests admitting probabilities (resp.\ internal nondeterminism). In such cases, the capability of making
copies of intermediate states of the processes under test arises, a fact that in general increases the
distinguishing power of testing equivalence, as pointed out in~\cite{Abr87}. In a probabilistic setting,
this may lead to questionable estimations of success probabilities (see~\cite{GA12} and the references
therein). Indeed, taking advantage of the increased discriminating power, in~\cite{DGHM08} it was shown that
the may-part of $\sbis{\textrm{PTe-}\sqcup\sqcap}$ coincides with a simulation equivalence akin to the one
in~\cite{LSV03} and the must-part coincides with a novel failure simulation equivalence. Moreover,
in~\cite{Seg96} it was shown that the may-part coincides with the coarsest congruence contained in the
probabilistic trace-distribution equivalence of~\cite{Seg95b} and the must-part coincides with the coarsest
congruence contained in a probabilistic failure-distribution equivalence.

	\begin{figure}[tp]

\input{Pictures/counterex_tepr_testing}
\caption{NPLTS models equated by $\sbis{\rm Te,fpr}$/$\sbis{\rm Te,rpr}$ and distinguished by
$\sbis{\textrm{PTe-}\sqcup\sqcap}$/$\sbis{\textrm{PTe-}\forall\exists}$}
\label{fig:counterex_tepr_testing}

	\end{figure}

	\begin{figure}

\input{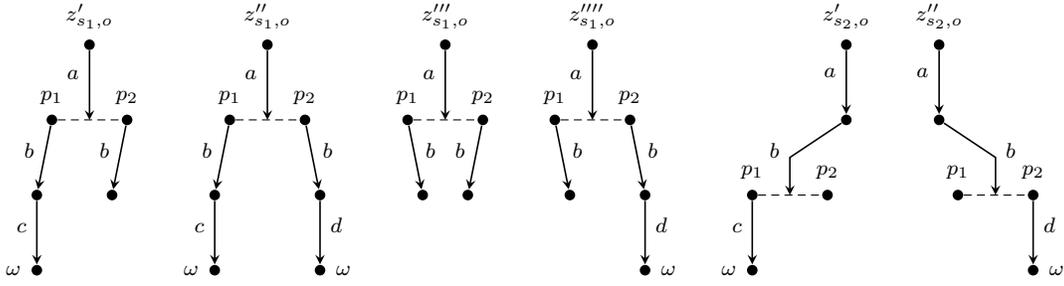}
\caption{Maximal resolutions of the two interaction systems in Fig.~\ref{fig:counterex_tepr_testing}.}
\label{fig:counterex_tepr_testing_max_res}

	\end{figure}

As observed in~\cite{JHY94,DGHMZ07b}, it is easy to see that there exist \emph{fully nondeterministic} NPLTS
models that are identified by $\sbis{\rm Te,fnd}$ but distinguished by $\sbis{\textrm{PTe-}\sqcup\sqcap}$
(and also by $\sbis{\textrm{PTe-}\forall\exists}$). Let us consider the two NPLTS models in
Fig.~\ref{fig:counterex_tefnd_testing}, which represent the classical example that illustrates the main
difference between testing semantics and bisimulation semantics in a nondeterministic setting. It turns out
that $s_{1} \sbis{\rm Te,fnd} s_{2}$ while $s_{1} \not\sbis{\textrm{PTe-}\sqcup\sqcap} s_{2}$ and $s_{1}
\not\sbis{\textrm{PTe-}\forall\exists} s_{2}$. The \emph{probabilistic} test in
Fig.~\ref{fig:counterex_tefnd_testing} distinguishes $s_{1}$ from $s_{2}$ with respect to
$\sbis{\textrm{PTe-}\sqcup\sqcap}$. Indeed, if we consider the two interaction systems also reported in
Fig.~\ref{fig:counterex_tefnd_testing} and their maximal resolutions shown in
Fig.~\ref{fig:counterex_tefnd_testing_max_res}, the supremum of the success probabilities of the four
maximal resolutions of $(s_{1}, o)$ is $1$ -- see the second maximal resolution of $(s_{1}, o)$ -- whereas
the supremum of the success probabilities of the two maximal resolutions of $(s_{2}, o)$ is equal to the
maximum between $p_{1}$ and $p_{2}$. The same test also distinguishes $s_{1}$ from~$s_{2}$ with respect to
$\sbis{\textrm{PTe-}\forall\exists}$ because the third maximal resolution of $(s_{1}, o)$ has a success
probability equal to~$0$ that is not matched by any of the two maximal resolutions of $(s_{2}, o)$, whose
success probabilities are $p_{1}$ and $p_{2}$, respectively.

Following~\cite{JHY94}, we can easily find also two \emph{fully/reactive probabilistic} NPLTS models that
are identified by $\sbis{\rm Te,fpr}$/$\sbis{\rm Te,rpr}$ and distinguished by
$\sbis{\textrm{PTe-}\sqcup\sqcap}$ (and also by $\sbis{\textrm{PTe-}\forall\exists}$). They are depicted in
Fig.~\ref{fig:counterex_tepr_testing} and constitute the classical example that differentiates probabilistic
testing semantics from probabilistic bisimulation semantics. We have that $s_{1} \sbis{\rm Te,fpr} s_{2}$
and $s_{1} \sbis{\rm Te,rpr} s_{2}$, while $s_{1} \not\sbis{\textrm{PTe-}\sqcup\sqcap} s_{2}$ and $s_{1}
\not\sbis{\textrm{PTe-}\forall\exists} s_{2}$. The \emph{fully nondeterministic} test in
Fig.~\ref{fig:counterex_tepr_testing} distinguishes $s_{1}$ from $s_{2}$ with respect to
$\sbis{\textrm{PTe-}\sqcup\sqcap}$ and $\sbis{\textrm{PTe-}\forall\exists}$, as can be seen from the two
interaction systems there reported and their maximal resolutions shown in
Fig.~\ref{fig:counterex_tepr_testing_max_res}.

Summing up, the relations $\sbis{\textrm{PTe-}\sqcup\sqcap}$ and $\sbis{\textrm{PTe-}\forall\exists}$ are
backward compatible with respect to testing equivalences defined over restricted classes of processes as
long as they only admit tests that belong to the same class as the processes under test.

	\begin{thm}\label{thm:testing_partial_compat}

Let $\call = (S, A, \! \arrow{}{} \!)$ be an NPLTS and $s_{1}, s_{2} \in S$.

		\begin{enumerate}

\item If $\call$ is fully nondeterministic and only fully nondeterministic tests are admitted,~then:
\cws{10}{\hspace*{-1.2cm} s_{1} \sbis{\textrm{{\rm PTe}-}\sqcup\sqcap} s_{2} \: \Longleftrightarrow \: s_{1}
\sbis{\textrm{{\rm PTe}-}\forall\exists} s_{2} \: \Longleftrightarrow \: s_{1} \sbis{\rm Te,fnd} s_{2}}

\item If $\call$ is fully probabilistic and only fully probabilistic tests are admitted, then:
\cws{10}{\hspace*{-1.2cm} s_{1} \sbis{\textrm{{\rm PTe}-}\sqcup\sqcap} s_{2} \: \Longleftrightarrow \: s_{1}
\sbis{\textrm{{\rm PTe}-}\forall\exists} s_{2} \: \Longleftrightarrow \: s_{1} \sbis{\rm Tr,fpr} s_{2}}

\item If $\call$ is reactive probabilistic and only reactive probabilistic tests are admitted, then:
\cws{10}{\hspace*{-1.2cm}\begin{array}{rcl}
s_{1} \sbis{\textrm{{\rm PTe}-}\sqcup\sqcap} s_{2} & \!\!\! \Longrightarrow \!\!\! & s_{1} \sbis{\rm Tr,rpr}
s_{2} \\
s_{1} \sbis{\textrm{{\rm PTe}-}\forall\exists} s_{2} & \!\!\! \Longrightarrow \!\!\! & s_{1} \sbis{\rm
Tr,rpr} s_{2} \\
\end{array}}

		\end{enumerate}

\proof
We proceed as follows:

		\begin{enumerate}

\item Suppose that $\call$ is fully nondeterministic and that only fully nondeterministic tests are
admitted, so that all the resulting interaction systems are fully nondeterministic too. We recall
from~\cite{DH84} that $s_{1} \sbis{\rm Te,fnd} s_{2}$ means that, for every test with initial state~$o$,
(i)~there exists a successful computation from $(s_{1}, o)$ iff there exists a successful computation from
$(s_{2}, o)$ and (ii)~all maximal computations from $(s_{1}, o)$ are successful iff all maximal computations
from $(s_{2}, o)$ are successful. The result is a straightforward consequence of the fact that the maximal
resolutions of each interaction system coincide with the maximal computations of the interaction system,
hence the probability of performing a successful computation within a maximal resolution of an interaction
system can only be $1$ or $0$.

\item Suppose that $\call$ is fully probabilistic and that only fully probabilistic tests are admitted, so
that all the resulting interaction systems are fully probabilistic too. We recall from~\cite{CDSY99} that
$s_{1} \sbis{\rm Te,fpr} s_{2}$ means that, for every test with initial state $o$, $\ms{prob}(\calsc(s_{1},
o)) = \ms{prob}(\calsc(s_{2}, o))$. The result is a straightforward consequence of the fact that each
interaction system has a single maximal resolution, which coincides with the interaction system itself.

\item Suppose that $\call$ is reactive probabilistic and that only reactive probabilistic tests are
admitted, so that all the resulting interaction systems are reactive probabilistic too. Taking inspiration
from~\cite{KN98}, $s_{1} \sbis{\rm Te,rpr} s_{2}$ means that, for every test with initial state~$o$,
$(s_{1}, o)$ and $(s_{2}, o)$ have the same suprema and infima of success probabilities over all of their
maximal traces. Success probabilities $\ms{prob}(\calsc_{\rm \alpha}(s, o))$ are viewed as being conditional
on selecting the maximal resolution of $(s, o)$ that contains all the $\alpha$-compatible computations from
$(s, o)$ (this resolution is unique because interaction systems are reactive probabilistic). The result
immediately follows by considering tests that reach success along a single trace.
\qed

		\end{enumerate}

	\end{thm}

We conclude with a remark about the four maximal resolutions of $(s_{1}, o)$ shown in
Figs.~\ref{fig:counterex_tefnd_testing_max_res} and~\ref{fig:counterex_tepr_testing_max_res}, whose success
probabilities are $p_{1}$, $1$, $0$, and $p_{2}$, respectively. The presence of all these resolutions is due
to a \emph{demonic} view of nondeterminism, which allows the considered \emph{almighty} schedulers to
perform different choices in different copies of the same state of the process under test. This is what
happens in the second and in the third maximal resolution, as graphically witnessed by the different
orientation of the two $b$-transitions. In order to be robust with respect to scheduling decisions, these
two resolutions cannot be ruled out and their success probabilities, $1$ and $0$, have to be taken into
account.

As pointed out in~\cite{CLSV06}, in a testing scenario schedulers come into play after the process has been
composed in parallel with the test, and hence can resolve both local and global nondeterministic choices.
This makes it possible for schedulers to make decisions in one component on the basis of the state of the
other component, as if there were an information leakage. However, under specific circumstances, one may
reasonably consider \emph{less powerful} schedulers ensuring that the choices they perform in different
copies of the same state are consistent with each other (see~\cite{GA12} and the references therein). In
that case, the two resolutions mentioned above would no longer make sense. As a consequence, values~$1$
and~$0$ would respectively become an overestimation and an underestimation of the success probability, and
in principle $s_{1}$ and $s_{2}$ could be identified by $\sbis{\textrm{PTe-}\sqcup\sqcap}$ and
$\sbis{\textrm{PTe-}\forall\exists}$. We will discuss again the power of schedulers at the end of
Sect.~\ref{sec:spectrum}.

%
%
\section{Trace-by-Trace Redefinition of Testing Equivalence}
\label{sec:tbt_testing_equiv}
%
%

In this section, we introduce a new testing equivalence for NPLTS models that is \emph{fully} backward
compatible with testing equivalences defined in the literature for restricted classes of processes. In order
to counterbalance the stronger discriminating power deriving from the copying capability enabled by tests
that do not belong to the class of processes under test, our basic idea is changing the definition of
$\sbis{\textrm{PTe-}\forall\exists}$ by considering success probabilities in a \emph{trace-by-trace fashion}
rather than cumulatively over all successful computations of the maximal resolutions.

In the following, given a state $s$ of an NPLTS, a state $o$ of an NPT, and a trace $\alpha \in A^{*}$, we
denote by $\ms{Res}_{{\rm max}, \alpha}(s, o)$ the set of resolutions $\calz \in \ms{Res}_{\rm max}(s, o)$
such that $\calcc_{\rm max}(z_{s, o}, \alpha) \neq \emptyset$, where $\calcc_{\rm max}(z_{s, o}, \alpha)$ is
the set of computations in $\calcc(z_{s, o}, \alpha)$ that are maximal. In other words, $\ms{Res}_{{\rm
max}, \alpha}(s, o)$ is the set of maximal resolutions of $z_{s, o}$ having at least one maximal computation
labeled with $\alpha$; the set $\ms{Res}_{{\rm max}, \alpha}^{\rm ct}(s, o)$ is defined similarly. Moreover,
for each resolution $\calz$ we denote by $\calscc(z_{s, o}, \alpha)$ the set of successful
$\alpha$-compatible computations from $z_{s, o}$.

	\begin{defi}\label{def:ptetbt}

Let $(S, A, \! \arrow{}{} \!)$ be an NPLTS. We say that $s_{1}, s_{2} \in S$ are \emph{probabilistic
trace-by-trace testing equivalent}, written $s_{1} \sbis{\textrm{PTe-tbt}} s_{2}$, iff for every NPT $\calt
= (O, A, \! \arrow{}{\calt} \!)$ with initial state $o \in O$ and \underline{for all $\alpha \in A^{*}$} it
holds that:

		\begin{itemize}

\item For each $\calz_{1} \in \ms{Res}_{{\rm max}, \alpha}(s_{1}, o)$ there exists $\calz_{2} \in
\ms{Res}_{{\rm max}, \alpha}(s_{2}, o)$ such that:
\cws{10}{\hspace*{-1.2cm} \ms{prob}(\calscc(z_{s_{1}, o}, \alpha)) \: = \: \ms{prob}(\calscc(z_{s_{2}, o},
\alpha))}

\item For each $\calz_{2} \in \ms{Res}_{{\rm max}, \alpha}(s_{2}, o)$ there exists $\calz_{1} \in
\ms{Res}_{{\rm max}, \alpha}(s_{1}, o)$ such that:
\cws{12}{\hspace*{-1.2cm} \ms{prob}(\calscc(z_{s_{2}, o}, \alpha)) \: = \: \ms{prob}(\calscc(z_{s_{1}, o},
\alpha))}

		\end{itemize}

\noindent
We denote by $\sbis{\textrm{PTe-tbt}}^{\rm ct}$ the coarser variant based on randomized schedulers.
\fullbox

	\end{defi}

If we consider again the two NPLTS models of Fig.~\ref{fig:counterex_tefnd_testing} (resp.\
Fig.~\ref{fig:counterex_tepr_testing}), it turns out that $s_{1} \sbis{\textrm{PTe-tbt}} s_{2}$, and hence
$s_{1} \sbis{\textrm{PTe-tbt}}^{\rm ct} s_{2}$. The interaction of the two processes with the test in the
same figure originates maximal computations from $(s_{1}, o)$ and $(s_{2}, o)$ that are all labeled with
traces $a \, b$, $a \, b \, c$, and $a \, b \, d$. It is easy to see that, in
Fig.~\ref{fig:counterex_tefnd_testing_max_res} (resp.\ Fig.~\ref{fig:counterex_tepr_testing_max_res}), for
each of these traces, say $\alpha$, the probability of performing a successful $\alpha$-compatible
computation in any of the four maximal resolutions of $(s_{1}, o)$ having a maximal $\alpha$-compatible
computation is matched by the probability of performing a successful $\alpha$-compatible computation in one
of the two maximal resolutions of $(s_{2}, o)$, and vice versa. As an example, the probability $p_{1}$
(resp.~$p_{2}$) of performing a successful computation compatible with $a \, b \, c$ (resp.~$a \, b \, d$)
in the second maximal resolution of $(s_{1}, o)$ is matched by the probability of performing a successful
computation compatible with that trace in the first (resp.\ second) maximal resolution of~$(s_{2}, o)$. As
another example, the probability $0$ of performing a successful computation compatible with $a \, b$ in the
third maximal resolution of $(s_{1}, o)$ is matched by the probability of performing a successful
computation compatible with that trace in any of the two maximal resolutions of $(s_{2}, o)$.

The examples of Figs.~\ref{fig:counterex_tefnd_testing} and~\ref{fig:counterex_tepr_testing} show that
$\sbis{\textrm{PTe-tbt}}$ and $\sbis{\textrm{PTe-tbt}}^{\rm ct}$ are included neither in
$\sbis{\textrm{PTe-}\sqcup\sqcap}$ nor in $\sbis{\textrm{PTe-}\forall\exists}$. On the other hand,
$\sbis{\textrm{PTe-}\sqcup\sqcap}$ is not included in $\sbis{\textrm{PTe-tbt}}$ as witnessed by the two
NPLTS models in Fig.~\ref{fig:counterex_ptesupinf_trace}, because the test in
Fig.~\ref{fig:test_pteallexists} distinguishes $s_{1}$ from~$s_{2}$ with respect to
$\sbis{\textrm{PTe-tbt}}$. In fact, the probability $0.25$ of performing a successful computation compatible
with $a \, b$ in the maximal resolution of $(s_{1}, o)$ beginning with the central $a$-transition is not
matched by the probability $0.5$ of performing a successful computation compatible with $a \, b$ in the only
maximal resolution of $(s_{2}, o)$ that has a maximal computation labeled with~$a \, b$. Thus,
$\sbis{\textrm{PTe-}\sqcup\sqcap}$ and $\sbis{\textrm{PTe-tbt}}$ are incomparable with each other. What
turns out is that $\sbis{\textrm{PTe-}\sqcup\sqcap}$ is (strictly) included in $\sbis{\textrm{PTe-tbt}}^{\rm
ct}$, while $\sbis{\textrm{PTe-}\forall\exists}$ is (strictly) included in $\sbis{\textrm{PTe-tbt}}$ and
hence in $\sbis{\textrm{PTe-tbt}}^{\rm ct}$.

	\begin{thm}\label{thm:pteallexists_incl_ptetbt}

Let $(S, A, \! \arrow{}{} \!)$ be an NPLTS and $s_{1}, s_{2} \in S$. Then:
\cws{10}{\begin{array}{rcl}
s_{1} \sbis{\textrm{{\rm PTe}-}\sqcup\sqcap} s_{2} & \!\!\! \Longrightarrow \!\!\! & s_{1}
\sbis{\textrm{\rm PTe-tbt}}^{\rm ct} s_{2} \\
s_{1} \sbis{\textrm{{\rm PTe}-}\forall\exists} s_{2} & \!\!\! \Longrightarrow \!\!\! & s_{1}
\sbis{\textrm{\rm PTe-tbt}} s_{2} \\
\end{array}}

\proof
Let us initially introduce the following behavioral equivalence: $s_{1} \sbis{\textrm{PTe-tbt},\sqcup\sqcap}
s_{2}$ iff for every NPT $\calt = (O, A, \! \arrow{}{\calt} \!)$ with initial state $o \in O$ and for all
$\alpha \in A^{*}$ it holds that $\ms{Res}_{{\rm max}, \alpha}(s_{1}, o) \neq \emptyset$ iff $\ms{Res}_{{\rm
max}, \alpha}(s_{2}, o) \neq \emptyset$ and:
\cws{0}{\begin{array}{rcl}
\bigsqcup\limits_{\calz_{1} \in \ms{Res}_{\rm max}(s_{1}, o)} \ms{prob}(\calscc(z_{s_{1}, o}, \alpha)) &
\!\!\! = \!\!\! & \bigsqcup\limits_{\calz_{2} \in \ms{Res}_{\rm max}(s_{2}, o)} \ms{prob}(\calscc(z_{s_{2},
o}, \alpha)) \\[0.4cm]
\bigsqcap\limits_{\calz_{1} \in \ms{Res}_{\rm max}(s_{1}, o)} \ms{prob}(\calscc(z_{s_{1}, o}, \alpha)) &
\!\!\! = \!\!\! & \bigsqcap\limits_{\calz_{2} \in \ms{Res}_{\rm max}(s_{2}, o)} \ms{prob}(\calscc(z_{s_{2},
o}, \alpha)) \\
\end{array}}
The proof of $s_{1} \sbis{\textrm{PTe-}\sqcup\sqcap} s_{2} \: \Longrightarrow \: s_{1}
\sbis{\textrm{PTe-tbt}}^{\rm ct} s_{2}$ is divided into two parts:

		\begin{itemize}

\item First, we show that $s_{1} \sbis{\textrm{PTe-}\sqcup\sqcap} s_{2} \: \Longrightarrow \: s_{1}
\sbis{\textrm{PTe-tbt},\sqcup\sqcap}^{\rm ct} s_{2}$. Suppose that $s_{1} \sbis{\textrm{PTe-}\sqcup\sqcap}
s_{2}$ and consider an arbitrary NPT $\calt = (O, A, \! \arrow{}{\calt} \!)$ with initial state $o \in O$.
Given $s \in S$ and $\calz \in \ms{Res}_{\rm max}(s, o)$, it holds that:
\cws{0}{\hspace*{-1.2cm} \ms{prob}(\calsc(z_{s, o})) \: = \: \sum\limits_{\alpha \in A^{*} \, {\rm s.t.} \,
\calcc_{\rm max}(z_{s, o}, \alpha) \neq \emptyset} \ms{prob}(\calscc(z_{s, o}, \alpha))}
If we further consider tests $\calt_{\rm \alpha}$, $\alpha \in A^{*}$, obtained from $\calt$ by making
unsuccessful all the successful computations of $\calt$ not compatible with $\alpha$, we have that for each
such test $\ms{prob}(\calsc^{\calt_{\alpha}}(z_{s, o}))$ reduces to $\ms{prob}(\calscc(z_{s, o}, \alpha))$.
As a consequence, from $s_{1} \sbis{\textrm{PTe-}\sqcup\sqcap} s_{2}$ we derive that for all $\alpha \in
A^{*}$ it holds that $\ms{Res}_{{\rm max}, \alpha}(s_{1}, o) \neq \emptyset$ iff $\ms{Res}_{{\rm max},
\alpha}(s_{2}, o) \neq \emptyset$ and:
\cws{0}{\hspace*{-1.2cm}\begin{array}{rcl}
\bigsqcup\limits_{\calz_{1} \in \ms{Res}_{\rm max}(s_{1}, o)} \ms{prob}(\calscc(z_{s_{1}, o}, \alpha)) &
\!\!\! = \!\!\! & \bigsqcup\limits_{\calz_{2} \in \ms{Res}_{\rm max}(s_{2}, o)} \ms{prob}(\calscc(z_{s_{2},
o}, \alpha)) \\[0.4cm]
\bigsqcap\limits_{\calz_{1} \in \ms{Res}_{\rm max}(s_{1}, o)} \ms{prob}(\calscc(z_{s_{1}, o}, \alpha)) &
\!\!\! = \!\!\! & \bigsqcap\limits_{\calz_{2} \in \ms{Res}_{\rm max}(s_{2}, o)} \ms{prob}(\calscc(z_{s_{2},
o}, \alpha)) \\
\end{array}}
which means that $s_{1} \sbis{\textrm{PTe-tbt},\sqcup\sqcap} s_{2}$. From this, it follows that $s_{1}
\sbis{\textrm{PTe-tbt},\sqcup\sqcap}^{\rm ct} s_{2}$.

\item Second, we show that $s_{1} \sbis{\textrm{PTe-tbt},\sqcup\sqcap}^{\rm ct} s_{2} \: \Longrightarrow \:
s_{1} \sbis{\textrm{PTe-tbt}}^{\rm ct} s_{2}$. Suppose $s_{1} \sbis{\textrm{PTe-tbt},\sqcup\sqcap}^{\rm ct}
s_{2}$ and consider an arbitrary trace $\alpha \in A^{*}$ for which there exists $\calz_{1} \in
\ms{Res}_{{\rm max}, \alpha}^{\rm ct}(s_{1})$ such that $\ms{prob}(\calscc(z_{s_{1}, o}, \alpha)) = p$.
Since $s_{1} \sbis{\textrm{PTe-tbt},\sqcup\sqcap}^{\rm ct} s_{2}$, we have $\ms{Res}_{{\rm max},
\alpha}^{\rm ct}(s_{2}) \neq \emptyset$ and there exist $\calz'_{2}, \calz''_{2} \in \ms{Res}_{{\rm max},
\alpha}^{\rm ct}(s_{2})$ such that $\ms{prob}(\calscc(z'_{s_{2}}, \alpha)) = p' \le p$ and
$\ms{prob}(\calscc(z''_{s_{2}}, \alpha)) = p'' \ge p$. \\
If $p' = p$ (resp.\ $p'' = p$), then $\calz_{1}$ is trivially matched by $\calz'_{2}$ (resp.\ $\calz''_{2}$)
with respect to $\sbis{\textrm{PTe-tbt}}^{\rm ct}$ when examining $\alpha$. \\
Assume that $p' < p < p''$ and consider the resolution $\calz_{2} = x \cdot \calz'_{2} + y \cdot
\calz''_{2}$ of $s_{2}$ defined as follows for $x, y \in \realns_{]0, 1]}$ such that $x + y = 1$. Since $p'
\neq p''$ and they both refer to the probability of performing a successful $\alpha$-compatible computation
from~$s_{2}$, the two resolutions $\calz'_{2}$ and $\calz''_{2}$ of $s_{2}$ differ at least in one point in
which the nondeterministic choice between two transitions labeled with the same action occurring in~$\alpha$
has been resolved differently. We obtain $\calz_{2}$ from $\calz'_{2}$ and $\calz''_{2}$ by combining the
two different transitions into a single one with coefficients $x$ and $y$ for their target distributions,
respectively, in the first of those points. When examining $\alpha$, if we take $x = \frac{p'' - p}{p'' -
p'}$ and $y = \frac{p - p'}{p'' - p'}$, then $\calz_{1}$ is matched by $\calz_{2}$ with respect to
$\sbis{\textrm{PTe-tbt}}^{\rm ct}$ because:
\cws{0}{\hspace*{-1.2cm}\begin{array}{rcl}
\ms{prob}(\calscc(z_{s_{2}, o}, \alpha)) & \!\!\! = \!\!\! & \frac{p'' - p}{p'' - p'} \cdot
\ms{prob}(\calscc(z'_{s_{2}, o}, \alpha)) + \frac{p - p'}{p'' - p'} \cdot \ms{prob}(\calscc(z''_{s_{2}, o},
\alpha)) \\
& \!\!\! = \!\!\! & \frac{p'' - p}{p'' - p'} \cdot p' + \frac{p - p'}{p'' - p'} \cdot p'' \: = \: \frac{p'
\cdot p'' - p \cdot p' + p \cdot p'' - p' \cdot p''}{p'' - p'} \\
& \!\!\! = \!\!\! & p \cdot \frac{p'' - p'}{p'' - p'} \: = \: p \: = \: \ms{prob}(\calscc(z_{s_{1}, o},
\alpha)) \\
\end{array}}
Due to the generality of $\alpha \in A^{*}$, it turns out that $s_{1} \sbis{\textrm{PTe-tbt}}^{\rm ct}
s_{2}$.

		\end{itemize}

\noindent
Suppose now that $s_{1} \sbis{\textrm{PTe-}\forall\exists} s_{2}$ and consider an arbitrary NPT $\calt = (O,
A, \! \arrow{}{} \!)$ with initial state $o \in O$. Then, in particular, for all variants $\calt_{\alpha} =
(O, A, \! \arrow{}{\calt_{\alpha}} \!)$ of $\calt$ in which all the successful computations of $\calt$ not
compatible with $\alpha$ are made unsuccessful, it holds that:

		\begin{itemize}

\item For each $\calz_{1} \in \ms{Res}_{\rm max}(s_{1}, o)$ there exists $\calz_{2} \in \ms{Res}_{\rm
max}(s_{2}, o)$ such that:
\cws{10}{\hspace*{-1.2cm} \ms{prob}(\calsc^{\calt_{\alpha}}(z_{s_{1}, o})) \: = \:
\ms{prob}(\calsc^{\calt_{\alpha}}(z_{s_{2}, o}))}

\item For each $\calz_{2} \in \ms{Res}_{\rm max}(s_{2}, o)$ there exists $\calz_{1} \in \ms{Res}_{\rm
max}(s_{1}, o)$ such that:
\cws{10}{\hspace*{-1.2cm} \ms{prob}(\calsc^{\calt_{\alpha}}(z_{s_{2}, o})) \: = \:
\ms{prob}(\calsc^{\calt_{\alpha}}(z_{s_{1}, o}))}

		\end{itemize}

\noindent
Since $\ms{prob}(\calsc^{\calt_{\alpha}}(z_{s, o})) = \ms{prob}(\calscc(z_{s, o}, \alpha))$ for all $s \in
S$ due to the structure of $\calt_{\alpha}$, we immediately derive that for all $\alpha \in A^{*}$ it holds
that:

		\begin{itemize}

\item For each $\calz_{1} \in \ms{Res}_{{\rm max}, \alpha}(s_{1}, o)$ there exists $\calz_{2} \in
\ms{Res}_{{\rm max}, \alpha}(s_{2}, o)$ such that:
\cws{10}{\hspace*{-1.2cm} \ms{prob}(\calscc(z_{s_{1}, o}, \alpha)) \: = \: \ms{prob}(\calscc(z_{s_{2}, o},
\alpha))}

\item For each $\calz_{2} \in \ms{Res}_{{\rm max}, \alpha}(s_{2}, o)$ there exists $\calz_{1} \in
\ms{Res}_{{\rm max}, \alpha}(s_{1}, o)$ such that:
\cws{10}{\hspace*{-1.2cm} \ms{prob}(\calscc(z_{s_{2}, o}, \alpha)) \: = \: \ms{prob}(\calscc(z_{s_{1}, o},
\alpha))}

		\end{itemize}

\noindent
This means that $s_{1} \sbis{\textrm{PTe-tbt}} s_{2}$.
\qed

	\end{thm}

Apart from the use of $\ms{prob}(\calscc(z_{s, o}, \alpha))$ values instead of $\ms{prob}(\calsc(z_{s, o}))$
values, another major difference between $\sbis{\textrm{PTe-tbt}}$ and $\sbis{\textrm{PTe-}\forall\exists}$
is the consideration of resolutions in $\ms{Res}_{{\rm max}, \alpha}$ rather than in $\ms{Res}_{{\rm max}}$.
In other words, the considered maximal resolutions are those having at least one $\alpha$-compatible
computation that corresponds to a maximal \linebreak $\alpha$-compatible computation in the interaction
system. The motivation behind this restriction is that it is not appropriate to match the $0$ success
probability of \emph{maximal} $\alpha$-compatible computations that are unsuccessful, with the $0$ success
probability of $\alpha$-compatible computations that are \emph{not maximal}, as may happen when considering
$\ms{Res}_{\rm max}$ instead of $\ms{Res}_{{\rm max}, \alpha}$.

\begin{figure}[tp]

\input{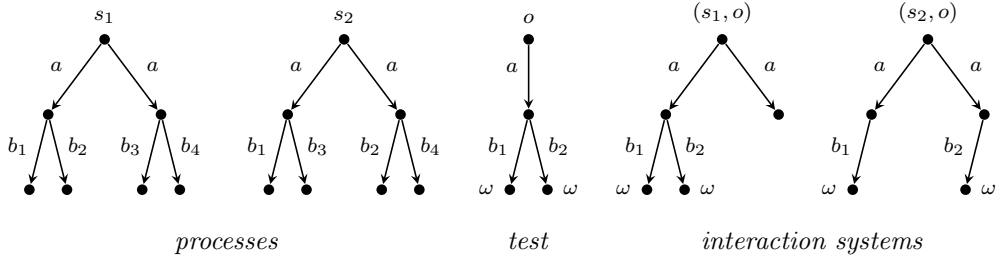}
\caption{NPLTS models distinguished by $\sbis{\textrm{PTe-tbt}}$ thanks to the restriction to
$\ms{Res}_{{\rm max}, \alpha}$}
\label{fig:max_comp_ptetbt}

	\end{figure}

Admitting all maximal resolutions would also cause $\sbis{\textrm{PTe-tbt}}$ not to be conservative with
respect to $\sbis{\rm Te,fnd}$ when restricting attention to fully nondeterministic tests. For example, if
we consider the two fully nondeterministic NPLTS models in Fig.~\ref{fig:max_comp_ptetbt}, it turns out that
$s_{1} \not\sbis{\rm Te,fnd} s_{2}$ because of the fully nondeterministic test in the same figure. In fact,
following the terminology of~\cite{DH84}, the second process must pass that test, while the first one is not
able to do so because the interaction system has a maximal computation labeled with $a$ that does not reach
success. In the setting of $\sbis{\textrm{PTe-tbt}}$, that computation in the first interaction system is
not matched by any computation labeled with $a$ in the second interaction system because of the restriction
to $\ms{Res}_{{\rm max}, a}$, thus correctly distinguishing the two processes. Notice that, under
$\ms{Res}_{\rm max}$, it would be matched by any of the two non-maximal computations labeled with $a$ in the
second interaction system.

We now investigate the inclusion and compatibility properties of
$\sbis{\textrm{PTe-tbt}}$/$\sbis{\textrm{\rm PTe-tbt}}^{\rm ct}$. Similar to
$\sbis{\textrm{PTe-}\forall\exists}$, they result in a testing semantics finer than trace semantics.

	\begin{thm}\label{thm:ptetbt_incl_ptr}

Let $(S, A, \! \arrow{}{} \!)$ be an NPLTS and $s_{1}, s_{2} \in S$. Then:
\cws{12}{\begin{array}{rcl}
s_{1} \sbis{\textrm{\rm PTe-tbt}} s_{2} & \!\!\! \Longrightarrow \!\!\! & s_{1} \sbis{\rm PTr} s_{2} \\
s_{1} \sbis{\textrm{\rm PTe-tbt}}^{\rm ct} s_{2} & \!\!\! \Longrightarrow \!\!\! & s_{1} \sbis{\rm PTr}^{\rm
ct} s_{2} \\
\end{array}}

\proof
If $s_{1} \sbis{\textrm{PTe-tbt}} s_{2}$, then in particular for every NPT $\calt_{\alpha} = (O, A, \!
\arrow{}{\calt_{\alpha}} \!)$ with initial state $o \in O$ having a single maximal computation that is
labeled with $\alpha \in A^{*}$ and reaches success, it holds that:

		\begin{itemize}

\item For each $\calz_{1} \in \ms{Res}_{{\rm max}, \alpha}(s_{1}, o)$ there exists $\calz_{2} \in
\ms{Res}_{{\rm max}, \alpha}(s_{2}, o)$ such that:
\cws{10}{\hspace*{-1.2cm} \ms{prob}(\calscc(z_{s_{1}, o}, \alpha)) \: = \: \ms{prob}(\calscc(z_{s_{2}, o},
\alpha))}

\item For each $\calz_{2} \in \ms{Res}_{{\rm max}, \alpha}(s_{2}, o)$ there exists $\calz_{1} \in
\ms{Res}_{{\rm max}, \alpha}(s_{1}, o)$ such that:
\cws{10}{\hspace*{-1.2cm} \ms{prob}(\calscc(z_{s_{2}, o}, \alpha)) \: = \: \ms{prob}(\calscc(z_{s_{1}, o},
\alpha))}

		\end{itemize}

\noindent
Since $\ms{prob}(\calscc^{\calz}(z_{s, o}, \alpha)) = \ms{prob}(\calcc^{\calz'}(z_{s}, \alpha))$ for all $s
\in S$ due to the structure of $\calt_{\alpha}$ -- where $\calz \in \ms{Res}_{{\rm max}, \alpha}(s, o)$ and
$\calz' \in \ms{Res}(s)$ originates $\calz$ in the interaction with $\calt_{\alpha}$ -- we immediately
derive that for all $\alpha \in A^{*}$ it holds that:

		\begin{itemize}

\item For each $\calz_{1} \in \ms{Res}(s_{1})$ there exists $\calz_{2} \in \ms{Res}(s_{2})$ such that:
\cws{10}{\hspace*{-1.2cm} \ms{prob}(\calcc(z_{s_{1}}, \alpha)) \: = \: \ms{prob}(\calcc(z_{s_{2}}, \alpha))}

\item For each $\calz_{2} \in \ms{Res}(s_{2})$ there exists $\calz_{1} \in \ms{Res}(s_{1})$ such that:
\cws{10}{\hspace*{-1.2cm} \ms{prob}(\calcc(z_{s_{2}}, \alpha)) \: = \: \ms{prob}(\calcc(z_{s_{1}}, \alpha))}

		\end{itemize}

\noindent
This means that $s_{1} \sbis{\rm PTr} s_{2}$. \\
The proof of $s_{1} \sbis{\textrm{PTe-tbt}}^{\rm ct} s_{2} \: \Longrightarrow \: s_{1} \sbis{\rm PTr}^{\rm
ct} s_{2}$ is analogous.
\qed

	\end{thm}

The inclusion of $\sbis{\textrm{PTe-tbt}}$ (resp.\ $\sbis{\textrm{PTe-tbt}}^{\rm ct}$) in $\sbis{\rm PTr}$
(resp.\ $\sbis{\rm PTr}^{\rm ct}$) is strict. For instance, the two NPLTS models in
Fig.~\ref{fig:counterex_pteallexists_trace} are not trace-by-trace testing equivalent. In fact, the test in
the same figure distinguishes $s_{1}$ from $s_{2}$ because -- looking at the two interaction systems in
Fig.~\ref{fig:counterex_pteallexists_trace} -- each of the two maximal resolutions of $(s_{2}, o)$ has a
maximal computation labeled with $a$ while the only maximal resolution of $(s_{1}, o)$ has not.

Unlike $\sbis{\textrm{PTe-}\sqcup\sqcap}$ and $\sbis{\textrm{PTe-}\forall\exists}$,
$\sbis{\textrm{PTe-tbt}}$/$\sbis{\textrm{\rm PTe-tbt}}^{\rm ct}$ result in a testing semantics that is
\emph{fully} (i.e., regardless of admitted tests) backward compatible with $\sbis{\rm Te,fnd}$, $\sbis{\rm
Te,fpr}$, and $\sbis{\rm Te,rpr}$. Concerning the two restricted classes of probabilistic processes, it is
worth recalling that bisimulation equivalence and trace equivalence were defined uniformly for fully
probabilistic processes~\cite{GJS90,JS90} and reactive probabilistic processes~\cite{LS91,Sei95}. In
contrast, testing equivalence for fully probabilistic processes was defined in~\cite{Chr90,CDSY99} in a way
that resembles $\sbis{\textrm{PTe-}\forall\exists}$, while for reactive probabilistic processes it was
defined in~\cite{KN98} in a way similar to $\sbis{\textrm{PTe-}\sqcup\sqcap}$. Our compatibility results
thus show that also testing equivalence could have been defined uniformly for both classes of probabilistic
processes without internal nondeterminism, by resorting to the trace-by-trace approach that we have
developed for NPLTS models.

	\begin{thm}\label{thm:ptetbt_compat}

Let $\call = (S, A, \! \arrow{}{} \!)$ be an NPLTS and $s_{1}, s_{2} \in S$.

		\begin{enumerate}

\item If $\call$ is fully nondeterministic, then:
\cws{10}{\hspace*{-1.2cm} s_{1} \sbis{\textrm{\rm PTe-tbt}} s_{2} \: \Longleftrightarrow \: s_{1}
\sbis{\textrm{\rm PTe-tbt}}^{\rm ct} s_{2} \: \Longleftrightarrow \: s_{1} \sbis{\rm Te,fnd} s_{2}}

\item If $\call$ is fully probabilistic, then:
\cws{10}{\hspace*{-1.2cm} s_{1} \sbis{\textrm{\rm PTe-tbt}} s_{2} \: \Longleftrightarrow \: s_{1}
\sbis{\textrm{\rm PTe-tbt}}^{\rm ct} s_{2} \: \Longleftrightarrow \: s_{1} \sbis{\rm Tr,fpr} s_{2}}

\item If $\call$ is reactive probabilistic, then:
\cws{10}{\hspace*{-1.2cm}\begin{array}{rcl}
s_{1} \sbis{\textrm{\rm PTe-tbt}} s_{2} & \!\!\! \Longrightarrow \!\!\! & s_{1} \sbis{\rm Tr,rpr} s_{2} \\
s_{1} \sbis{\textrm{\rm PTe-tbt}}^{\rm ct} s_{2} & \!\!\! \Longrightarrow \!\!\! & s_{1} \sbis{\rm Tr,rpr}
s_{2} \\
\end{array}}

		\end{enumerate}

\proof
We proceed as follows:

		\begin{enumerate}

\item Suppose that $\call$ is fully nondeterministic. We recall from~\cite{DH84} that $s_{1} \sbis{\rm
Te,fnd} s_{2}$ means that for every \emph{fully nondeterministic} NPT $\calt = (O, A, \! \arrow{}{\calt}
\!)$ with initial state $o \in O$ it holds that:

			\begin{itemize}

\item There exists a successful computation from $(s_{1}, o)$ iff there exists a successful computation from
$(s_{2}, o)$.

\item All maximal computations from $(s_{1}, o)$ are successful iff all maximal computations from $(s_{2},
o)$ are successful.

			\end{itemize}

\noindent
In this setting, randomized schedulers are not important because, due to the absence of probabilistic
choices, the model cannot contain submodels that arise from convex combinations of other submodels. Thus, we
can concentrate on $\sbis{\textrm{PTe-tbt}}$. Suppose that $s_{1} \sbis{\textrm{PTe-tbt}} s_{2}$. Then, in
particular, for every \emph{fully nondeterministic} NPT $\calt = (O, A, \! \arrow{}{\calt} \!)$ with initial
state $o \in O$ and for all $\alpha \in A^{*}$ it holds that:

			\begin{itemize}

\item For each $\calz_{1} \in \ms{Res}_{{\rm max}, \alpha}(s_{1}, o)$ there exists $\calz_{2} \in
\ms{Res}_{{\rm max}, \alpha}(s_{2}, o)$ such that:
\cws{10}{\hspace*{-2.4cm} \ms{prob}(\calscc(z_{s_{1}, o}, \alpha)) \: = \: \ms{prob}(\calscc(z_{s_{2}, o},
\alpha))}

\item For each $\calz_{2} \in \ms{Res}_{{\rm max}, \alpha}(s_{2}, o)$ there exists $\calz_{1} \in
\ms{Res}_{{\rm max}, \alpha}(s_{1}, o)$ such that:
\cws{10}{\hspace*{-2.4cm} \ms{prob}(\calscc(z_{s_{2}, o}, \alpha)) \: = \: \ms{prob}(\calscc(z_{s_{1}, o},
\alpha))}

			\end{itemize}

\noindent
Since the NPLTS under test and the considered tests are all fully nondeterministic, the resulting
interaction systems are fully nondeterministic too, and hence their maximal resolutions coincide with their
maximal computations and each of the probability values above is either $1$ or $0$. As a consequence, the
previous relationships among maximal resolutions can be rephrased as follows:

			\begin{itemize}

\item For each maximal $\alpha$-compatible computation from $(s_{1}, o)$ there exists a maximal
$\alpha$-compatible computation from $(s_{2}, o)$ such that the two computations are both successful or both
unsuccessful.

\item For each maximal $\alpha$-compatible computation from $(s_{2}, o)$ there exists a maximal
$\alpha$-compatible computation from $(s_{1}, o)$ such that the two computations are both successful or both
unsuccessful.

			\end{itemize}

\noindent
From this, we immediately derive that:

			\begin{itemize}

\item There exists a successful computation from $(s_{1}, o)$ iff there exists a successful computation from
$(s_{2}, o)$.

\item All maximal computations from $(s_{1}, o)$ are successful iff all maximal computations from $(s_{2},
o)$ are successful. In fact, assume that all maximal computations from, e.g., $(s_{1}, o)$ are successful.
Then at least one maximal computation from $(s_{2}, o)$ is successful. Assume that $(s_{2}, o)$ has at least
two maximal computations and that one of them is not successful. Then at least one maximal computation from
$(s_{1}, o)$ would not be successful, thus contradicting the assumption that all maximal computations from
$(s_{1}, o)$ are successful. Therefore, whenever all maximal computations from $(s_{1}, o)$ are successful,
then all maximal computations from $(s_{2}, o)$ are successful. Likewise, whenever all maximal computations
from $(s_{2}, o)$ are successful, then all maximal computations from $(s_{1}, o)$ are successful.

			\end{itemize}

\noindent
This means that $s_{1} \sbis{\rm Te,fnd} s_{2}$. \\
Suppose now that $s_{1} \sbis{\rm Te,fnd} s_{2}$ and consider an \emph{arbitrary} NPT $\calt = (O, A, \!
\arrow{}{\calt} \!)$ with initial state $o \in O$, an arbitrary trace $\alpha \in A^{*}$ such that
$\ms{Res}_{{\rm max}, \alpha}(s_{1}, o) \neq \emptyset$, and an arbitrary resolution $\calz_{1} \in
\ms{Res}_{{\rm max}, \alpha}(s_{1}, o)$. \\
Assume that $\ms{Res}_{{\rm max}, \alpha}(s_{2}, o) = \emptyset$, i.e., assume that for all $\calz_{2} \in
\ms{Res}_{\rm max}(s_{2}, o)$ it holds that $\calcc_{\rm max}(z_{s_{2}, o}, \alpha) = \emptyset$. Let
$\calt_{\alpha} = (O, A, \! \arrow{}{\calt_{\alpha}} \!)$ be a \emph{fully nondeterministic} NPT obtained
from $\calt$ in which (i)~only the maximal $\alpha$-compatible computations reach $\omega$ and (ii)~each
transition $o' \arrow{a}{\calt} \cald$ such that the set $O' = \{ o'' \in O \mid \cald(o'') > 0 \}$ has
cardinality greater than $1$ is transformed into $|O'|$ transitions $o' \arrow{a}{\calt_{\alpha}}
\cald_{o''}$, $o'' \in O'$, where $\cald_{o''}(o'') = 1$ and $\cald_{o''}(o''') = 0$ for all $o''' \in O
\setminus \{ o'' \}$. Observing that $\calt_{\alpha}$ yields the same $\alpha$-compatible computations as
$\calt$ in the interaction systems, the test $\calt_{\alpha}$ would violate $s_{1} \sbis{\rm Te,fnd} s_{2}$
because at least one maximal computation from $(s_{1}, o)$ is successful whilst there are no maximal
computations from $(s_{2}, o)$ that are successful. We have thus deduced that, whenever $s_{1} \sbis{\rm
Te,fnd} s_{2}$, then the existence of $\calz_{1} \in \ms{Res}_{{\rm max}, \alpha}(s_{1}, o)$ implies the
existence of $\calz_{2} \in \ms{Res}_{{\rm max}, \alpha}(s_{2}, o)$. \\
Assume now that for all $\calz_{2} \in \ms{Res}_{{\rm max}, \alpha}(s_{2}, o)$ it holds that:
\cws{0}{\hspace*{-1.2cm} \ms{prob}(\calscc(z_{s_{1}, o}, \alpha)) \: \neq \: \ms{prob}(\calscc(z_{s_{2}, o},
\alpha))}
Observing that $\calt$ must have a successful $\alpha$-compatible computation -- otherwise it would hold
that $\ms{prob}(\calscc(z_{s_{1}, o}, \alpha)) = 0 = \ms{prob}(\calscc(z_{s_{2}, o}, \alpha))$ for all
$\calz_{2} \in \ms{Res}_{{\rm max}, \alpha}(s_{2}, o)$ -- from $\calcc_{\rm max}(z_{s_{1}, o}, \alpha) \neq
\emptyset$ and $\calcc_{\rm max}(z_{s_{2}, o}, \alpha) \neq \emptyset$ we derive that
$\ms{prob}(\calscc(z_{s_{1}, o}, \alpha)) > 0$ and $\ms{prob}(\calscc(z_{s_{2}, o}, \alpha)) > 0$. Denoting
by $\calz'_{1}$ the element of $\ms{Res}_{\rm max}(s_{1})$ that originates $\calz_{1}$, we would then have
that for each $\calz'_{2} \in \ms{Res}_{\rm max}(s_{2})$ originating $\calz_{2}$:
\cws{0}{\hspace*{-1.2cm}\begin{array}{rcccl}
\ms{prob}(\calcc(z'_{s_{1}}, \alpha)) & \!\!\! = \!\!\! & \ms{prob}(\calscc(z_{s_{1}, o}, \alpha)) / p &
\!\!\! \neq \!\!\! & \\
& \!\!\! \neq \!\!\! & \ms{prob}(\calscc(z_{s_{2}, o}, \alpha)) / p & \!\!\! = \!\!\! &
\ms{prob}(\calcc(z'_{s_{2}}, \alpha)) \\
\end{array}}
where $p$ is the probability of performing a successful $\alpha$-compatible computation in the element
$\calz$ of $\ms{Res}_{\rm max}(o)$ that originates $\calz_{1}$. However, since the NPLTS under test is fully
nondeterministic, $\calz'_{1}$ and $\calz'_{2}$ boil down to two $\alpha$-compatible computations and it
holds that:
\cws{0}{\hspace*{-1.2cm} \ms{prob}(\calcc(z'_{s_{1}}, \alpha)) \: = \: 1 \: = \:
\ms{prob}(\calcc(z'_{s_{2}}, \alpha))}
which contradicts what established before. \\
In conclusion, whenever $s_{1} \sbis{\rm Te,fnd} s_{2}$, then for each $\calz_{1} \in \ms{Res}_{{\rm max},
\alpha}(s_{1}, o)$ there exists $\calz_{2} \in \ms{Res}_{{\rm max}, \alpha}(s_{2}, o)$ such that:
\cws{0}{\hspace*{-1.2cm} \ms{prob}(\calscc(z_{s_{1}, o}, \alpha)) \: = \: \ms{prob}(\calscc(z_{s_{2}, o},
\alpha))}
With a similar argument, we can prove that, whenever $s_{1} \sbis{\rm Te,fnd} s_{2}$, then for each
$\calz_{2} \in \ms{Res}_{{\rm max}, \alpha}(s_{2}, o)$ there exists $\calz_{1} \in \ms{Res}_{{\rm max},
\alpha}(s_{1}, o)$ such that:
\cws{0}{\hspace*{-1.2cm} \ms{prob}(\calscc(z_{s_{2}, o}, \alpha)) \: = \: \ms{prob}(\calscc(z_{s_{1}, o},
\alpha))}
This means that $s_{1} \sbis{\textrm{PTe-tbt}} s_{2}$.

\item Suppose that $\call$ is fully probabilistic. We recall from~\cite{CDSY99} that $s_{1} \sbis{\rm
Te,fpr} s_{2}$ means that for every \emph{fully probabilistic} NPT $\calt = (O, A, \! \arrow{}{\calt} \!)$
with initial state $o \in O$ it holds that:
\cws{0}{\hspace*{-1.2cm} \ms{prob}(\calsc(s_{1}, o)) \: = \: \ms{prob}(\calsc(s_{2}, o))}
In this setting, schedulers are not important because there is no nondeterminism. Thus, we can concentrate
on $\sbis{\textrm{PTe-tbt}}$. Suppose that $s_{1} \sbis{\textrm{PTe-tbt}} s_{2}$. Then, in particular, for
every \emph{fully probabilistic} NPT $\calt = (O, A, \! \arrow{}{\calt} \!)$ with initial state $o \in O$
and for all $\alpha \in A^{*}$ it holds that:

			\begin{itemize}

\item For each $\calz_{1} \in \ms{Res}_{{\rm max}, \alpha}(s_{1}, o)$ there exists $\calz_{2} \in
\ms{Res}_{{\rm max}, \alpha}(s_{2}, o)$ such that:
\cws{10}{\hspace*{-2.4cm} \ms{prob}(\calscc(z_{s_{1}, o}, \alpha)) \: = \: \ms{prob}(\calscc(z_{s_{2}, o},
\alpha))}

\item For each $\calz_{2} \in \ms{Res}_{{\rm max}, \alpha}(s_{2}, o)$ there exists $\calz_{1} \in
\ms{Res}_{{\rm max}, \alpha}(s_{1}, o)$ such that:
\cws{10}{\hspace*{-2.4cm} \ms{prob}(\calscc(z_{s_{2}, o}, \alpha)) \: = \: \ms{prob}(\calscc(z_{s_{1}, o},
\alpha))}

			\end{itemize}

\noindent
Since the NPLTS under test and the considered tests are all fully probabilistic, the resulting interaction
systems are fully probabilistic too, and hence each of them has a single maximal resolution that coincides
with the interaction system itself. As a consequence, the previous relationships among maximal resolutions
can be rephrased by saying that for all $\alpha \in A^{*}$:
\cws{0}{\hspace*{-1.2cm} \ms{prob}(\calscc((s_{1}, o), \alpha)) \: = \: \ms{prob}(\calscc((s_{2}, o),
\alpha))}
From this, we immediately derive that:
\cws{0}{\hspace*{-1.2cm}\begin{array}{rcccl}
\ms{prob}(\calsc(s_{1}, o)) & \!\!\! = \!\!\! & \sum\limits_{\alpha \in A^{*}} \ms{prob}(\calscc((s_{1}, o),
\alpha)) & \!\!\! = \!\!\! & \\[0.4cm]
& \!\!\! = \!\!\! & \sum\limits_{\alpha \in A^{*}} \ms{prob}(\calscc((s_{2}, o), \alpha)) & \!\!\! = \!\!\!
& \ms{prob}(\calsc(s_{2}, o)) \\
\end{array}}
which means that $s_{1} \sbis{\rm Te,fpr} s_{2}$. \\
Suppose now that $s_{1} \sbis{\rm Te,fpr} s_{2}$ and consider an \emph{arbitrary} NPT $\calt = (O, A, \!
\arrow{}{\calt} \!)$ with initial state $o \in O$, an arbitrary trace $\alpha \in A^{*}$ such that
$\ms{Res}_{{\rm max}, \alpha}(s_{1}, o) \neq \emptyset$, and an arbitrary resolution $\calz_{1} \in
\ms{Res}_{{\rm max}, \alpha}(s_{1}, o)$. \\
Assume that $\ms{Res}_{{\rm max}, \alpha}(s_{2}, o) = \emptyset$, i.e., assume that for all $\calz_{2} \in
\ms{Res}_{\rm max}(s_{2}, o)$ it holds that $\calcc_{\rm max}(z_{s_{2}, o}, \alpha) = \emptyset$. Let
$\calt_{\alpha} = (O, A, \! \arrow{}{\calt_{\alpha}} \!)$ be a \emph{fully probabilistic} NPT obtained from
$\calt$ in which (i)~only the maximal $\alpha$-compatible computations reach $\omega$, (ii)~each state $o'
\in O$ having at most one outgoing transition $o' \arrow{a}{\calt} \cald$ retains all of its transitions,
and (iii)~any other state in $O$ retains among its transitions only one of those that are instrumental to
preserve the original $\alpha$-compatible computations of~$\calt$. Observing that $\calt_{\alpha}$ yields at
least one of the $\alpha$-compatible computations of $\calt$ in the interaction systems, the test
$\calt_{\alpha}$ would violate $s_{1} \sbis{\rm Te,fpr} s_{2}$ because at least one maximal computation from
$(s_{1}, o)$ is successful whilst there are no maximal computations from $(s_{2}, o)$ that are successful.
We have thus deduced that, whenever $s_{1} \sbis{\rm Te,fpr} s_{2}$, then the existence of $\calz_{1} \in
\ms{Res}_{{\rm max}, \alpha}(s_{1}, o)$ implies the existence of $\calz_{2} \in \ms{Res}_{{\rm max},
\alpha}(s_{2}, o)$. \\
Assume now that for all $\calz_{2} \in \ms{Res}_{{\rm max}, \alpha}(s_{2}, o)$ it holds that:
\cws{0}{\hspace*{-1.2cm} \ms{prob}(\calscc(z_{s_{1}, o}, \alpha)) \: \neq \: \ms{prob}(\calscc(z_{s_{2}, o},
\alpha))}
Observing that $\calt$ must have a successful $\alpha$-compatible computation -- otherwise it would hold
that $\ms{prob}(\calscc(z_{s_{1}, o}, \alpha)) = 0 = \ms{prob}(\calscc(z_{s_{2}, o}, \alpha))$ for all
$\calz_{2} \in \ms{Res}_{{\rm max}, \alpha}(s_{2}, o)$ -- from $\calcc_{\rm max}(z_{s_{1}, o}, \alpha) \neq
\emptyset$ and $\calcc_{\rm max}(z_{s_{2}, o}, \alpha) \neq \emptyset$ we derive that
$\ms{prob}(\calscc(z_{s_{1}, o}, \alpha)) > 0$ and $\ms{prob}(\calscc(z_{s_{2}, o}, \alpha)) > 0$. Denoting
by $\calz'_{1}$ the element of $\ms{Res}_{\rm max}(s_{1})$ that originates $\calz_{1}$, we would then have
that for each $\calz'_{2} \in \ms{Res}_{\rm max}(s_{2})$ originating $\calz_{2}$:
\cws{0}{\hspace*{-1.2cm}\begin{array}{rcccl}
\ms{prob}(\calcc(z'_{s_{1}}, \alpha)) & \!\!\! = \!\!\! & \ms{prob}(\calscc(z_{s_{1}, o}, \alpha)) / p &
\!\!\! \neq \!\!\! & \\
& \!\!\! \neq \!\!\! & \ms{prob}(\calscc(z_{s_{2}, o}, \alpha)) / p & \!\!\! = \!\!\! &
\ms{prob}(\calcc(z'_{s_{2}}, \alpha)) \\
\end{array}}
where $p$ is the probability of performing a successful $\alpha$-compatible computation in the element
$\calz$ of $\ms{Res}_{\rm max}(o)$ that originates $\calz_{1}$. However, since the NPLTS under test is fully
probabilistic, it holds that:
\cws{0}{\hspace*{-1.2cm}\begin{array}{rcl}
\ms{prob}(\calcc(z'_{s_{1}}, \alpha)) & \!\!\! = \!\!\! & \ms{prob}(\calcc(s_{1}, \alpha)) \\
\ms{prob}(\calcc(z'_{s_{2}}, \alpha)) & \!\!\! = \!\!\! & \ms{prob}(\calcc(s_{2}, \alpha)) \\
\end{array}}
where:
\cws{0}{\hspace*{-1.2cm} \ms{prob}(\calcc(s_{1}, \alpha)) \: = \: \ms{prob}(\calcc(s_{2}, \alpha))}
because otherwise $s_{1} \sbis{\rm Te,fpr} s_{2}$ would be violated by a test having a single maximal
computation that is labeled with $\alpha$ and reaches $\omega$. Thus:
\cws{0}{\hspace*{-1.2cm} \ms{prob}(\calcc(z'_{s_{1}}, \alpha)) \: = \: \ms{prob}(\calcc(z'_{s_{2}},
\alpha))}
which contradicts what established before. \\
In conclusion, whenever $s_{1} \sbis{\rm Te,fpr} s_{2}$, then for each $\calz_{1} \in \ms{Res}_{{\rm max},
\alpha}(s_{1}, o)$ there exists $\calz_{2} \in \ms{Res}_{{\rm max}, \alpha}(s_{2}, o)$ such that:
\cws{0}{\hspace*{-1.2cm} \ms{prob}(\calscc(z_{s_{1}, o}, \alpha)) \: = \: \ms{prob}(\calscc(z_{s_{2}, o},
\alpha))}
With a similar argument, we can prove that, whenever $s_{1} \sbis{\rm Te,fpr} s_{2}$, then for each
$\calz_{2} \in \ms{Res}_{{\rm max}, \alpha}(s_{2}, o)$ there exists $\calz_{1} \in \ms{Res}_{{\rm max},
\alpha}(s_{1}, o)$ such that:
\cws{0}{\hspace*{-1.2cm} \ms{prob}(\calscc(z_{s_{2}, o}, \alpha)) \: = \: \ms{prob}(\calscc(z_{s_{1}, o},
\alpha))}
This means that $s_{1} \sbis{\textrm{PTe-tbt}} s_{2}$.

\item Suppose that $\call$ is reactive probabilistic. Taking inspiration from~\cite{KN98}, $s_{1} \sbis{\rm
Te,rpr} s_{2}$ means that for every \emph{reactive probabilistic} NPT $\calt = (O, A, \! \arrow{}{\calt}
\!)$ with initial state $o \in O$ it holds that:
\cws{0}{\hspace*{-1.2cm}\begin{array}{rcl}
\bigsqcup\limits_{\alpha \in \ms{Tr}_{\rm max}(s_{1}, o)} \ms{prob}(\calscc((s_{1}, o), \alpha)) & \!\!\! =
\!\!\! & \bigsqcup\limits_{\alpha \in \ms{Tr}_{\rm max}(s_{2}, o)} \ms{prob}(\calscc((s_{2}, o), \alpha))
\\[0.4cm]
\bigsqcap\limits_{\alpha \in \ms{Tr}_{\rm max}(s_{1}, o)} \ms{prob}(\calscc((s_{1}, o), \alpha)) & \!\!\! =
\!\!\! & \bigsqcap\limits_{\alpha \in \ms{Tr}_{\rm max}(s_{2}, o)} \ms{prob}(\calscc((s_{2}, o), \alpha)) \\
\end{array}}
Given $s \in S$, the set $\ms{Tr}_{\rm max}(s, o)$ contains all the traces labeling the maximal computations
from $(s, o)$, while success probabilities $\ms{prob}(\calscc((s, o), \alpha))$ are viewed as being
conditional on selecting the maximal resolution of $(s, o)$ that contains all the $\alpha$-compatible
computations from $(s, o)$ (this resolution is unique because interaction systems are reactive
probabilistic). \\
Suppose that $s_{1} \sbis{\textrm{PTe-tbt}} s_{2}$. Then, in particular, for every \emph{reactive
probabilistic} NPT $\calt = (O, A, \! \arrow{}{\calt} \!)$ with initial state $o \in O$ and for all $\alpha
\in A^{*}$ it holds that:

			\begin{itemize}

\item For each $\calz_{1} \in \ms{Res}_{{\rm max}, \alpha}(s_{1}, o)$ there exists $\calz_{2} \in
\ms{Res}_{{\rm max}, \alpha}(s_{2}, o)$ such that:
\cws{10}{\hspace*{-2.4cm} \ms{prob}(\calscc(z_{s_{1}, o}, \alpha)) \: = \: \ms{prob}(\calscc(z_{s_{2}, o},
\alpha))}

\item For each $\calz_{2} \in \ms{Res}_{{\rm max}, \alpha}(s_{2}, o)$ there exists $\calz_{1} \in
\ms{Res}_{{\rm max}, \alpha}(s_{1}, o)$ such that:
\cws{10}{\hspace*{-2.4cm} \ms{prob}(\calscc(z_{s_{2}, o}, \alpha)) \: = \: \ms{prob}(\calscc(z_{s_{1}, o},
\alpha))}

			\end{itemize}

\noindent
Since the NPLTS under test and the considered tests are all reactive probabilistic, the resulting
interaction systems are reactive probabilistic too, and hence in each of them there is a unique maximal
resolution that collects all the computations compatible with a given maximal trace. As a consequence, from
the previous relationships among maximal resolutions we derive that for all $\alpha \in A^{*}$:
\cws{0}{\hspace*{-1.2cm} \ms{prob}(\calscc((s_{1}, o), \alpha)) \: = \: \ms{prob}(\calscc((s_{2}, o),
\alpha))}
From this, we immediately derive that:
\cws{0}{\hspace*{-1.2cm}\begin{array}{rcl}
\bigsqcup\limits_{\alpha \in \ms{Tr}_{\rm max}(s_{1}, o)} \ms{prob}(\calscc((s_{1}, o), \alpha)) & \!\!\! =
\!\!\! & \bigsqcup\limits_{\alpha \in \ms{Tr}_{\rm max}(s_{2}, o)} \ms{prob}(\calscc((s_{2}, o), \alpha))
\\[0.4cm]
\bigsqcap\limits_{\alpha \in \ms{Tr}_{\rm max}(s_{1}, o)} \ms{prob}(\calscc((s_{1}, o), \alpha)) & \!\!\! =
\!\!\! & \bigsqcap\limits_{\alpha \in \ms{Tr}_{\rm max}(s_{2}, o)} \ms{prob}(\calscc((s_{2}, o), \alpha)) \\
\end{array}}
which means that $s_{1} \sbis{\rm Te,rpr} s_{2}$. \\
The proof that $s_{1} \sbis{\textrm{PTe-tbt}}^{\rm ct} s_{2}$ implies $s_{1} \sbis{\rm Te,rpr} s_{2}$ is
similar.
\qed

		\end{enumerate}

	\end{thm}

\noindent In~\cite{DGHM08}, it was shown that $\sbis{\textrm{PTe-}\sqcup\sqcap}$ is a congruence with respect to
parallel composition. To conclude, we prove that also the trace-by-trace approach results in a compositional
testing semantics.

	\begin{thm}\label{thm:ptetbt_compos}

Let $\call_{k} = (S_{k}, A, \! \arrow{}{k} \!)$ be an NPLTS for $k = 0, 1, 2$ and consider $\call_{1}
\pco{\cala} \call_{0}$ and $\call_{2} \pco{\cala} \call_{0}$ for $\cala \subseteq A$. Let $s_{k} \in S_{k}$
for $k = 0, 1, 2$. Then:
\cws{10}{\begin{array}{rcl}
s_{1} \sbis{\textrm{\rm PTe-tbt}} s_{2} & \!\!\! \Longrightarrow \!\!\! & (s_{1}, s_{0}) \sbis{\textrm{\rm
PTe-tbt}} (s_{2}, s_{0}) \\
s_{1} \sbis{\textrm{\rm PTe-tbt}}^{\rm ct} s_{2} & \!\!\! \Longrightarrow \!\!\! & (s_{1}, s_{0})
\sbis{\textrm{\rm PTe-tbt}}^{\rm ct} (s_{2}, s_{0}) \\
\end{array}}

\proof
Given an arbitrary NPT $\calt = (O, A, \! \arrow{}{\calt} \!)$ with initial state $o \in O$, first of all we
observe that $\call_{0} \pco{} \calt$ is still an NPT, with initial state $(s_{0}, o) \in S_{0} \times O$.
\\
If $s_{1} \sbis{\textrm{PTe-tbt}} s_{2}$, then in particular for all $\alpha \in A^{*}$ it holds that:

		\begin{itemize}

\item For each $\calz_{1} \in \ms{Res}_{{\rm max}, \alpha}(s_{1}, (s_{0}, o))$ there exists $\calz_{2} \in
\ms{Res}_{{\rm max}, \alpha}(s_{2}, (s_{0}, o))$ such that:
\cws{10}{\hspace*{-1.2cm} \ms{prob}(\calscc(z_{s_{1}, (s_{0}, o)}, \alpha)) \: = \:
\ms{prob}(\calscc(z_{s_{2}, (s_{0}, o)}, \alpha))}

\item For each $\calz_{2} \in \ms{Res}_{{\rm max}, \alpha}(s_{2}, (s_{0}, o))$ there exists $\calz_{1} \in
\ms{Res}_{{\rm max}, \alpha}(s_{1}, (s_{0}, o))$ such that:
\cws{10}{\hspace*{-1.2cm} \ms{prob}(\calscc(z_{s_{2}, (s_{0}, o)}, \alpha)) \: = \:
\ms{prob}(\calscc(z_{s_{1}, (s_{0}, o)}, \alpha))}

		\end{itemize}

\noindent
For $h = 1, 2$, we note that $(s_{h}, (s_{0}, o))$ is a configuration of $\call_{h} \pco{} (\call_{0} \pco{}
\calt)$ while $((s_{h}, s_{0}), o)$ is a configuration of $(\call_{h} \pco{\cala} \call_{0}) \pco{} \calt$,
hence $\ms{Res}_{{\rm max}, \alpha}(s_{h}, (s_{0}, o)) \subseteq \ms{Res}_{{\rm max}, \alpha}((s_{h},
s_{0}), o)$ because $\call_{h} \pco{} (\call_{0} \pco{} \calt)$ is fully synchronous. There are three cases.
\\
If $\cala = A$, then $(\call_{h} \pco{\cala} \call_{0}) \pco{} \calt = (\call_{h} \pco{} \call_{0}) \pco{}
\calt$ and we can exploit associativity of $\pco{}$ to establish that $\ms{Res}_{{\rm max}, \alpha}(s_{h},
(s_{0}, o)) = \ms{Res}_{{\rm max}, \alpha}((s_{h}, s_{0}), o)$ for $h = 1, 2$. \\
If $\cala \subset A$ and $\call_{1}$ and $\call_{2}$ have no transitions labeled with actions not in
$\cala$, then for $h = 1, 2$ it holds that all transitions of $\call_{h}$ must synchronize with transitions
of $\calt$ both in $\call_{h} \pco{} (\call_{0} \pco{} \calt)$ and in $(\call_{h} \pco{\cala} \call_{0})
\pco{} \calt$, hence possible resolutions in $\ms{Res}_{{\rm max}, \alpha}((s_{h}, s_{0}), o)$ that do not
belong to $\ms{Res}_{{\rm max}, \alpha}(s_{h}, (s_{0}, o))$ are due to transitions of $\call_{0}$ not
labeled with actions in $\cala$ that synchronize with transitions of $\calt$. \\
If $\cala \subset A$ and $\call_{1}$ and $\call_{2}$ have transitions labeled with actions not in $\cala$,
then these transitions (which originate resolutions in $\ms{Res}_{{\rm max}, \alpha}((s_{h}, \! s_{0}), o)$
that do not belong to $\ms{Res}_{{\rm max}, \alpha}(s_{h}, (s_{0}, \! o))$ for $h = 1, 2$) must occur in
corresponding points of $\call_{1}$ and $\call_{2}$ (otherwise we could find a test that distinguishes
$s_{1}$ from $s_{2}$ with respect to $\sbis{\textrm{PTe-tbt}}$) and must synchronize with transitions of
$\calt$ in order for them to emerge in the interaction systems. \\
In each of the three cases, for all $\alpha \in A^{*}$ it holds that:

		\begin{itemize}

\item For each $\calz_{1} \in \ms{Res}_{{\rm max}, \alpha}((s_{1}, s_{0}), o)$ there exists $\calz_{2} \in
\ms{Res}_{{\rm max}, \alpha}((s_{2}, s_{0}), o)$ such that:
\cws{10}{\hspace*{-1.2cm} \ms{prob}(\calscc(z_{(s_{1}, s_{0}), o}, \alpha)) \: = \:
\ms{prob}(\calscc(z_{(s_{2}, s_{0}), o}, \alpha))}

\item For each $\calz_{2} \in \ms{Res}_{{\rm max}, \alpha}((s_{2}, s_{0}), o)$ there exists $\calz_{1} \in
\ms{Res}_{{\rm max}, \alpha}((s_{1}, s_{0}), o)$ such that:
\cws{10}{\hspace*{-1.2cm} \ms{prob}(\calscc(z_{(s_{2}, s_{0}), o}, \alpha)) \: = \:
\ms{prob}(\calscc(z_{(s_{1}, s_{0}), o}, \alpha))}

		\end{itemize}

\noindent
This means that $(s_{1}, s_{0}) \sbis{\textrm{PTe-tbt}} (s_{2}, s_{0})$ because $\calt$ is an arbitrary NPT.
\\
The proof of compositionality for $\sbis{\textrm{PTe-tbt}}^{\rm ct}$ is analogous.
\qed

	\end{thm}

\section{Placing Trace and Testing Equivalences in a Spectrum}
\label{sec:spectrum}
%
%

In this section, we investigate the relationships between the various equivalences that we have recalled
from the literature ($\sbis{\rm PTr,dis}$ and $\sbis{\textrm{PTe-}\sqcup\sqcap}$) or introduced for the
first time ($\sbis{\rm PTr}$, $\sbis{\textrm{PTe-}\forall\exists}$, and $\sbis{\textrm{PTe-tbt}}$) together
with their variants based on randomized schedulers. Some inclusion, coincidence, and incomparability results
have already been established in Thms.~\ref{thm:ptrdis_incl_ptr}, \ref{thm:ptesupinf_sched},
\ref{thm:pteallexists_incl_ptesupinf}, \ref{thm:pteallexists_incl_ptr}, \ref{thm:pteallexists_incl_ptetbt},
and~\ref{thm:ptetbt_incl_ptr}.

We start by providing a surprising characterization of the finest relation considered so far, i.e.,
$\sbis{\textrm{PTe-}\forall\exists}$, that will be useful later on to establish a connection with failure
semantics. The characterization is expressed in terms of a variant of $\sbis{\textrm{PTe-tbt}}$, denoted by
$\sbis{\textrm{PTe-tbt,dis}}$, that is inspired by $\sbis{\rm PTr,dis}$ and hence considers successful trace
distributions.

	\begin{defi}\label{def:ptetbtdis}

Let $(S, A, \! \arrow{}{} \!)$ be an NPLTS. We say that $s_{1}, s_{2} \in S$ are \emph{probabilistic
trace-by-trace-distribution testing equivalent}, written $s_{1} \sbis{\textrm{PTe-tbt,dis}} s_{2}$, iff for
every NPT $\calt = (O, A, \! \arrow{}{\calt} \!)$ with initial state $o \in O$ it holds that:

		\begin{itemize}

\item For each $\calz_{1} \in \ms{Res}_{\rm max}(s_{1}, o)$ there exists $\calz_{2} \in \ms{Res}_{\rm
max}(s_{2}, o)$ such that \underline{for all $\alpha \in A^{*}$} it holds that $\calcc_{\rm max}(z_{s_{1},
o}, \alpha) \neq \emptyset$ implies $\calcc_{\rm max}(z_{s_{2}, o}, \alpha) \neq \emptyset$ and:
\cws{10}{\hspace*{-1.2cm} \ms{prob}(\calscc(z_{s_{1}, o}, \alpha)) \: = \: \ms{prob}(\calscc(z_{s_{2}, o},
\alpha))}

\item For each $\calz_{2} \in \ms{Res}_{\rm max}(s_{2}, o)$ there exists $\calz_{1} \in \ms{Res}_{\rm
max}(s_{1}, o)$ such that \underline{for all $\alpha \in A^{*}$} it holds that $\calcc_{\rm max}(z_{s_{2},
o}, \alpha) \neq \emptyset$ implies $\calcc_{\rm max}(z_{s_{1}, o}, \alpha) \neq \emptyset$ and:
\cws{12}{\hspace*{-1.2cm} \ms{prob}(\calscc(z_{s_{2}, o}, \alpha)) \: = \: \ms{prob}(\calscc(z_{s_{1}, o},
\alpha))}

		\end{itemize}

\noindent
We denote by $\sbis{\textrm{PTe-tbt,dis}}^{\rm ct}$ the coarser variant based on randomized schedulers.
\fullbox

	\end{defi}

	\begin{thm}\label{thm:pteallexists_eq_ptetbtdis}

Let $(S, A, \! \arrow{}{} \!)$ be an NPLTS and $s_{1}, s_{2} \in S$. Then:
\cws{10}{\begin{array}{rcl}
s_{1} \sbis{\textrm{\rm PTe-}\forall\exists} s_{2} & \!\!\! \Longleftrightarrow \!\!\! & s_{1}
\sbis{\textrm{\rm PTe-tbt,dis}} s_{2} \\
s_{1} \sbis{\textrm{\rm PTe-}\forall\exists}^{\rm ct} s_{2} & \!\!\! \Longleftrightarrow \!\!\! & s_{1}
\sbis{\textrm{\rm PTe-tbt,dis}}^{\rm ct} s_{2} \\
\end{array}}

\proof
Let us prove the contrapositive of $s_{1} \sbis{\textrm{PTe-}\forall\exists} s_{2} \Longrightarrow s_{1}
\sbis{\textrm{PTe-tbt,dis}} s_{2}$. Thus, suppose that $s_{1} \not\sbis{\textrm{PTe-tbt,dis}} s_{2}$. This
means that there exist an NPT $\calt = (O, A, \! \arrow{}{\calt} \!)$ with initial state $o \in O$ and, say,
a resolution $\calz_{1} \in \ms{Res}_{\rm max}(s_{1}, o)$ such that for each $\calz_{2} \in \ms{Res}_{\rm
max}(s_{2}, o)$ there exists $\alpha_{2} \in A^{*}$ such that $\calcc_{\rm max}(z_{s_{1}, o}, \alpha_{2})
\neq \emptyset$ and (i)~$\calcc_{\rm max}(z_{s_{2}, o}, \alpha_{2}) = \emptyset$ or
(ii)~$\ms{prob}(\calscc(z_{s_{1}, o}, \alpha_{2})) \neq \ms{prob}(\calscc(z_{s_{2}, o}, \alpha_{2}))$. We
show that from this fact it follows that $s_{1} \not\sbis{\textrm{PTe-}\forall\exists} s_{2}$ by proceeding
by induction on the number $n$ of traces labeling the successful computations from~$o$ (note that $n$ is
finite -- because $\calt$ is finite -- and greater than~$0$ -- otherwise $\calt$ cannot distinguish $s_{1}$
from $s_{2}$ with respect to $\sbis{\textrm{PTe-tbt,dis}}$):

		\begin{itemize}

\item Let $n = 1$ and denote by $\alpha$ the only trace labeling the successful computations from~$o$. Then
$\calcc_{\rm max}(z_{s_{1}, o}, \alpha) \neq \emptyset$ and (i)~$\calcc_{\rm max}(z_{s_{2}, o}, \alpha) =
\emptyset$ in which case:
\cws{0}{\hspace*{-1.2cm} \ms{prob}(\calsc(z_{s_{1}, o})) \: > \: 0 \: = \: \ms{prob}(\calsc(z_{s_{2}, o}))}
or (ii)~it holds that:
\cws{0}{\hspace*{-1.2cm}\begin{array}{rcccl}
\ms{prob}(\calsc(z_{s_{1}, o})) & \!\!\! = \!\!\! & \ms{prob}(\calscc(z_{s_{1}, o}, \alpha)) & \!\!\! \neq
\!\!\! & \\
& \!\!\! \neq \!\!\! & \ms{prob}(\calscc(z_{s_{2}, o}, \alpha)) & \!\!\! = \!\!\! &
\ms{prob}(\calsc(z_{s_{2}, o})) \\
\end{array}}
As a consequence, in both cases $s_{1} \not\sbis{\textrm{PTe-}\forall\exists} s_{2}$.

\item Let $n \in \natns_{> 1}$ and suppose that the result holds for all $m = 1, \dots, n - 1$. Given a
trace $\alpha$ labeling some of the successful computations from~$o$, we denote by $\calt_{\downarrow
\alpha}$ the NPT obtained from $\calt$ by transforming into a normal terminal state every success state
reached by a maximal $\alpha$-compatible computation, and by $\calt_{\uparrow \alpha}$ the NPT obtained from
$\calt$ by transforming into a normal terminal state every success state reached by a maximal computation
not compatible with $\alpha$. Since $\calt$ distinguishes $s_{1}$ from $s_{2}$ with respect to
$\sbis{\textrm{PTe-tbt,dis}}$, $\calt_{\downarrow \alpha}$ and $\calt_{\uparrow \alpha}$ have the same
structure as $\calt$, and $\alpha$ labels some of the successful computations of $\calt$, either
$\calt_{\downarrow \alpha}$ or $\calt_{\uparrow \alpha}$ still distinguishes $s_{1}$ from $s_{2}$ with
respect to $\sbis{\textrm{PTe-tbt,dis}}$. Since $\calt_{\downarrow \alpha}$ has $n - 1$ traces labeling its
successful computations and $\calt_{\uparrow \alpha}$ has a single trace labeling its successful
computations, by the induction hypothesis it follows that $s_{1} \not\sbis{\textrm{PTe-}\forall\exists}
s_{2}$.

		\end{itemize}

\noindent
Suppose now that $s_{1} \sbis{\textrm{PTe-tbt,dis}} s_{2}$ and consider an arbitrary NPT $\calt = (O, A, \!
\arrow{}{\calt} \!)$ with initial state $o \in O$. Since for all $s \in S$ and $\calz \in \ms{Res}_{\rm
max}(s, o)$ it holds that:
\cws{0}{\ms{prob}(\calsc(z_{s, o})) \: = \: \sum\limits_{\alpha \in A^{*} \, {\rm s.t.} \, \calcc_{\rm
max}(z_{s, o}, \alpha) \neq \emptyset} \ms{prob}(\calscc(z_{s, o}, \alpha))}
from $s_{1} \sbis{\textrm{PTe-tbt,dis}} s_{2}$ it follows that:

		\begin{itemize}

\item For each $\calz_{1} \in \ms{Res}_{\rm max}(s_{1}, o)$ there exists $\calz_{2} \in \ms{Res}_{\rm
max}(s_{2}, o)$ such that:
\cws{0}{\hspace*{-1.2cm}\begin{array}{rcccl}
\ms{prob}(\calsc(z_{s_{1}, o})) & \!\!\! = \!\!\! & \sum\limits_{\alpha \in A^{*} \, {\rm s.t.} \,
\calcc_{\rm max}(z_{s_{1}, o}, \alpha) \neq \emptyset} \ms{prob}(\calscc(z_{s_{1}, o}, \alpha)) & \!\!\! =
\!\!\! & \\
& \!\!\! = \!\!\! & \sum\limits_{\alpha \in A^{*} \, {\rm s.t.} \, \calcc_{\rm max}(z_{s_{2}, o}, \alpha)
\neq \emptyset} \ms{prob}(\calscc(z_{s_{2}, o}, \alpha)) & \!\!\! = \!\!\! & \ms{prob}(\calsc(z_{s_{2}, o}))
\\
\end{array}}

\item For each $\calz_{2} \in \ms{Res}_{\rm max}(s_{2}, o)$ there exists $\calz_{1} \in \ms{Res}_{\rm
max}(s_{1}, o)$ such that:
\cws{4}{\hspace*{-1.2cm}\begin{array}{rcccl}
\ms{prob}(\calsc(z_{s_{2}, o})) & \!\!\! = \!\!\! & \sum\limits_{\alpha \in A^{*} \, {\rm s.t.} \,
\calcc_{\rm max}(z_{s_{2}, o}, \alpha) \neq \emptyset} \ms{prob}(\calscc(z_{s_{2}, o}, \alpha)) & \!\!\! =
\!\!\! & \\
& \!\!\! = \!\!\! & \sum\limits_{\alpha \in A^{*} \, {\rm s.t.} \, \calcc_{\rm max}(z_{s_{1}, o}, \alpha)
\neq \emptyset} \ms{prob}(\calscc(z_{s_{1}, o}, \alpha)) & \!\!\! = \!\!\! & \ms{prob}(\calsc(z_{s_{1}, o}))
\\
\end{array}}

		\end{itemize}

\noindent
This means that $s_{1} \sbis{\textrm{PTe-}\forall\exists} s_{2}$. \\
The fact that $\sbis{\textrm{PTe-}\forall\exists}^{\rm ct}$ and $\sbis{\textrm{PTe-tbt,dis}}^{\rm ct}$
coincide immediately follows.
\qed

	\end{thm}

We know from~\cite{DeN87} that for fully nondeterministic processes there is a strong connection between the
testing semantics of~\cite{DH84} and the failure semantics of~\cite{BHR84}. Thus, for a more complete
comparison of the various trace and testing equivalences, we also present failure semantics for NPLTS
models. In particular, we consider two variants $\sbis{\rm PF,dis}$/$\sbis{\rm PF,dis}^{\rm ct}$ of the
probabilistic failure-distribution equivalence defined in~\cite{Seg96} on the basis of the pattern of
$\sbis{\rm PTr,dis}^{\rm ct}$~\cite{Seg95b}, and we introduce two variants $\sbis{\rm PF}$/$\sbis{\rm
PF}^{\rm ct}$ of a novel probabilistic failure equivalence by taking inspiration from the pattern of
$\sbis{\rm PTr}$. We shall see that $\sbis{\textrm{PTe-}\forall\exists}$ (i.e.,
$\sbis{\textrm{PTe-tbt,dis}}$) is strictly finer than $\sbis{\rm PF,dis}$/$\sbis{\rm PF,dis}^{\rm ct}$,
while $\sbis{\textrm{PTe-tbt}}$ and $\sbis{\textrm{PTe-tbt}}^{\rm ct}$ are strictly coarser than $\sbis{\rm
PF}$ and $\sbis{\rm PF}^{\rm ct}$, respectively.

In the following, we call \emph{failure pair} an element $\varphi \in A^{*} \times 2^{A}$ formed by a trace
$\alpha$ and a failure set $F$. Given a state $s$ of an NPLTS $\call$, a resolution $\calz$ of $s$, and a
computation \linebreak $c \in \calc_{\rm fin}(z_{s})$, we say that $c$ is compatible with $\varphi$ iff $c
\in \calcc(z_{s}, \alpha)$ and the state in $\call$ corresponding to the last state reached by~$c$ has no
outgoing transitions in $\call$ labeled with an action in $F$. We denote by $\calfcc(z_{s}, \varphi)$ the
set of $\varphi$-compatible computations from~$z_{s}$.

	\begin{defi}\label{def:pfdis}

Let $(S, A, \! \arrow{}{} \!)$ be an NPLTS. We say that $s_{1}, s_{2} \in S$ are \emph{probabilistic
failure-distribution equivalent}, written $s_{1} \sbis{\rm PF,dis} s_{2}$, iff:

		\begin{itemize}

\item For each $\calz_{1} \in \ms{Res}(s_{1})$ there exists $\calz_{2} \in \ms{Res}(s_{2})$ such that
\underline{for all $\varphi \in A^{*} \times 2^{A}$}:
\cws{10}{\hspace*{-1.2cm} \ms{prob}(\calfcc(z_{s_{1}}, \varphi)) \: = \: \ms{prob}(\calfcc(z_{s_{2}},
\varphi))}

\item For each $\calz_{2} \in \ms{Res}(s_{2})$ there exists $\calz_{1} \in \ms{Res}(s_{1})$ such that
\underline{for all $\varphi \in A^{*} \times 2^{A}$}:
\cws{12}{\hspace*{-1.2cm} \ms{prob}(\calfcc(z_{s_{2}}, \varphi)) \: = \: \ms{prob}(\calfcc(z_{s_{1}},
\varphi))}

		\end{itemize}

\noindent
We denote by $\sbis{\rm PF,dis}^{\rm ct}$ the coarser variant based on randomized schedulers.
\fullbox

	\end{defi}

	\begin{defi}\label{def:pf}

Let $(S, A, \! \arrow{}{} \!)$ be an NPLTS. We say that $s_{1}, s_{2} \in S$ are \emph{probabilistic failure
equivalent}, written $s_{1} \sbis{\rm PF} s_{2}$, iff \underline{for all $\varphi \in A^{*} \times 2^{A}$}
it holds that:

		\begin{itemize}

\item For each $\calz_{1} \in \ms{Res}(s_{1})$ there exists $\calz_{2} \in \ms{Res}(s_{2})$ such that:
\cws{10}{\hspace*{-1.2cm} \ms{prob}(\calfcc(z_{s_{1}}, \varphi)) \: = \: \ms{prob}(\calfcc(z_{s_{2}},
\varphi))}

\item For each $\calz_{2} \in \ms{Res}(s_{2})$ there exists $\calz_{1} \in \ms{Res}(s_{1})$ such that:
\cws{12}{\hspace*{-1.2cm} \ms{prob}(\calfcc(z_{s_{2}}, \varphi)) \: = \: \ms{prob}(\calfcc(z_{s_{1}},
\varphi))}

		\end{itemize}

\noindent
We denote by $\sbis{\rm PF}^{\rm ct}$ the coarser variant based on randomized schedulers.
\fullbox

	\end{defi}

	\begin{figure}[tp]

\input{Pictures/counterex_pfdis_pf}
\caption{NPLTS models distinguished by $\sbis{\rm PF,dis}$/$\sbis{\rm PF,dis}^{\rm ct}$ and identified by
$\sbis{\rm PF}$/$\sbis{\rm PF}^{\rm ct}$}
\label{fig:counterex_pfdis_pf}

	\end{figure}

	\begin{thm}\label{thm:pfdis_incl_pf}

Let $(S, A, \! \arrow{}{} \!)$ be an NPLTS and $s_{1}, s_{2} \in S$. Then:
\cws{10}{\begin{array}{rcl}
s_{1} \sbis{\rm PF,dis} s_{2} & \!\!\! \Longrightarrow \!\!\! & s_{1} \sbis{\rm PF} s_{2} \\
s_{1} \sbis{\rm PF,dis}^{\rm ct} s_{2} & \!\!\! \Longrightarrow \!\!\! & s_{1} \sbis{\rm PF}^{\rm ct} s_{2}
\\
\end{array}}

\proof
If $s_{1} \sbis{\rm PF,dis} s_{2}$ (resp.\ $s_{1} \sbis{\rm PF,dis}^{\rm ct} s_{2}$), then $s_{1} \sbis{\rm
PF} s_{2}$ (resp. $s_{1} \sbis{\rm PF}^{\rm ct} s_{2}$) follows by taking the same fully matching
resolutions considered for $\sbis{\rm PF,dis}$ (resp.\ $\sbis{\rm PF,dis}^{\rm ct}$).
\qed

	\end{thm}

The inclusion of $\sbis{\rm PF,dis}$ (resp.\ $\sbis{\rm PF,dis}^{\rm ct}$) in $\sbis{\rm PF}$ (resp.\
$\sbis{\rm PF}^{\rm ct}$) is strict, because the initial states of the two NPLTS models in
Fig.~\ref{fig:counterex_pfdis_pf} are equated by the latter equivalence and told apart by the former.
Moreover, Figs.~\ref{fig:counterex_pfdis_pf} and~\ref{fig:counterex_ptesupinf_trace} together show that
$\sbis{\rm PF}$ and $\sbis{\rm PF,dis}^{\rm ct}$ are incomparable with each other.

	\begin{thm}\label{thm:ptetbtdis_incl_pfdis}

Let $(S, A, \! \arrow{}{} \!)$ be an NPLTS and $s_{1}, s_{2} \in S$. Then:
\cws{12}{s_{1} \sbis{\textrm{\rm PTe-tbt,dis}} s_{2} \: \Longrightarrow \: s_{1} \sbis{\rm PF,dis} s_{2}}

\proof
Firstly, we prove that $s_{1} \sbis{\textrm{PTe-tbt,dis}} s_{2} \: \Longrightarrow \: s_{1} \sbis{\rm
PRTr,dis} s_{2}$ where $\sbis{\rm PRTr,dis}$ is defined as follows. We call \emph{ready trace} an element
$\rho \in (A \times 2^{A})^{*}$ given by a sequence of $n \in \natns$ pairs of the form $(a_{i}, R_{i})$.
Given $s \in S$, $\calz \in \ms{Res}(s)$, and $c \in \calc_{\rm fin}(z_{s})$, we say that $c$ is compatible
with $\rho$ iff $c \in \calcc(z_{s}, a_{1} \dots a_{n})$ and, denoting by $z_{i}$ the state reached by~$c$
after the $i$-th step for all $i = 1, \dots, n$, the set of actions labeling the transitions in $\call$
departing from the state in $\call$ corresponding to $z_{i}$ is precisely~$R_{i}$. We denote by
$\calrtcc(z_{s}, \rho)$ the set of $\rho$-compatible computations from $z_{s}$. We say that $s_{1}$ and
$s_{2}$ are \emph{probabilistic ready-trace-distribution equivalent}, written $s_{1} \sbis{\rm PRTr,dis}
s_{2}$, iff for each $\calz_{1} \in \ms{Res}(s_{1})$ there exists $\calz_{2} \in \ms{Res}(s_{2})$ such that
for all $\rho \in (A \times 2^{A})^{*}$:
\cws{0}{\ms{prob}(\calrtcc(z_{s_{1}}, \rho)) \: = \: \ms{prob}(\calrtcc(z_{s_{2}}, \rho))}
and symmetrically for each $\calz_{2} \in \ms{Res}(s_{2})$. \\
We show that $s_{1} \sbis{\textrm{PTe-tbt,dis}} s_{2}$ implies $s_{1} \sbis{\rm PRTr,dis} s_{2}$ by building
a test that permits to reason about all ready traces at once for each resolution of $s_{1}$ and $s_{2}$. We
start by deriving a new NPLTS $(S_{\rm r}, A_{\rm r}, \! \arrow{}{\rm r} \!)$ that is isomorphic to the
given one up to transition labels and terminal states. A transition $s \arrow{a}{} \cald$ becomes $s_{\rm r}
\arrow{a \triangleleft R}{\rm r} \cald_{\rm r}$ where $R \subseteq A$ is the set of actions labeling the
outgoing transitions of $s$ and $\cald_{\rm r}(s_{\rm r}) = \cald(s)$ for all $s \in S$. If $s$ is a
terminal state, i.e., it has no outgoing transitions, then we add a transition $s_{\rm r} \arrow{\circ
\triangleleft \emptyset}{\rm r} \delta_{s_{\rm r}}$ where $\delta_{s_{\rm r}}(s_{\rm r}) = 1$ and
$\delta_{s_{\rm r}}(s'_{\rm r}) = 0$ for all $s' \in S \setminus \{ s \}$. Transition relabeling preserves
$\sbis{\textrm{PTe-tbt,dis}}$, i.e., $s_{1} \sbis{\textrm{PTe-tbt,dis}} s_{2}$ implies $s_{1, \rm r}
\sbis{\textrm{PTe-tbt,dis}} s_{2, \rm r}$, because $\sbis{\textrm{PTe-tbt,dis}}$ is able to distinguish a
state that has a single $\alpha$-compatible computation reaching a state with a nondeterministic branching
formed by a $b$-transition and a $c$-transition, from a state that has two $\alpha$-compatible computations
such that one of them reaches a state with only one outgoing transition labeled with $b$ and the other one
reaches a state with only one outgoing transition labeled with $c$ (e.g., use a test that has a single
$\alpha$-compatible computation whose last step leads to a distribution whose support contains only a state
with only one outgoing transition labeled with $b$ that reaches success and a state with only one outgoing
transition labeled with~$c$ that reaches success). \\
For each $\alpha_{\rm r} \in (A_{\rm r})^{*}$ and $R \subseteq A$, we build an NPT $\calt_{\alpha_{\rm r},
R} = (O_{\alpha_{\rm r}, R}, A_{\rm r}, \! \arrow{}{\alpha_{\rm r}, R} \!)$ having a single $\alpha_{\rm
r}$-compatible computation that goes from the initial state $o_{\alpha_{\rm r}, R}$ to a state having a
single transition to $\omega$ labeled with (i) $\circ \triangleleft \emptyset$ if $R = \emptyset$ or (ii)
$\_ \triangleleft R$ if $R \neq \emptyset$. Since we compare individual states (like $s_{1}$ and $s_{2}$)
rather than state distributions, the distinguishing power of $\sbis{\textrm{PTe-tbt,dis}}$ does not change
if we additionally consider tests starting with a single \linebreak $\tau$-transition that can initially
evolve autonomously in any interaction system. We thus build a further NPT $\calt = (O, A_{\rm r}, \!
\arrow{}{\calt} \!)$ that has an initial $\tau$-transition and then behaves as one of the tests
$\calt_{\alpha_{\rm r}, R}$, i.e., its initial $\tau$-transition goes from the initial state $o$ to a state
distribution whose support is the set $\{ o_{\alpha_{\rm r}, R} \mid \alpha_{\rm r} \in (A_{\rm r})^{*}
\land R \subseteq A \}$, with the probability $p_{\alpha_{\rm r}, R}$ associated with $o_{\alpha_{\rm r},
R}$ being taken from the distribution whose values are of the form $1 / 2^{i}$, $i \in \natns_{> 0}$. Note
that $\calt$ is not finite state, but this affects only the initial step, whose only purpose is to
internally select a specific ready trace. \\
After this step, $\calt$ interacts with the process under test. Let $\rho \in (A \times 2^{A})^{*}$ be a
ready trace of the form $(a_{1}, R_{1}) \dots (a_{n}, R_{n})$, where $n \in \natns$. Given $s \in S$,
consider the trace $\alpha_{\rho, \rm r} \in (A_{\rm r})^{*}$ of length $n + 1$ in which the first element
is $a_{1} \triangleleft R$, with $R \subseteq A$ being the set of actions labeling the outgoing transitions
of $s$, the subsequent elements are of the form $a_{i} \triangleleft R_{i - 1}$ for $i = 2, \dots, n$, and
the last element is (i) $\circ \triangleleft \emptyset$ if $R_{n} = \emptyset$ or (ii) $\_ \triangleleft
R_{n}$ if $R_{n} \neq \emptyset$. Then for all $\calz \in \ms{Res}(s)$ it holds that:
\cws{0}{\ms{prob}(\calrtcc(z_{s}, \rho)) \: = \: 0}
if there is no $a_{1} \dots a_{n}$-compatible computation from $z_{s}$, otherwise:
\cws{0}{\ms{prob}(\calrtcc(z_{s}, \rho)) \: = \: \ms{prob}(\calscc(z_{s_{\rm r}, o}, \alpha_{\rho, \rm r}))
/ p_{\alpha'_{\rho, \rm r}, R_{n}}}
where $\alpha'_{\rho, \rm r}$ is $\alpha_{\rho, \rm r}$ without its last element. \\
Suppose that $s_{1} \sbis{\textrm{PTe-tbt,dis}} s_{2}$, which implies that $s_{1}$ and $s_{2}$ have the same
set $R$ of actions labeling their outgoing transitions and $s_{1, \rm r} \sbis{\textrm{PTe-tbt,dis}} s_{2,
\rm r}$. Then:

		\begin{itemize}

\item For each $\calz_{1} \in \ms{Res}(s_{1})$ there exists $\calz_{2} \in \ms{Res}(s_{2})$ such that for
all ready traces $\rho = (a_{1}, R_{1}) \dots (a_{n}, R_{n}) \in (A \times 2^{A})^{*}$ either:
\cws{0}{\hspace*{-1.2cm} \ms{prob}(\calrtcc(z_{s_{1}}, \rho)) \: = \: 0 \: = \:
\ms{prob}(\calrtcc(z_{s_{2}}, \rho))}
or:
\cws{4}{\hspace*{-1.2cm}\begin{array}{rcccl}
\ms{prob}(\calrtcc(z_{s_{1}}, \rho)) & \!\!\! = \!\!\! & \ms{prob}(\calscc(z_{s_{1, \rm r}, o},
\alpha_{\rho, \rm r})) / p_{\alpha'_{\rho, \rm r}, R_{n}} & \!\!\! = \!\!\! & \\
& \!\!\! = \!\!\! & \ms{prob}(\calscc(z_{s_{2, \rm r}, o}, \alpha_{\rho, \rm r})) / p_{\alpha'_{\rho, \rm
r}, R_{n}} & \!\!\! = \!\!\! & \ms{prob}(\calrtcc(z_{s_{2}}, \rho)) \\
\end{array}}

\item Symmetrically for each $\calz_{2} \in \ms{Res}(s_{2})$.

		\end{itemize}

\noindent
This means that $s_{1} \sbis{\rm PRTr,dis} s_{2}$. \\
Secondly, we prove that $s_{1} \sbis{\rm PRTr,dis} s_{2} \: \Longrightarrow \: s_{1} \sbis{\rm PFTr,dis}
s_{2}$ where $\sbis{\rm PFTr,dis}$ is defined as follows. We call \emph{failure trace} an element $\phi \in
(A \times 2^{A})^{*}$ given by a sequence of $n \in \natns$ pairs of the form $(a_{i}, F_{i})$. Given $s \in
S$, $\calz \in \ms{Res}(s)$, and $c \in \calc_{\rm fin}(z_{s})$, we say that $c$ is compatible with $\phi$
iff $c \in \calcc(z_{s}, a_{1} \dots a_{n})$ and, denoting by $z_{i}$ the state reached by~$c$ after the
$i$-th step for all $i = 1, \dots, n$, the state in $\call$ corresponding to $z_{i}$ has no outgoing
transitions in $\call$ labeled with an action in $F_{i}$. We denote by $\calftcc(z_{s}, \phi)$ the set of
$\phi$-compatible computations from $z_{s}$. We say that $s_{1}$ and~$s_{2}$ are \emph{probabilistic
failure-trace-distribution equivalent}, written $s_{1} \sbis{\rm PFTr,dis} s_{2}$, iff for each $\calz_{1}
\in \ms{Res}(s_{1})$ there exists $\calz_{2} \in \ms{Res}(s_{2})$ such that for all $\phi \in (A \times
2^{A})^{*}$:
\cws{0}{\ms{prob}(\calftcc(z_{s_{1}}, \phi)) \: = \: \ms{prob}(\calftcc(z_{s_{2}}, \phi))}
and symmetrically for each $\calz_{2} \in \ms{Res}(s_{2})$. \\
Suppose that $s_{1} \sbis{\rm PRTr,dis} s_{2}$. Since for all $s \in S$, $\calz \in \ms{Res}(s)$, $n \in
\natns$, $\alpha = a_{1} \dots a_{n} \in A^{*}$, and $F_{1}, \dots, F_{n} \in 2^{A}$ it holds that:
\cws{0}{\begin{array}{l}
\ms{prob}(\calftcc(z_{s}, (a_{1}, F_{1}) \dots (a_{n}, F_{n}))) \: = \\
\hspace*{2.5cm} \sum_{R'_{1}, \dots, R'_{n} \in 2^{A} \, {\rm s.t.} \, R'_{i} \cap F_{i} = \emptyset \, {\rm
for \hspace{0.08cm} all} \, i = 1, \dots, n} \ms{prob}(\calrtcc(z_{s}, (a_{1}, R'_{1}) \dots (a_{n},
R'_{n}))) \\
\end{array}}
we immediately derive that:

		\begin{itemize}

\item For each $\calz_{1} \in \ms{Res}(s_{1})$ there exists $\calz_{2} \in \ms{Res}(s_{2})$ such that for
all failure traces\\  $(a_{1}, F_{1}) \dots (a_{n}, F_{n}) \in (A \times 2^{A})^{*}$:
\cws{0}{\hspace*{-1.2cm}\begin{array}{l}
\ms{prob}(\calftcc(z_{s_{1}}, (a_{1}, F_{1}) \dots (a_{n}, F_{n}))) \: = \\
\hspace*{0.8cm} = \: \sum_{R'_{1}, \dots, R'_{n} \in 2^{A} \, {\rm s.t.} \, R'_{i} \cap F_{i} = \emptyset \,
{\rm for \hspace{0.08cm} all} \, i = 1, \dots, n} \ms{prob}(\calrtcc(z_{s_{1}}, (a_{1}, R'_{1}) \dots
(a_{n}, R'_{n}))) \\[0.1cm]
\hspace*{0.8cm} = \: \sum_{R'_{1}, \dots, R'_{n} \in 2^{A} \, {\rm s.t.} \, R'_{i} \cap F_{i} = \emptyset \,
{\rm for \hspace{0.08cm} all} \, i = 1, \dots, n} \ms{prob}(\calrtcc(z_{s_{2}}, (a_{1}, R'_{1}) \dots
(a_{n}, R'_{n}))) \\[0.1cm]
\hspace*{0.8cm} = \: \ms{prob}(\calftcc(z_{s_{2}}, (a_{1}, F_{1}) \dots (a_{n}, F_{n}))) \\
\end{array}}

\item Symmetrically for each $\calz_{2} \in \ms{Res}(s_{2})$.

		\end{itemize}

\noindent
This means that $s_{1} \sbis{\rm PFTr,dis} s_{2}$. \\
Thirdly, we prove that $s_{1} \sbis{\rm PFTr,dis} s_{2} \: \Longrightarrow \: s_{1} \sbis{\rm PF,dis}
s_{2}$. Suppose that $s_{1} \sbis{\rm PFTr,dis} s_{2}$. Since for all $s \in S$, $\calz \in \ms{Res}(s)$, $n
\in \natns$, $\alpha = a_{1} \dots a_{n} \in A^{*}$, and $F \in 2^{A}$ it holds that:
\cws{0}{\ms{prob}(\calfcc(z_{s}, (\alpha, F))) \: = \: \ms{prob}(\calftcc(z_{s}, (a_{1}, \emptyset) \dots
(a_{n - 1}, \emptyset) (a_{n}, F)))}
we immediately derive that:

		\begin{itemize}

\item For each $\calz_{1} \in \ms{Res}(s_{1})$ there exists $\calz_{2} \in \ms{Res}(s_{2})$ such that for
all failure pairs $(a_{1} \dots a_{n}, F) \in A^{*} \times 2^{A}$:
\cws{0}{\hspace*{-1.2cm}\begin{array}{rcl}
\ms{prob}(\calfcc(z_{s_{1}}, (a_{1} \dots a_{n}, F))) & \!\!\! = \!\!\! & \ms{prob}(\calftcc(z_{s_{1}},
(a_{1}, \emptyset) \dots (a_{n - 1}, \emptyset) (a_{n}, F))) \\
& \!\!\! = \!\!\! & \ms{prob}(\calftcc(z_{s_{2}}, (a_{1}, \emptyset) \dots (a_{n - 1}, \emptyset) (a_{n},
F))) \\
& \!\!\! = \!\!\! & \ms{prob}(\calfcc(z_{s_{2}}, (a_{1} \dots a_{n}, F))) \\
\end{array}}

\item Symmetrically for each $\calz_{2} \in \ms{Res}(s_{2})$.

		\end{itemize}

\noindent
This means that $s_{1} \sbis{\rm PF,dis} s_{2}$.
\qed

	\end{thm}

The inclusion of $\sbis{\textrm{PTe-tbt,dis}}$ in $\sbis{\rm PF,dis}$ is strict, because for the two NPLTS
models in Fig.~\ref{fig:counterex_tefnd_testing} it holds that $s_{1} \sbis{\rm PF,dis} s_{2}$ while $s_{1}
\not\sbis{\textrm{PTe-tbt,dis}} s_{2}$ as witnessed by the test in the same figure (see the maximal
resolutions of the interaction systems in Fig.~\ref{fig:counterex_tefnd_testing_max_res}).

	\begin{thm}\label{thm:pfdis_incl_ptrdis}

Let $(S, A, \! \arrow{}{} \!)$ be an NPLTS and $s_{1}, s_{2} \in S$. Then:
\cws{10}{\begin{array}{rcl}
s_{1} \sbis{\rm PF,dis} s_{2} & \!\!\! \Longrightarrow \!\!\! & s_{1} \sbis{\rm PTr,dis} s_{2} \\
s_{1} \sbis{\rm PF,dis}^{\rm ct} s_{2} & \!\!\! \Longrightarrow \!\!\! & s_{1} \sbis{\rm PTr,dis}^{\rm ct}
s_{2} \\
\end{array}}

\proof
Suppose that $s_{1} \sbis{\rm PF,dis} s_{2}$. Then $s_{1} \sbis{\rm PTr,dis} s_{2}$ because for all $s \in
S$, $\calz \in \ms{Res}(s)$, and $\alpha \in A^{*}$ it holds that:
\cws{0}{\ms{prob}(\calcc(z_{s}, \alpha)) \: = \: \ms{prob}(\calfcc(z_{s}, (\alpha, \emptyset)))}
and hence:

		\begin{itemize}

\item For each $\calz_{1} \in \ms{Res}(s_{1})$ there exists $\calz_{2} \in \ms{Res}(s_{2})$ such that for
all $\alpha \in A^{*}$:
\cws{0}{\hspace*{-1.2cm}\begin{array}{rcccl}
\ms{prob}(\calcc(z_{s_{1}}, \alpha)) & \!\!\! = \!\!\! & \ms{prob}(\calfcc(z_{s_{1}}, (\alpha, \emptyset)))
& \!\!\! = \!\!\! & \\
& \!\!\! = \!\!\! & \ms{prob}(\calfcc(z_{s_{2}}, (\alpha, \emptyset))) & \!\!\! = \!\!\! &
\ms{prob}(\calcc(z_{s_{2}}, \alpha)) \\
\end{array}}

\item Symmetrically for each $\calz_{2} \in \ms{Res}(s_{2})$.

		\end{itemize}

\noindent
The proof that $s_{1} \sbis{\rm PF,dis}^{\rm ct} s_{2}$ implies $s_{1} \sbis{\rm PTr,dis}^{\rm ct} s_{2}$ is
similar.
\qed

	\end{thm}

The inclusion of $\sbis{\rm PF,dis}$ (resp.\ $\sbis{\rm PF,dis}^{\rm ct}$) in $\sbis{\rm PTr,dis}$ (resp.\
$\sbis{\rm PTr,dis}^{\rm ct}$) is strict, because the initial states of the two NPLTS models in
Fig.~\ref{fig:counterex_pteallexists_trace} are equated by the latter equivalence and told apart by the
former.

	\begin{thm}\label{thm:pf_incl_ptetbt}

Let $(S, A, \! \arrow{}{} \!)$ be an NPLTS and $s_{1}, s_{2} \in S$. Then:
\cws{10}{\begin{array}{rcl}
s_{1} \sbis{\rm PF} s_{2} & \!\!\! \Longrightarrow \!\!\! & s_{1} \sbis{\textrm{\rm PTe-tbt}} s_{2} \\
s_{1} \sbis{\rm PF}^{\rm ct} s_{2} & \!\!\! \Longrightarrow \!\!\! & s_{1} \sbis{\textrm{\rm PTe-tbt}}^{\rm
ct} s_{2} \\
\end{array}}

\proof
Let us prove the contrapositive of the first result, i.e., $s_{1} \not\sbis{\textrm{PTe-tbt}} s_{2} \:
\Longrightarrow \: s_{1} \not\sbis{\rm PF} s_{2}$. Thus, suppose that $s_{1} \not\sbis{\textrm{PTe-tbt}}
s_{2}$. This means that there exist an NPT $\calt = (O, A, \! \arrow{}{\calt} \!)$ with initial state $o \in
O$, a trace $\alpha \in A^{*}$, and, say, a resolution $\calz_{1} \in \ms{Res}_{{\rm max}, \alpha}(s_{1},
o)$ such that $\ms{Res}_{{\rm max}, \alpha}(s_{2}, o) = \emptyset$ or for all $\calz_{2} \in \ms{Res}_{{\rm
max}, \alpha}(s_{2}, o)$ it holds that:
\cws{0}{\ms{prob}(\calscc(z_{s_{1}, o}, \alpha)) \: \neq \: \ms{prob}(\calscc(z_{s_{2}, o}, \alpha))}
Observing that $\ms{Res}_{{\rm max}, \alpha}(s_{1}, o) \neq \emptyset$, in the case that $\ms{Res}_{{\rm
max}, \alpha}(s_{2}, o) = \emptyset$ either $s_{2}$ cannot perform $\alpha$ at all -- let $\varphi =
(\alpha, \emptyset)$ -- or, after performing $\alpha$, the states reached by $s_{2}$ can always synchronize
with the states reached by $o$ on a set $F$ of actions whereas the states reached by $s_{1}$ cannot -- let
$\varphi = (\alpha, F)$. The failure pair $\varphi$ shows that $s_{1} \not\sbis{\rm PF} s_{2}$ in this case
because, denoting by $\calz'_{1}$ the element of $\ms{Res}(s_{1})$ that originates~$\calz_{1}$, we have that
for all $\calz'_{2} \in \ms{Res}(s_{2})$:
\cws{0}{\ms{prob}(\calfcc(z'_{s_{1}}, \varphi)) \: > \: 0 \: = \: \ms{prob}(\calfcc(z'_{s_{2}}, \varphi))}
In the case that $\ms{Res}_{{\rm max}, \alpha}(s_{2}, o) \neq \emptyset$, the failure pair $\varphi =
(\alpha, \emptyset)$ shows that $s_{1} \not\sbis{\rm PF} s_{2}$. In fact, without loss of generality we can
assume that the only $\alpha$-compatible computations in~$\calt$ are the ones exercised by $\calz_{1}$ --
note that they must belong to the same element $\calz$ of $\ms{Res}(o)$ -- as the only effect of this
assumption is that of possibly reducing the number of resolutions in $\ms{Res}_{{\rm max}, \alpha}(s_{2},
o)$. At least one of these computations must be successful -- and hence maximal -- in~$\calt$ because
otherwise the success probabilities of the considered resolutions would all be equal to~$0$. Denoting by
$\calz'_{1}$ the element of $\ms{Res}(s_{1})$ that originates~$\calz_{1}$, we then have that for all
$\calz'_{2} \in \ms{Res}(s_{2})$ originating some $\calz_{2} \in \ms{Res}_{{\rm max}, \alpha}(s_{2}, o)$:
\cws{0}{\begin{array}{rcccl}
\ms{prob}(\calfcc(z'_{s_{1}}, \varphi)) & \!\!\! = \!\!\! & \ms{prob}(\calscc(z_{s_{1}, o}, \alpha)) / p &
\!\!\! \neq \!\!\! & \\
& \!\!\! \neq \!\!\! & \ms{prob}(\calscc(z_{s_{2}, o}, \alpha)) / p & \!\!\! = \!\!\! &
\ms{prob}(\calfcc(z'_{s_{2}}, \varphi)) \\
\end{array}}
where $p$ is the probability of performing the $\alpha$-compatible computations in the only element $\calz$
of~$\ms{Res}(o)$ that originates $\calz_{1}$ and all the resolutions $\calz_{2}$. \\
The proof that $s_{1} \sbis{\rm PF}^{\rm ct} s_{2}$ implies $s_{1} \sbis{\textrm{PTe-tbt}}^{\rm ct} s_{2}$
is similar.
\qed

	\end{thm}

The inclusion of $\sbis{\rm PF}$ (resp.\ $\sbis{\rm PF}^{\rm ct}$) in $\sbis{\textrm{PTe-tbt}}$ (resp.\
$\sbis{\textrm{PTe-tbt}}^{\rm ct}$) is strict, because the initial states of the two NPLTS models in
Fig.~\ref{fig:counterex_ptrdis_ptr} are equated by the latter equivalence and told apart by the former.
For instance, the rightmost maximal resolution of $s_{1}$ has probability~$1$ of performing a computation
compatible with the failure pair $(a, \{ b_{1}, b_{2} \})$, whilst each of the two maximal resolutions of
$s_{2}$ has probability~$0.5$.

	\begin{figure}[tp]

\centerline{\includegraphics{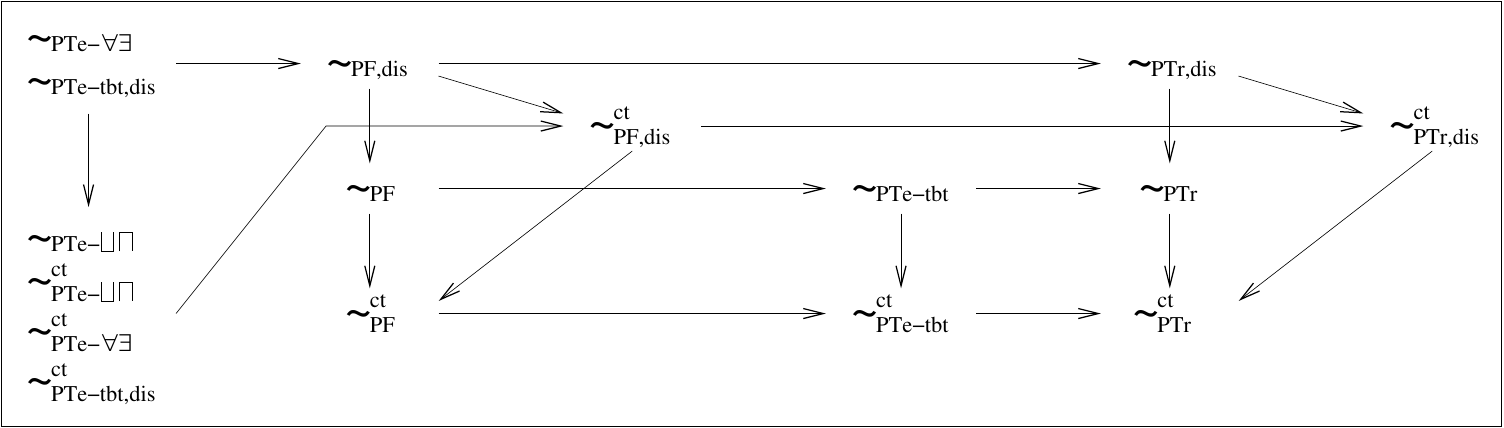}}
\caption{The spectrum of testing, failure, and trace equivalences for NPLTS models}
\label{fig:spectrum}

	\end{figure}

The relationships among the various probabilistic testing, failure, and trace equivalences for NPLTS models
are summarized in Fig.~\ref{fig:spectrum}. Arrows represent the more-discriminating-than partial order,
equivalences close to each other coincide, and incomparability is denoted by the absence of (chains of)
arrows. The various relationships have been established in this paper, except for the arrow from
$\sbis{\textrm{PTe-}\sqcup\sqcap}$ to $\sbis{\rm PF,dis}^{\rm ct}$ that is due to~\cite{Seg96}.

We observe that $\sbis{\textrm{PTe-}\sqcup\sqcap}$ is incomparable not only with $\sbis{\textrm{PTe-tbt}}$
as established right before Thm.~\ref{thm:pteallexists_incl_ptetbt}, but also with $\sbis{\rm PF,dis}$,
$\sbis{\rm PF}$, $\sbis{\rm PTr,dis}$, and $\sbis{\rm PTr}$. In fact, in
Fig.~\ref{fig:counterex_ptesupinf_trace} it holds that $s_{1} \sbis{\textrm{PTe-}\sqcup\sqcap} s_{2}$ while
$s_{1} \not\sbis{\rm PF,dis} s_{2}$, $s_{1} \not\sbis{\rm PF} s_{2}$, $s_{1} \not\sbis{\rm PTr,dis} s_{2}$,
and $s_{1} \not\sbis{\rm PTr} s_{2}$. On the other hand, in Fig.~\ref{fig:counterex_tefnd_testing} it holds
that $s_{1} \not\sbis{\textrm{PTe-}\sqcup\sqcap} s_{2}$ while $s_{1} \sbis{\rm PF,dis} s_{2}$, $s_{1}
\sbis{\rm PF} s_{2}$, $s_{1} \sbis{\rm PTr,dis} s_{2}$, and $s_{1} \sbis{\rm PTr} s_{2}$.

Likewise, $\sbis{\rm PF,dis}^{\rm ct}$ is incomparable not only with $\sbis{\rm PF}$ as established right
after Thm.~\ref{thm:pfdis_incl_pf}, but also with $\sbis{\rm PTr,dis}$, $\sbis{\textrm{PTe-tbt}}$, and
$\sbis{\rm PTr}$. Indeed, in Fig.~\ref{fig:counterex_ptesupinf_trace} it holds that $s_{1} \sbis{\rm
PF,dis}^{\rm ct} s_{2}$ while $s_{1} \not\sbis{\rm PTr,dis} s_{2}$, $s_{1} \not\sbis{\textrm{PTe-tbt}}
s_{2}$, and $s_{1} \not\sbis{\rm PTr} s_{2}$. In contrast, in Fig.~\ref{fig:counterex_pfdis_pf} it holds
that $s_{1} \not\sbis{\rm PF,dis}^{\rm ct} s_{2}$ while $s_{1} \sbis{\rm PTr,dis} s_{2}$, $s_{1}
\sbis{\textrm{PTe-tbt}} s_{2}$, and $s_{1} \sbis{\rm PTr} s_{2}$. Moreover, $\sbis{\rm PF}^{\rm ct}$ is
incomparable with $\sbis{\rm PTr,dis}$ and $\sbis{\rm PTr,dis}^{\rm ct}$. In fact, in
Fig.~\ref{fig:counterex_pfdis_pf} it holds that $s_{1} \sbis{\rm PF}^{\rm ct} s_{2}$ while $s_{1}
\not\sbis{\rm PTr,dis} s_{2}$ and $s_{1} \not\sbis{\rm PTr,dis}^{\rm ct} s_{2}$. On the other hand, in
Fig.~\ref{fig:counterex_pteallexists_trace} it holds that $s_{1} \not\sbis{\rm PF}^{\rm ct} s_{2}$ while
$s_{1} \sbis{\rm PTr,dis} s_{2}$ and $s_{1} \sbis{\rm PTr,dis}^{\rm ct} s_{2}$. Additionally, $\sbis{\rm
PF}^{\rm ct}$ is incomparable with $\sbis{\rm PTr}$ because in Fig.~\ref{fig:counterex_ptesupinf_trace} we
have that $s_{1} \sbis{\rm PF}^{\rm ct} s_{2}$ and $s_{1} \not\sbis{\rm PTr} s_{2}$, whereas in
Fig.~\ref{fig:counterex_pteallexists_trace} we have that $s_{1} \not\sbis{\rm PF}^{\rm ct} s_{2}$ and $s_{1}
\sbis{\rm PTr} s_{2}$. Furthermore, $\sbis{\rm PF}^{\rm ct}$ is incomparable also with
$\sbis{\textrm{PTe-tbt}}$ because in Fig.~\ref{fig:counterex_ptesupinf_trace} we have that $s_{1} \sbis{\rm
PF}^{\rm ct} s_{2}$ and $s_{1} \not\sbis{\textrm{PTe-tbt}} s_{2}$, whilst in
Fig.~\ref{fig:counterex_ptrdis_ptr} we have that $s_{1} \not\sbis{\rm PF}^{\rm ct} s_{2}$ and $s_{1}
\sbis{\textrm{PTe-tbt}} s_{2}$.

Analogously, $\sbis{\rm PTr,dis}^{\rm ct}$ is incomparable not only with $\sbis{\rm PTr}$ as established
right after Thm.~\ref{thm:ptrdis_incl_ptr}, but also with $\sbis{\rm PF}$, $\sbis{\textrm{PTe-tbt}}$, and
$\sbis{\textrm{PTe-tbt}}^{\rm ct}$. It holds that $s_{1} \sbis{\rm PTr,dis}^{\rm ct} s_{2}$ and $s_{1}
\not\sbis{\rm PF} s_{2}$, $s_{1} \not\sbis{\textrm{PTe-tbt}} s_{2}$, and $s_{1}
\not\sbis{\textrm{PTe-tbt}}^{\rm ct} s_{2}$ in Fig.~\ref{fig:counterex_pteallexists_trace}, while $s_{1}
\not\sbis{\rm PTr,dis}^{\rm ct} s_{2}$ and $s_{1} \sbis{\rm PF} s_{2}$, $s_{1} \sbis{\textrm{PTe-tbt}}
s_{2}$, and $s_{1} \sbis{\textrm{PTe-tbt}}^{\rm ct} s_{2}$ in Fig.~\ref{fig:counterex_pfdis_pf}. The same
two figures show that also $\sbis{\rm PTr,dis}$ is incomparable with $\sbis{\rm PF}$,
$\sbis{\textrm{PTe-tbt}}$, and $\sbis{\textrm{PTe-tbt}}^{\rm ct}$. Finally, we have that $\sbis{\rm PTr}$ is
incomparable with $\sbis{\textrm{PTe-tbt}}^{\rm ct}$ because in Fig.~\ref{fig:counterex_pteallexists_trace}
it holds that $s_{1} \sbis{\rm PTr} s_{2}$ and $s_{1} \not\sbis{\textrm{PTe-tbt}}^{\rm ct} s_{2}$, whereas
in Fig.~\ref{fig:counterex_ptesupinf_trace} it holds that $s_{1} \not\sbis{\rm PTr} s_{2}$ and $s_{1}
\sbis{\textrm{PTe-tbt}}^{\rm ct} s_{2}$.

We conclude by recalling another probabilistic testing equivalence that has been recently proposed
in~\cite{GA12}, where a probabilistic model significantly different from ours is considered. Unfortunately,
the differences prevent us from placing that equivalence in the spectrum we have just presented. However,
that testing equivalence shares with our $\sbis{\textrm{PTe-tbt}}$ motivations and intuitions concerning the
power of schedulers and the estimation of success probabilities that call for further comments.

The model considered in~\cite{GA12} has three types of transitions: action transitions, internal
transitions, and probabilistic transitions. Since each state can have only one type of outgoing transitions,
also states are divided into three classes: action states, nondeterministic states, and probabilistic
states. Action states cannot have two identically labeled action transitions, so this model can be viewed as
a variant of reactive probabilistic processes in which states of different classes can alternate along a
computation. Notice that our NPLTS model is non-alternating, because there is a single class of states and
probabilistic choices are somehow embedded within each single transition.

In order to make the proposed testing theory insensitive to the exact moment in which internal choices
occur, in~\cite{GA12} internal transitions are decorated with so-called internal labels. Similar to action
states, nondeterministic states cannot have two identically labeled internal transitions. Moreover, given
two nondeterministic states, either they share the same set of internal labels decorating their outgoing
transitions, or the sets of internal labels of their outgoing transitions are disjoint. Internal labels are
meant to provide precisely the information that schedulers should use to resolve internal choices, so that
internal choices relying on the same information are resolved in the same way. For example, continuing the
discussion done in the last two paragraphs of Sect.~\ref{sec:testing_equiv}, with the approach
of~\cite{GA12} the two internal choices between the two $b$-transitions in the interaction system with
initial configuration $(s_{1}, o)$ of Figs.~\ref{fig:counterex_tefnd_testing}
and~\ref{fig:counterex_tepr_testing} would be identically tagged, say with $b_{l}$ and~$b_{r}$ based on the
orientation of the arrows. As a consequence, the only allowed maximal resolutions of that interaction system
among the four shown in Figs.~\ref{fig:counterex_tefnd_testing_max_res}
and~\ref{fig:counterex_tepr_testing_max_res} would be the first one (choice of $b_{l}$) and the fourth one
(choice of~$b_{r}$), thus excluding success probabilities~$1$ and~$0$.

An important technical point made in~\cite{GA12} is that, in the presence of cycles of transitions within
the model, the same internal choice may occur several times along a computation. This is not due to the
copying capability that arises when composing in parallel a process and a test, which -- as we have recalled
above -- is dealt with by labeling in the same way the internal transitions departing from all the copies of
the cloned state and by forcing schedulers to perform consistent choices in all the copies (we will refer to
the resulting fully probabilistic models as \emph{consistent resolutions}). Replications of the same
internal choice at different \emph{unfolding depths} of a cycle are independent of each other and are thus
given additional labels that keep them distinct from depth to depth. Notice that, in contrast, our approach
based on $\sbis{\textrm{PTe-tbt}}$ is not invasive at all, as it does not require any label massaging on the
model to restrict the power of schedulers.

	\begin{figure}[tp]

\input{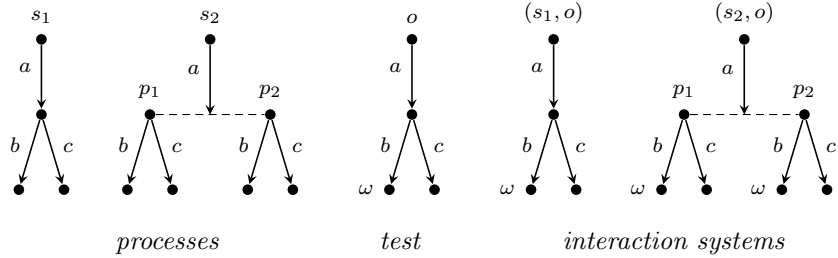}
\caption{NPLTS models equated by~\cite{GA12} and distinguished by $\sbis{\textrm{PTe-tbt}}$}
\label{fig:counterex_ptetbt_ga}

	\end{figure}

	\begin{figure}

\input{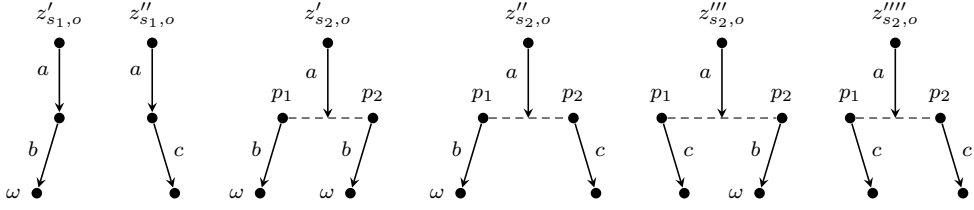}
\caption{Maximal resolutions of the two interaction systems in Fig.~\ref{fig:counterex_ptetbt_ga}}
\label{fig:counterex_ptetbt_ga_max_res}

	\end{figure}

Two processes are equated by the testing equivalence proposed in~\cite{GA12} iff, for each test, every
consistent resolution at unfolding depth $m$ of a suitably labeled version of the first interaction system
that reaches success with probability $p$, is matched by a consistent resolution at the same unfolding depth
of a suitably labeled version of the second interaction system that reaches success with the same
probability. This equivalence cannot be directly applied to NPLTS models. Since a major difference with
$\sbis{\textrm{PTe-tbt}}$ is the use of restricted schedulers, an adaptation of the testing equivalence
of~\cite{GA12} to a common model should lead to an equivalence that is coarser than
$\sbis{\textrm{PTe-tbt}}$.

It can however be shown that the two equivalences are different if attention is restricted to a common
submodel that does not permit internal nondeterminism. Indeed, absence of internal nondeterminism makes
label massaging unnecessary, and we have that reactive probabilistic processes constitute the largest
submodel common to the model of~\cite{GA12} and NPLTS. Consider the two reactive probabilistic processes
depicted as NPLTS models in Fig.~\ref{fig:counterex_ptetbt_ga}, and suppose that what is called
synchronization nondeterminism in~\cite{GA12} is handled without using $\tau$ inside the labels of the
transitions of the interaction systems. The two processes are discriminated by $\sbis{\textrm{PTe-tbt}}$
because, if we consider the test in the same figure and the maximal resolutions shown in
Fig.~\ref{fig:counterex_ptetbt_ga_max_res} of the interaction systems, the success probability $p_{1}$ of
trace $a \, b$ in the second maximal resolution of $(s_{2}, o)$ is not matched by the success probability
$1$ of the only maximal resolution of $(s_{1}, o)$ having a maximal computation labeled with $a \, b$. In
contrast, the testing equivalence of~\cite{GA12} cannot distinguish the two processes. \linebreak Whenever
they remain in the interaction system with an arbitrary test, the two identical choices between $b$ and $c$
in the second process must be resolved in the same way by any restricted scheduler that can only yield
consistent resolutions. For instance, the only maximal resolutions of $(s_{2}, o)$ that are consistent among
the four shown in Fig.~\ref{fig:counterex_ptetbt_ga_max_res} are the first one (choice of $b$) and the
fourth one (choice of~$c$), and their respective success probabilities $1$ and $0$ are precisely matched by
those of the only two maximal resolutions of $(s_{1}, o)$.



%
%
\section{Conclusion}
\label{sec:concl}
%
%

In this paper, we have proposed two variants of trace and testing equivalences, respectively denoted by
$\sbis{\rm PTr}$ and $\sbis{\textrm{PTe-tbt}}$, for the general class of nondeterministic and probabilistic
processes, which enjoy desirable properties like:

	\begin{enumerate}

\item being preserved by parallel composition, 

\item being fully conservative extensions of the corresponding equivalences studied for nondeterministic
processes and for probabilistic processes, and

\item guaranteeing that trace equivalence is coarser than testing equivalence.

	\end{enumerate}

\noindent
For both equivalences, we have assumed history-independent centralized schedulers. In particular, we have
considered the impact of employing deterministic schedulers or randomized schedulers to resolve
nondeterminism. We have denoted by $\sbis{\rm PTr}^{\rm ct}$ and $\sbis{\textrm{PTe-tbt}}^{\rm ct}$ the
equivalence variants based on randomized schedulers.

The most studied trace and testing equivalences known in the literature of nondeterministic and
probabilistic processes, namely the probabilistic trace-distribution equivalence $\sbis{\rm PTr,dis}^{\rm
ct}$ investigated in~\cite{Seg95b,CSV07,LSV03,PS04,CLSV06} and the probabilistic testing equivalence
$\sbis{\textrm{PTe-}\sqcup\sqcap}$ investigated in~\cite{YL92,JY95,Seg96,DGHM08}, do not fulfill all of
these properties. In particular, $\sbis{\rm PTr,dis}^{\rm ct}$ is not a congruence with respect to parallel
composition and $\sbis{\textrm{PTe-}\sqcup\sqcap}$ is not a fully conservative extension of the testing
equivalences defined in~\cite{DH84} for fully nondeterministic processes, in~\cite{CDSY99} for generative
probabilistic processes, and in~\cite{KN98} for reactive probabilistic processes. Moreover, while the
discriminating power of $\sbis{\textrm{PTe-}\sqcup\sqcap}$ is independent from the use of deterministic of
randomized schedulers, the inclusion of this testing equivalence in the trace-distribution equivalence
heavily depends on the use of randomized schedulers when defining the trace semantics. Specifically, we have
that $\sbis{\textrm{PTe-}\sqcup\sqcap}$ is contained in $\sbis{\rm PTr,dis}^{\rm ct}$ but not in $\sbis{\rm
PTr,dis}$, being the former based on randomized schedulers and the latter on deterministic schedulers.

The main idea behind the new trace equivalence $\sbis{\rm PTr}$ that we have proposed is that of comparing
the execution probabilities of single traces rather than entire trace distributions, so as to avoid
debatable distinctions such as the one made by $\sbis{\rm PTr,dis}$ in Fig.~\ref{fig:counterex_ptrdis_ptr}.
This requires a shift from considering fully matching resolutions to considering partially matching
resolutions, which opens the way to compositionality under centralized schedulers.

The main ideas behind the new testing equivalence $\sbis{\textrm{PTe-tbt}}$ are: (i)~matching all
resolutions on the basis of their success probabilities, rather than taking into account only maximal and
minimal success probabilities, and (ii) considering success probabilities in a trace-by-trace fashion,
rather than cumulatively on entire resolutions. It is the trace-by-trace approach that annihilates the
impact of the copying capability introduced by observers not of the same nature as the processes under test,
and thus permits defining an equivalence that is fully conservative with respect to classical testing
equivalences. Remarkably, we have seen in Thm.~\ref{thm:ptetbt_compat} that our new approach, when
restricted to fully nondeterministic processes, generative probabilistic processes, and reactive
probabilistic processes, yields the same testing equivalences longly studied in the literature.

In order to get to the trace-by-trace approach, it has been important to pass through an additional testing
semantics, $\sbis{\textrm{PTe-}\forall\exists}$, which is not fully backward compatible with testing
semantics for restricted classes of processes but, unlike $\sbis{\textrm{PTe-}\sqcup\sqcap}$, it implies
trace semantics. This testing semantics does act as a trait d'union between the testing semantics focussing
only on extremal success probabilities -- because $\sbis{\textrm{PTe-}\forall\exists}^{\rm ct}$ coincides
with $\sbis{\textrm{PTe-}\sqcup\sqcap}$ -- and our new fully backward compatible testing semantics comparing
success probabilities trace-by-trace -- because $\sbis{\textrm{PTe-}\forall\exists}$ coincides with
$\sbis{\textrm{PTe-tbt,dis}}$.

Another interesting result about testing semantics is that using randomized schedulers to resolve
nondeterminism annihilates the difference between many equivalences. Indeed, we have that
$\sbis{\textrm{PTe-tbt,dis}}^{\rm ct}$ coincides with $\sbis{\textrm{PTe-}\forall\exists}^{\rm ct}$ and with
$\sbis{\textrm{PTe-}\sqcup\sqcap}^{\rm ct}$, which in turn coincides with
$\sbis{\textrm{PTe-}\sqcup\sqcap}$, its variant based on deterministic schedulers. Thus,
$\sbis{\textrm{PTe-tbt,dis}}^{\rm ct}$ constitutes an alternative characterization of
$\sbis{\textrm{PTe-}\sqcup\sqcap}$, a fact that reconciles the testing equivalence deeply investigated in
the literature with the three approaches recently explored in~\cite{BDL14} to the definition of behavioral
relations for NPLTS models.

We would like to mention that $\sbis{\rm PTr}$ and $\sbis{\textrm{PTe-tbt}}$ did pop up when working in the
framework of \ultras~\cite{BDL13a}. This is a parametric model encompassing many others such as labeled
transition systems, discrete-/continuous-time Markov chains, and discrete-/continuous-time Markov decision
processes without/with internal nondeterminism. On this unifying model, we have defined trace, testing, and
bisimulation equivalences in an abstract way and shown that they induce new equivalences (like $\sbis{\rm
PTr}$ and $\sbis{\textrm{PTe-tbt}}$) different from those known in the literature (like $\sbis{\rm
PTr,dis}^{\rm ct}$ and $\sbis{\textrm{PTe-}\sqcup\sqcap}$) when instantiating the model to the NPLTS case.

In this paper, we have also studied the relationships between our new testing semantics and previously
defined failure semantics for nondeterministic and probabilistic processes. While in the fully
nondeterministic case the two semantics coincide~\cite{DeN87}, we have shown that
$\sbis{\textrm{PTe-tbt,dis}}$ is strictly finer than $\sbis{\rm PF,dis}$, while $\sbis{\textrm{PTe-tbt}}$ is
strictly coarser than $\sbis{\rm PF}$. We conjecture that the former two equivalences and the latter two
equivalences respectively coincide if, in the trace-by-trace approach, we compare not only trace-based
probabilities of reaching success, but also failure probabilities, i.e., the probabilities of performing
maximal computations compatible with a certain trace that do not reach success.

As future work, we plan to study equational and logical characterizations of the new trace and testing
equivalences that we have introduced in this paper.

\section*{Acknowledgement}
We would like to thank the anonymous referees for their stimulating comments and Marco Tinacci for his
useful suggestions on the comparison with~\cite{GA12}. This work has been partially supported by the
FP7-IST-FET Project ASCENS, grant no.~257414, by the EU Project QUANTICOL, grant no.~600708, and by the MIUR
project CINA.

\bibliographystyle{plain}
\bibliography{lmcs}

\vspace{-30 pt}
\end{document}